\PassOptionsToPackage{sort&compress}{natbib}
\documentclass[preprint,12pt]{elsarticle}

\usepackage{amssymb}
\usepackage{amsmath}
\usepackage{graphicx}
\usepackage{subcaption}
\usepackage{amssymb}
\usepackage{hyperref}
\usepackage{tabularray}
\usepackage{physics}
\usepackage[dvipsnames]{xcolor}
\usepackage{{booktabs}}
\usepackage{siunitx}
\usepackage{mathtools}
\usepackage{fullpage}

\usepackage[colorinlistoftodos]{todonotes}
\presetkeys{todonotes}{inline}{}

\usepackage{listings}

\newcommand{\bt}[1]{\textbf{\textrm{#1}}}

\definecolor{mygreen}{RGB}{0,0,0}
\definecolor{mygreen2}{RGB}{0,0,0}

\begin{document}

\begin{frontmatter}



\title{Similarity Equivariant Graph Neural Networks for Homogenization of Metamaterials}


\author[label1,label4]{Fleur Hendriks}
\author[label2,label4]{Vlado Menkovski}
\author[label3]{Martin Do\v{s}k\'{a}\v{r}}
\author[label1,label4]{Marc G.D. Geers}
\author[label1,label4]{Ond\v{r}ej Roko\v{s}}

\affiliation[label1]{organization={Mechanics of Materials, Department of Mechanical Engineering, Eindhoven University of Technology},
            addressline={Postbus 513},
            postcode={5600 MB},
            city={Eindhoven},
            country={The Netherlands}}
\affiliation[label2]{organization={Data Mining, Department of Mathematics and Computer Science, Eindhoven University of Technology},
            addressline={Postbus 513},
            postcode={5600 MB},
            city={Eindhoven},
            country={The Netherlands}}
\affiliation[label3]{organization={Department of Mechanics, Faculty of Civil Engineering, Czech Technical University in Prague},
            addressline={Thákurova 7},
            postcode={166 29},
            city={Prague 6},
            country={Czech Republic}}
\affiliation[label4]{organization={Eindhoven AI Systems Institute (EAISI)},
            addressline={PO Box 513},
            postcode={5600 MB},
            city={Eindhoven},
            country={The Netherlands}}

\begin{abstract}
Soft, porous mechanical metamaterials exhibit pattern transformations that may have important applications in soft robotics, sound reduction and biomedicine.
To design these innovative materials, it is important to be able to simulate them accurately and quickly, in order to tune their mechanical properties. Since conventional simulations using the finite element method entail a high computational cost, in this article we aim to develop a machine learning-based approach that scales favorably to serve as a surrogate model.
To ensure that the model is also able to handle various microstructures, including those not encountered during training, we include the microstructure as part of the network input. Therefore, we introduce a graph neural network that predicts global quantities (energy, stress\textcolor{mygreen}{,} stiffness) as well as the pattern transformations that occur (the kinematics) \textcolor{mygreen}{in hyperelastic, two-dimensional, microporous materials}. \textcolor{mygreen}{Predicting these pattern transformations means predicting the displacement field.}
To make our model as accurate and data-efficient as possible, various symmetries are incorporated into the model.
The starting point is an $E(n)$-equivariant graph neural network (which respects translation, rotation and reflection) that has periodic boundary conditions (i.e., it is in-/equivariant with respect to the choice of RVE), is scale in-/equivariant, can simulate large deformations, and can predict scalars, vectors as well as second and fourth order tensors (specifically energy, stress and stiffness).
The incorporation of scale equivariance makes the model equivariant with respect to the similarities group, of which the Euclidean group $E(n)$ is a subgroup. We show that this network is more accurate and data-efficient than graph neural networks with fewer symmetries.
To create an efficient graph representation of the finite element discretization, we use only the internal geometrical hole boundaries from the finite element mesh to achieve a better speed-up and scaling with the mesh size.

\end{abstract}


\begin{highlights}
\item Development of a Similarity-Equivariant Graph Neural Network (SimEGNN) for homogenization of metamaterials
\item $E(n)$-Equivariant Graph Neural Networks extended to higher-order tensors
\item Incorporation of all relevant symmetry groups to achieve similarity in-/equivariance
\item Formulation \& implementation of Representative Volume Element in-/equivariance (periodic boundary conditions) in a graph neural network
\item Efficient graph representation of the finite element mesh; speed-up and better scaling with mesh size
\end{highlights}

\begin{keyword}
Graph Neural Networks
\sep Similarity Equivariance
\sep $E(n)$-equivariance
\sep Periodicity
\sep Computational Homogenization
\sep Mechanical Metamaterials



\end{keyword}

\end{frontmatter}


\section{Introduction}
	Metamaterials are materials with a microstructure designed to exhibit special properties. For instance, mechanical metamaterials have unusual mechanical properties \cite{Bertoldi2017, Yu2018}, such as a negative Poisson's ratio \cite{Babaee2013} or negative compressibility \cite{Nicolaou2012}. Here, we are specifically interested in flexible, porous mechanical metamaterials that have a tunable stiffness as a result of a pattern transformation \cite{Boyce2008, Overvelde2014} that can be activated, for example, by mechanical loading, pneumatics \cite{Yang2015} or magnetic fields \cite{Kim2018}.
	Such pattern transformations can also change the acoustic properties, allowing the design of mechanically tunable acoustic metamaterials \cite{GuellIzard2020, Ning2021, Montgomery2021, Wu2021}, or they can be used to control the shape of the material, with applications in soft robotics \cite{Yang2015, Terryn2017, Kim2019}. They also have applications in biomedicine \cite{Veerabagu2022}.

	\label{par:designing_mechmetamat}
	These kinds of pattern-transforming materials provide a large design space to be explored, because there are a lot of possibilities regarding geometry (shape and number of holes and inclusions), electromagnetic, chemical and mechanical properties of the base material, and different means of loading and activation.
	Moreover, thanks to the recent advancements in 3D printing \cite{Jiang2016, DAlessandro2016, Shi2017, Ren2018, Montgomery2020, Bodaghi2020}, the broad design space is no longer inherently restricted by manufacturing concerns. Consequently, being able to design mechanical metamaterials for a specific target response (i.e., having a prescribed stiffness and buckling exactly when needed) is highly appealing.
	\label{par:topology_optimization}
	This can be achieved with shape \cite{Wang2017, Medina2023} and/or topology optimization \cite{Bendsøe1988, Bendsøe2004} on the Representative Volume Element (RVE) \cite{Sigmund1994}. Topology optimization can be used to design a microstructure for various homogenized properties \cite{Andreassen2014, Chen2018d}, including the buckling strength of materials \cite{Thomsen2018, Wang2021, Xue2022b}.
	Because topology and/or shape optimization are almost always iterative, being able to rapidly evaluate the performance of new designs using numerical simulations is critical for accelerating the design process.

	\label{par:homogenization_difficult}
	However, the behavior of the metamaterials of interest is complex, highly nonlinear, and involves large deformations, which requires repeated evaluation of the constitutive law of the matrix material. The buckling behavior that enables pattern transformations makes the system even more complicated to solve (here avoided by using a perturbation). Moreover, to capture all fine microstructural details, the discretization needs to have a sufficiently high resolution. For these reasons, the numerical simulations are typically expensive to evaluate, which makes them too inefficient to explore the large design space. Therefore, the main objective of this article is to build an efficient, accurate and fast-to-evaluate machine learning-based surrogate model for the modeling of periodic elastomeric mechanical metamaterials with a highly non-linear response.

	Given the goal is to develop a model enabling the design of new mechanical metamaterials, the model must have the potential to generalize to unseen, arbitrary microstructures, even including topology changes. This implies that the surrogate model should be able to take a description of the full geometry as input. This rules out other approaches, such as analytical (closed-form) homogenization methods (e.g. mixture methods \cite{Hill1963} or methods based on Eshelby's \cite{Eshelby1957} analytical result for an ellipsoidal inclusion in infinite matrix \cite{Hill1965, Mori1973}), reduced order models \cite{Lucia2004, Kerschen2005, ElHalabi2013,Yvonnet2007, Hernandez2014,Gogu2014, Guo2018, Guo2021a, Guo2022}, data-driven approaches \cite{Kirchdoerfer2016}, self-consistent clustering analysis \cite{Liu2016}, relatively simple machine learning approaches that map deformation gradient (and optionally some microstructural parameters) to global quantities \cite{Liang2008,Le2015a, Linka2021}, and graph networks (not to be confused with graph neural networks) \cite{Xue2022}, which model porous mechanical metamaterials as rigid crosses connected by neural network-modeled springs.

	These methods are certainly valuable in their developed context, but only work very well in the linear-elastic regime or only for a fixed geometry and topology.
	One reason neural networks are promising is the fact that they are universal approximators \cite{Cybenko1989a, Hornik1991, Leshno1992}, which means they can approximate any arbitrary function and are therefore not constrained to any specific form of macroscopic constitutive law. Specifically, we are interested in deep learning solutions, since deep learning excels at using high-dimensional data \cite{Peng2021a}, which allows us to use a full description of the geometry of the RVE as an input and which also scales favorably (often linearly) with the size of the input and the number of parameters. Especially in 3D, the scaling advantage could be considerable.

	\label{par:constraints}
	However, creating an efficient deep learning model is not trivial.
	Specifically, because the following properties need to be incorporated into the architecture of the model:
    \begin{itemize}
		\item The model should be able to handle large deformations (to cover buckling), high strains \textcolor{mygreen}{\emph{and}} rotations.
		\item If the input is a mesh, the model should be in-/equivariant under permutation of nodes (the order of nodes should not matter). The concepts of invariance and equivariance are defined and explained in detail in Section~\ref{sec:symmetries}.
        \item Similarity in-/equivariance ($E(n)$ in-/equivariance, i.e., rotation, translation, reflection symmetry), to satisfy material objectivity. These transformations correspond to choosing a different coordinate system. This is also one of the requirements of the principle of material frame indifference, also called objectivity \cite[p.195]{Tadmor2012}.\footnote{Usually, when discussing material frame indifference, reflection symmetry is omitted. However, it still applies, because changing the orientation of the coordinate system does not change the behavior of the material, and that is why, for completeness, we include it here as well.}
        \item Scale in-/equivariance; since we limit ourselves to first-order computational homogenization of a hyperelastic material (Bertoldi-Boyce \cite{Boyce2008}, Equation~\eqref{eq:bertoldiboyce}), a change in scale of the RVE should not change its behavior.
        \item Periodicity, which implies in-/equivariance with respect to a shift of the RVE (i.e., RVE window translation) and with respect to merging multiple RVEs into a new, bigger one. This assumes the RVE is already big enough to capture all relevant buckling patterns. If not, a larger RVE could allow for a new buckling pattern, which breaks the in-/equivariance with respect to merging multiple RVEs.
        \item Prediction of scalars (energy), vectors (displacement), and higher-order tensors (stress and stiffness).
    \end{itemize}

	To the best of our knowledge, there are no existing machine-learning based models that satisfy all these constraints. See Table~\ref{tab:modelcomparison} for an overview of the constraints respected by different types of models.

	\textcolor{mygreen}{
		The in-/equivariances we mention above are all related to geometric symmetries. However, there are other possible constraints that can be incorporated, such as (poly-)convexity as formulated by Ball \cite{Ball}, symmetry of the stress tensor and the condition that in the reference configuration, the stress tensor is zero. There has been a lot of work in this area already \cite{Fernandez2021, Linka2021, Linden2023, Linka2023}, specifically concerning polyconvexity, usually implemented using approaches based on input-convex neural networks (ICNNs) by Amos et al \cite{Amos2017}, such as the approach by Klein et al. \cite{Klein2022,  Klein2023}. These all concern constitutive relations where the deformation gradient $\bt F$ or right Cauchy-Green deformation tensor $\bt C = \bt F^T \cdot \bt F$ is mapped to stress or stiffness. These models do not take the full geometry as input, but could possibly be incorporated into a model that does. However, in our case, our potential energy density is not convex, because of the presence of instabilities due to buckling.
	}

	\begin{table}
		\scriptsize
		\centering
		\caption{Overview of GNN architectures and the constraints they address, along with the presently developed EGNN and similarity-equivariant GNN (SimEGNN).
		\label{tab:modelcomparison}}
		\begin{tblr}{
  			hline{2} = {3-8}{},
  			hline{3} = {7-8}{},
			width = \linewidth,
			colspec = {Q[l,150]Q[l,140]Q[120]Q[98]Q[98]Q[98]Q[104]Q[100]},
			cell{1}{3} = {c=6}{0.53\linewidth,c},
			cell{2}{7} = {c=2}{0.17\linewidth,c},
			cell{8}{1} = {c=8}{\linewidth},
			cell{12}{1} = {c=8}{\linewidth},
			cell{13}{1} = {c=8}{\linewidth},
		}
			\hline
			&  & Invariances/equivariances &  &  &  &  & \\
			Architecture & Updates node positions? & Translation & Rotation & Reflection & Scale & Periodicity & \\
			& &  &  &  &   & Shifted RVE & Extended RVE\\
			\hline
			CNN* & {\color{red}\sffamily X} & {\color{ForestGreen}\checkmark} & {\color{red}\sffamily X} & {\color{red}\sffamily X} & {\color{ForestGreen}\sffamily {\color{ForestGreen}\checkmark}} & {\color{ForestGreen}\checkmark} & {\color{ForestGreen}\checkmark}\\
			GNN & \sffamily {\color{ForestGreen}\checkmark} & {\color{ForestGreen}\checkmark} & {\color{red}\sffamily X} & {\color{red}\sffamily X} & {\color{red}\sffamily X} & {\color{ForestGreen}\checkmark} & {\color{ForestGreen}\checkmark}\\
			GNN** & {\color{red}\sffamily X} & {\color{ForestGreen}\checkmark} & {\color{ForestGreen}\checkmark} & {\color{ForestGreen}\checkmark} & {\color{red}\sffamily X} & {\color{ForestGreen}\checkmark} & {\color{ForestGreen}\checkmark}\\
			Original EGNN & {\color{ForestGreen}\checkmark} & {\color{ForestGreen}\checkmark} & {\color{ForestGreen}\checkmark} & {\color{ForestGreen}\checkmark} & {\color{red}\sffamily X} & {\color{red}\sffamily X} & {\color{red}\sffamily X}\\
			\hline
			\textcolor{mygreen}{\textbf{From the present study:}} & & & & & & & \\
			GNN & {\color{ForestGreen}\checkmark} & {\color{ForestGreen}\checkmark} & {\color{red}\sffamily X} & {\color{red}\sffamily X} & {\color{red}\sffamily X} & {\color{ForestGreen}\checkmark} & {\color{ForestGreen}\checkmark}\\
			EGNN & {\color{ForestGreen}\checkmark} & {\color{ForestGreen}\checkmark} & {\color{ForestGreen}\checkmark} & {\color{ForestGreen}\checkmark} & {\color{red}\sffamily X} & {\color{ForestGreen}\checkmark} & {\color{ForestGreen}\checkmark}\\
			SimEGNN & {\color{ForestGreen}\checkmark} & {\color{ForestGreen}\checkmark} & {\color{ForestGreen}\checkmark} & {\color{ForestGreen}\checkmark} & {\color{ForestGreen}\checkmark} & {\color{ForestGreen}\checkmark} & {\color{ForestGreen}\checkmark} \\
			\hline
			\textcolor{mygreen}{
			{}*periodicity and translation in-/equivariance are only exactly respected if there is no pooling or dense layers (which is possible but unusual), otherwise the CNN is only approximately translation in-/equivariant. Periodicity additionally needs circular padding.}& & & & & & &\\
			\textcolor{mygreen}{
			{}**assuming only $E(n)$-invariant attributes such as distances and angles are used.
			}& & & & & & &
		\end{tblr}
	\end{table}

	\label{par:CNNs}
	\textcolor{mygreen}{The convolutional layers of convolutional neural networks (CNNs) respect translation in-/equivariance and can respect periodicity when used with circular/periodic padding. CNNs applied to 2D or 3D images of microstructures \cite{Frankel2019, Wilt2020a, Pandey2021, Yang2021, Mianroodi2021, Mianroodi2022, Khorrami2023} therefore approximately respect these symmetries.} \footnote{\textcolor{mygreen}{CNN with circular padding will not be exactly in-/equivariant with respect to shifting the RVE, because of the max pooling and the flattening that usually happens at the end of the convolutional layers, which is fed into the dense layers afterwards.}} They also respect scale in-/equivariance, because the input picture does not change in that case. They can be made to respect rotation and reflection in-/equivariance \cite{Cohen2016}, but not for any arbitrary rotation or reflection; only with respect to rotations that are multiples of \ang{90} and reflections that are horizontal or vertical. In addition, they can only predict vectors and higher-order tensors by treating them as an arbitrary list of numbers (that therefore do not necessarily transform as tensors), they cannot track the movement of nodes, and their geometry input is constrained to a square grid.

	\label{par:our_goal}
	A promising alternative is a graph neural network (GNN): a type of neural network that takes a graph as input. The mesh representation used for the finite element simulations is well-suited for a graph representation.
	GNNs also respect permutation in-/equivariance, which is advantageous, because any description of the geometry of the material microstructure has to respect permutation in-/equivariance with respect to the geometrical features (i.e., the order in which the nodes, elements or entire holes are described should be irrelevant).
	GNNs have been used before \cite{Vlassis2020, Thomas2023, Karapiperis, Pfaff2020} for the simulation of materials. However, these GNNs (i) only predict scalars (e.g. energy, mean pressure, deviatoric stress, fracture probability, damage), (ii) operate typically in the small-strain regime, and (iii) are neither $E(n)$-in-/equivariant nor (iv) scale in-/equivariant. This makes them unsuitable for the goal of simulating the specific metamaterials of interest here.
	MeshGraphNets \cite{Pfaff2020} using translation equivariant GNNs to recreate results from physical simulations \emph{do} allow for large strains, as demonstrated by a simulation of a body made of a hyperelastic material. However, these models are neither $E(n)$-equivariant nor scale in-/equivariant.

	\label{par:equivariant_features}
	$E(n)$-in-/equivariance in GNNs can be achieved simply by using only $E(n)$-invariant features such as distances or angles, as many GNNs for molecular property prediction do \cite{Schutt2017, Schutt, Schutt2018, Lubbers2018, Unke2019, Gasteiger}. In addition, periodicity can be incorporated by merging nodes from one side of the RVE with their corresponding side (left to right, top to bottom), or equivalently, have `ghost nodes' \cite{Xie2018}.
	Unfortunately, these molecular GNNs tend to be suitable only for graphs with a small graph diameter, whereas a finite element (FE) mesh is usually much larger with a large graph diameter, requiring a high number of message passing steps,  which are the iterations used by a GNN to update the graph, see Section~\ref{sec:GNN}.
	Moreover, using these $E(n)$-invariant features precludes any dependence on a global coordinate system. Consequently, the node positions cannot be updated directly.

	\label{par:other_equivariant_GNNs}
	There exist other ways to update node positions in a global coordinate system such that the updates are independent of the global coordinate system, using, e.g., expansions in spherical harmonics.
	Relying on spherical harmonics allows for updating positions independently of a global coordinate system, albeit at increased computational cost \cite{Thomas2018,Batatia2022a}.
	Alternatively, GVP-GNNs (Geometric Vector Perceptron Graph Neural Networks) \cite{Jing2021}, PaiNN (polarizable atom interaction neural network) \cite{Schutt2021} and GemNet (geometric message passing neural network) \cite{Gasteiger2021} can predict vector quantities as well, without the use of spherical harmonics, which makes them also suitable to predict deformation. Here, we choose another approach named $E(n)$-equivariant graph neural networks (EGNNs) \cite{Satorras2021a}, due to their simplicity, which updates node positions by `pulling' or `pushing' them along their edges.

	\label{par:EGNN_drawbacks}
	EGNNs currently do not respect periodicity, because the distances are recomputed from the new coordinates of the nodes after each message passing step. This approach does not work for edges that wrap around from one side of the RVE to another (i.e., for periodic edges). We show that it is relatively easy to adjust an EGNN such that it is periodic (referred to as `EGNN' in this paper). EGNNs are also not scale invariant, because there is a dependence on the distances between nodes. We show this can also be fixed (SimEGNN).

	\label{par:paper_contents}
	The paper is structured as follows: Section~\ref{sec:dataset} covers the essentials of first-order computational homogenization and describes the generation of the training dataset. Section~\ref{sec:symmetries} describes the concept of in-/equivariance, and how it applies here. Section~\ref{sec:GNN} details the developed GNN architecture.
	Numerical results and discussion are provided in Section~\ref{sec:results}, while Section~\ref{sec:conclusion} concludes this paper. All the code can be found at: \url{https://github.com/FHendriks11/SimEGNN}, and all the data is available in a Zenodo resitory here: \url{https://zenodo.org/records/14229619}.

	\label{par:notation}
	Throughout the paper we use the following notation: scalars (except for the strain energy density $\mathfrak{W}$) are in italics (e.g. $r_{ij}$), abstract `vectors' (neural network-generated lists of numbers without a physical meaning), which can have any dimension, are in bold (e.g. $\bt h_i$), physical/geometric vectors, which are always 2D or 3D, are in italics with an arrow (e.g. $\vec{x}_{i}$), geometric tensors are in bold capitals (e.g. $\bt P$), of which the 4\textsuperscript{th}-order tensors have a superscript 4 in front (e.g. ${}^4 \bt D$). We use the following definitions of the tensor product (also called open product or dyadic product)
	\begin{align}
		\bt A &= \vec{a} \otimes \vec{b} \quad \Longleftrightarrow \quad A_{ij} = a_i b_j\\
		{}^4 \bt C & = \bt A \otimes \bt B \quad \Longleftrightarrow \quad C_{ijkl} = A_{ij} B_{kl}
	\end{align}
	 and the dot product
	\begin{align}
		\vec{b} &= \vec{a} \cdot \bt A \quad \Longleftrightarrow \quad b_{j} = \sum_i a_i B_{ij}\\
		\vec{b} & = \bt A \cdot \vec{a} \quad \Longleftrightarrow \quad b_{i} = \sum_j A_{ij} a_{j}\\
		\bt C & = \bt A \cdot \bt B \quad \Longleftrightarrow \quad C_{ik} = \sum_j A_{ij} B_{jk},
	\end{align}
	\textcolor{mygreen}{
	as well as the double dot product
	\begin{align}
		\bt C & = {}^4\bt A : \bt B \quad \Longleftrightarrow \quad C_{ij} = \sum_k \sum_l A_{ijkl} B_{lk}.
	\end{align}
	}
	\textcolor{mygreen}{
	The divergence of a second order tensor field with respect to the reference configuration is defined as
	\begin{equation}
		\vec{a} = \vec{\nabla}_0 \cdot \bt A
		\quad \Longleftrightarrow \quad
		a_{j} = \sum_i \pdv{A_{ij}}{x_i^{\textrm{ref}}}.
	\end{equation}
	}
	\textcolor{mygreen}{Note that for the definitions in the equations above, the indices correspond to the different components of tensors or vectors. In the rest of the paper, however, the subscripts indicate how many nodes are relevant to that quantity, e.g., $\bt h_i$ is an abstract vector specific to one node $i$, $r_{ij}$ is a scalar quantity specific to an edge $ij$ between node $i$ and $j$.}

	\section{Data Set Generation}\label{sec:dataset}
	To create a training data set, finite element simulations are exploited, in combination with the computational homogenization scheme to compute representative volume element (RVE) responses and their effective properties.
	Specifically, we construct a dataset of the following mapping from the microscopic RVE (geometry and material) and the macroscopic deformation gradient tensor $\bt F$ to the microscopic deformation and the macroscopic target variables
	\begin{equation}
		\begin{rcases}
			\textrm{RVE geometry \& material}\\
			\bt F
		\end{rcases}
		\quad \rightarrow \quad
		\begin{cases}
			\vec{x}\\
			\mathfrak{W}\\
			\bt P\\
			{}^4 \bt D
		\end{cases},\label{eq:constitutive_mapping}
	\end{equation}
	where $\mathfrak{W}$ is the effective (macroscopic) strain energy density, $\bt P$ is the effective first Piola-Kirchhoff stress tensor and ${}^4 \bt D$ is the effective stiffness tensor and $\vec{x}$ is the deformed configuration, meaning the final position of each node.
	This stiffness is defined through the incremental stiffness relationship
	\begin{equation}
		\delta \bt P = {}^4 \bt D : \delta \bt F.
	\end{equation}
	Even though $\vec{x}$ is not strictly necessary, we want to obtain it as well, to make the eventual machine learning model more interpretable, and to offer the possibility of observing the various pattern transformations directly. This dataset will later be used to train a machine learning model.

	\subsection{Computational Homogenization}
\begin{figure}
		\centering
        \includegraphics[width=0.6\textwidth]{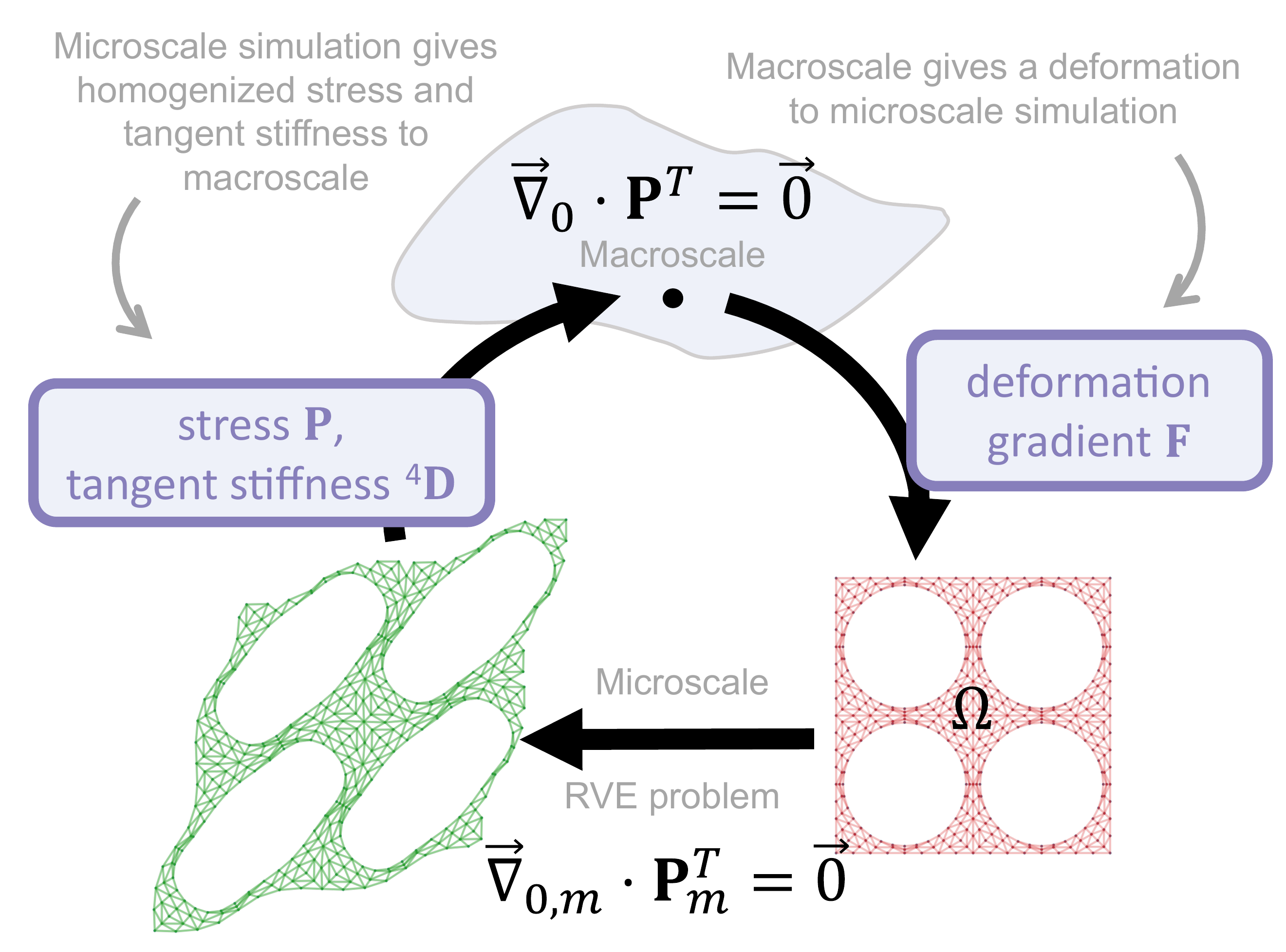}
        \caption{Schematic overview of the first-order computational homogenization procedure.}
        \label{fig:micromacro}
    \end{figure}

	We focus on calculations pertinent to computational homogenization, which is a workhorse in shape or topology optimization methods of materials. It replaces an explicit constitutive law at the macroscale with a boundary value problem posed on an RVE at the microscale, loaded by a deformation gradient $\bt F$, which is extracted from the macroscale (for an overview of the basics and extensions of computational homogenization, see \cite{Geers2010}).
	The macroscale (global scale) is the scale over which the mechanical loading of the entire system varies, whereas the microscale is the scale of the heterogeneities in the microstructure of the underlying material. In this section, the microscopic quantities will be denoted by the subscript $m$.
	See Figure~\ref{fig:micromacro} for a schematic representation of the computational homogenization procedure.

	We focus here on the microscale computation, which is a finite element (FE) simulation of an RVE.
	Since the microstructural features in our materials are quite fine, the micro-scale FE mesh must be adequately small as well, which makes the micro-scale simulations expensive.

	Even though the nature of architectured materials often calls for a generalized continuum at a macroscale (\textcolor{mygreen}{e.g.,} second-order \cite{Kouznetsova2004,Kaczmarczyk2008} or micromorphic \cite{Rokos2019, VanBree2020, Rokos2020}), we limit ourselves here to the standard first-order computational homogenization which leads to a classical Cauchy continuum at the macroscale, assuming separation of scales.

	\subsection{Microstructural RVE problem}
	\begin{figure}
		\centering
		\begin{subfigure}[b]{0.49\textwidth}
			\centering
			\includegraphics[width=\textwidth]{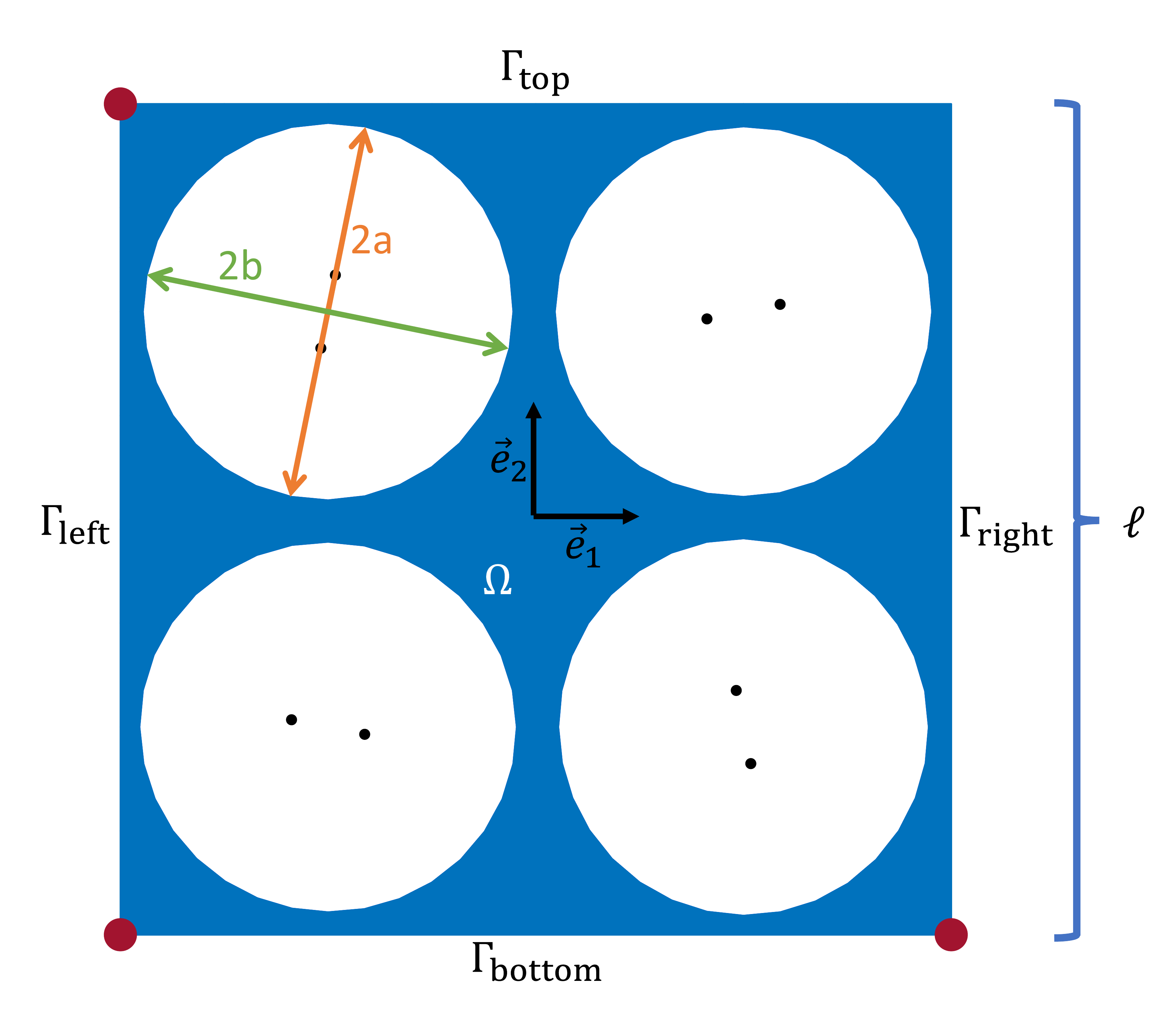}
			\caption{}
			\label{fig:geometry}
		\end{subfigure}
		\hfill
		\begin{subfigure}[b]{0.49\textwidth}
			\centering
			\includegraphics[width=\textwidth]{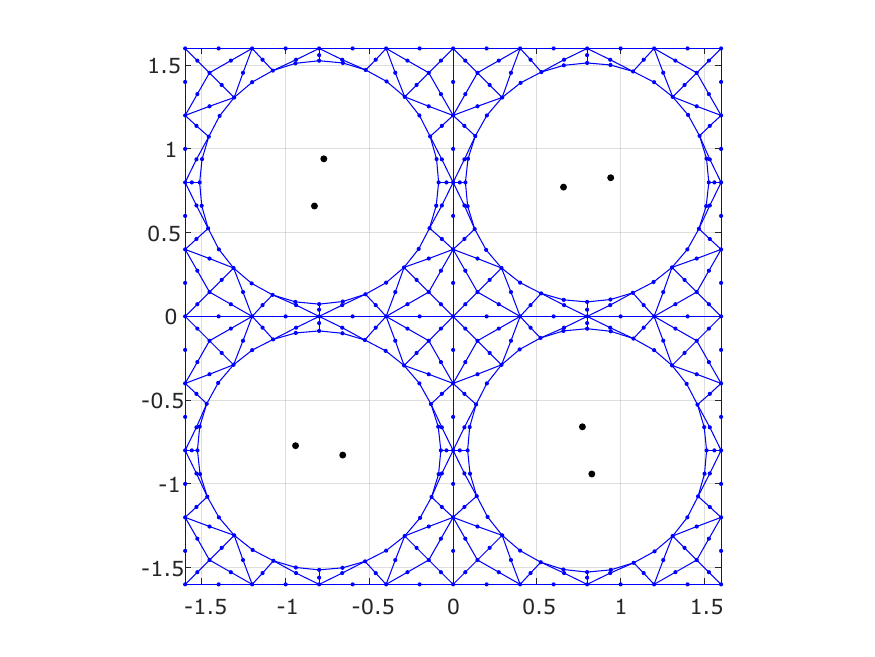}
			\caption{}
			\label{fig:mesh}
		\end{subfigure}
		\caption{(a) The RVE geometry, with four holes, slightly flattened by 1\% into ellipses, with $2a$ the major axis and $2b$ the minor axis. The periodic boundaries are indicated by $\Gamma$, the control points by red dots, and the foci of the ellipses by black dots. The length and width of the RVE is $\ell$ and the domain is $\Omega$. The directions of the basis vectors are indicated by $\vec{e}_1$ and $\vec{e}_2$. (b) The finite element discretization of this geometry.}
		\label{fig:geommesh}
	\end{figure}

	\label{par:geometry}
	The underlying pattern-transforming 2D metamaterial consists of a flexible, hyperelastic matrix with circular (or, in our case, almost circular) holes arranged in a square grid as shown in Figure~\ref{fig:geometry} \cite{Bertoldi2010,Rokos2019}. This material exhibits auxetic behavior after buckling. We assume a hyperelastic, finite-strain model at the microscale.
	One RVE has four holes, which has been confirmed to be of a sufficient size to capture the most important pattern transformation when buckling \textcolor{mygreen}{under prescribed macroscopic strain (e.g., \cite{Boyce2008}), although it should be noted that it is possible for a $2\times 2$ RVE to be inadequate in other cases. For example, Polukhov et al.\cite{Polukhov2018} show that in a electrostatically activated version of the same material sometimes a $2\times 3$ RVE is necessary. Ideally, Bloch analysis should be used to determine the appropriate RVE size  \cite{Geymonat1993a, Triantafyllidis2006, Boyce2008, Zhang2021a}, although this is computationally expensive.}

	Each hole has a diameter 0.45 times the side length $\ell$ of the RVE. \textcolor{mygreen}{In the extended data set, the diameter is also varied; all holes have the same diameter, which can be $0.4\ell$, $0.425\ell$, $0.45\ell$, or $0.475\ell$. This results in ligament thicknesses of $0.1\ell$, $0.075\ell$, $0.05\ell$, and $0.025\ell$, respectively.}

\begin{figure}
		\centering
        \includegraphics[width=0.85\textwidth]{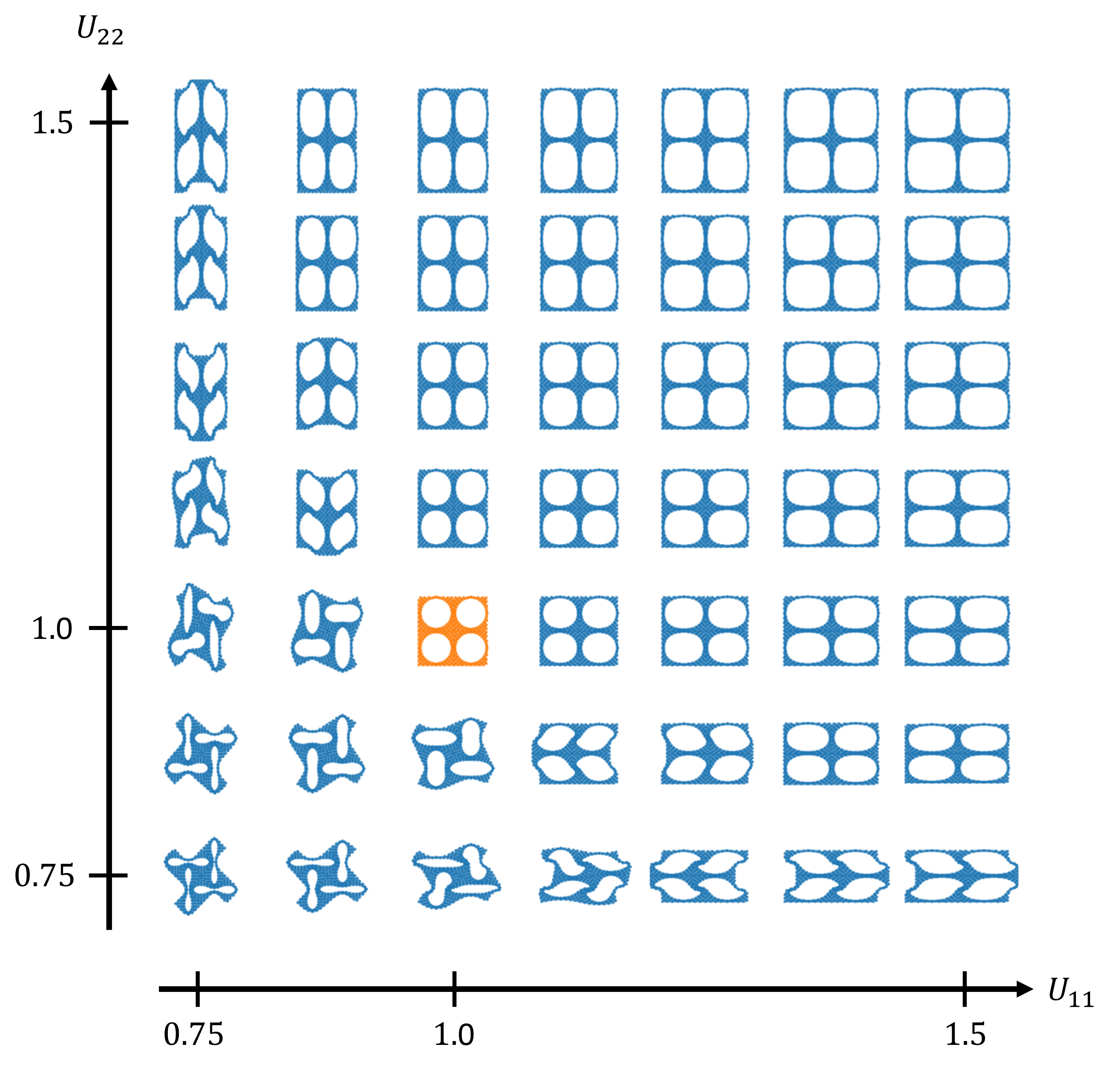}
        \caption{
			The behavior of the metamaterial RVE under various loading conditions. In this plot, $U_{ij}$ are components of the right stretch tensor, see Section~\ref{subsec:sampling}. The components $U_{11}$ and $U_{22}$ are varied, while $U_{12}$ is constrained at zero. The reference configuration (i.e., the unloaded state) is plotted in orange.
		}
        \label{fig:bucklingcases}
    \end{figure}

\begin{figure}
		\centering
		\begin{subfigure}[b]{0.49\textwidth}
			\centering
			\includegraphics[width=\textwidth]{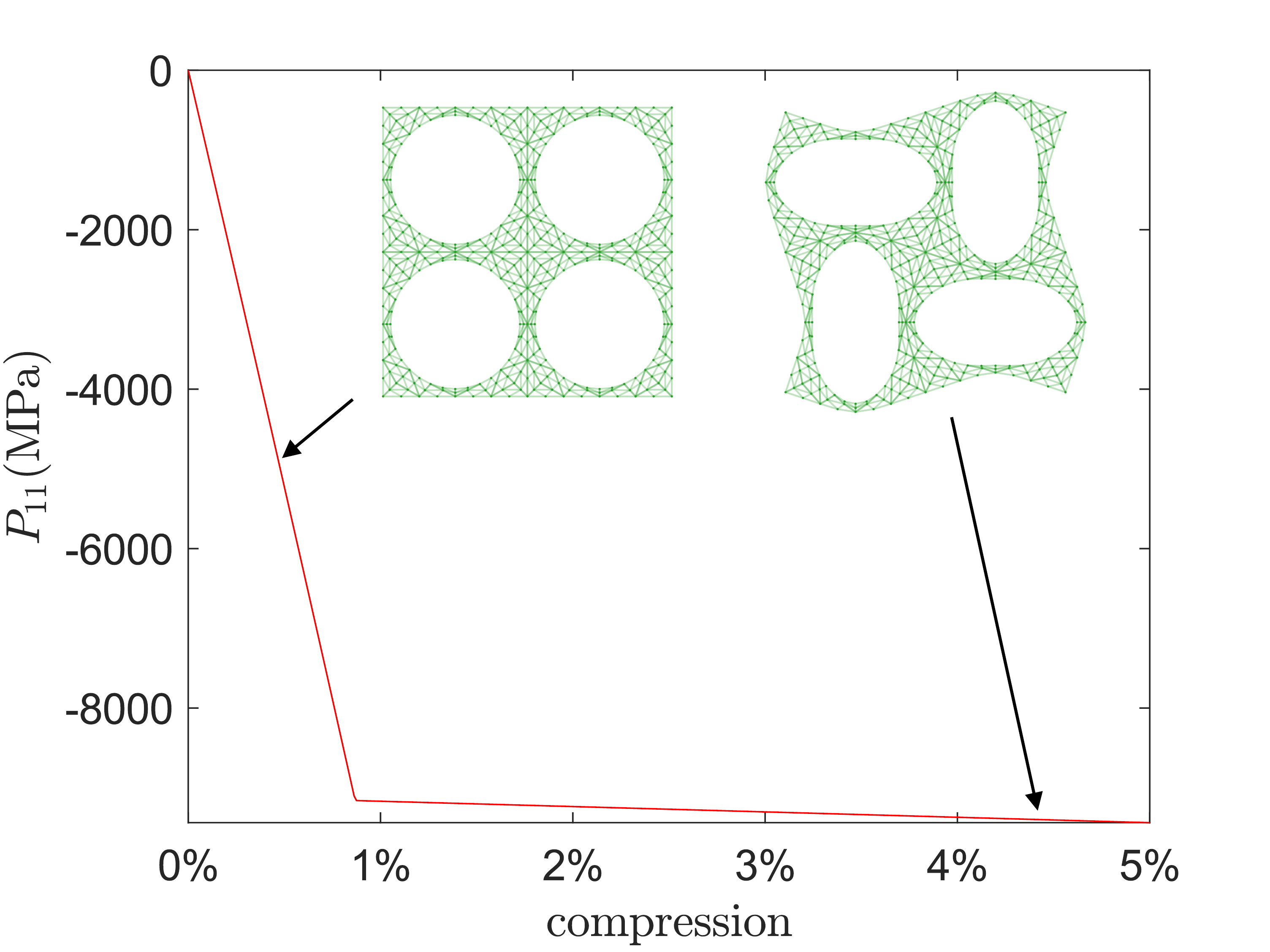}
			\caption{}
			\label{fig:bucklingP11}
		\end{subfigure}
		\hfill
		\begin{subfigure}[b]{0.49\textwidth}
			\centering
			\includegraphics[width=\textwidth]{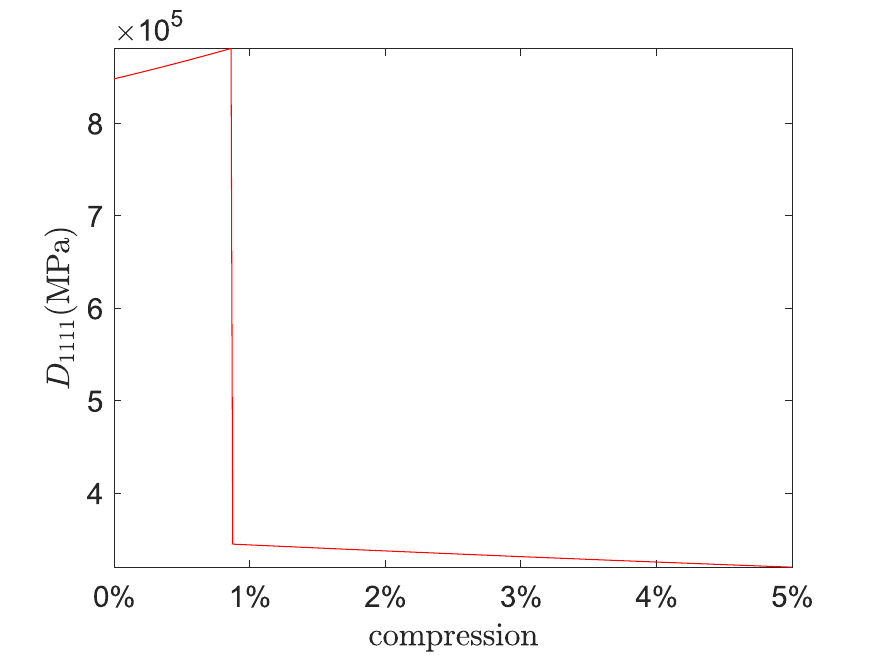}
			\caption{}
			\label{fig:bucklingD1111}
		\end{subfigure}
		\caption{Change in (a) stress component $P_{11}$ and (b) stiffness component $D_{1111}$ as a function of the biaxial compression.}
		\label{fig:buckling}
	\end{figure}

	The above-described metamaterial constitutes a challenging test case, because of the multiple buckling modes it shows, and the strong dependence of effective properties on the deformation state.
	Depending on the loading conditions, i.e., prescribed macroscopic $\bt F$, a bifurcation breaking the RVE's symmetry can occur. There are two bifurcation modes: the microstructure can either (i) buckle clockwise or counterclockwise -- referred to in this paper as `rotational bifurcation' --, or (ii) bulge out left or right (or up or down if $F_{11} < F_{22}$) -- `left/right bifurcation',\footnote{This left/right buckling mode would actually correspond to a global buckling mode, and the RVE with $2\times 2$ holes we use therefore does not capture it accurately. However, because here the system only serves as a test case for the GNN, we ignore this fact for now. } or even both situations can occur one after another \textcolor{mygreen}{when increasing the loading}, see also Figure~\ref{fig:bucklingcases}. Buckling significantly alters the homogenized stress and stiffness; Figure~\ref{fig:buckling} shows how the normal stress $P_{11}$ and stiffness component $D_{1111}$ change under biaxial compression.

	\label{par:bulk_material}
	We use the Bertoldi-Boyce constitutive law for the polymer, given by the strain energy density function \cite{Boyce2008}
	\begin{equation}
		\mathfrak{W}_m(\bt F_m) = \textcolor{mygreen}{c_1} \left(I_1 - 2\right) + \textcolor{mygreen}{c_2} \left(I_1 - 2\right)^2 - 2\textcolor{mygreen}{c_1} \ln\left(J_{\textcolor{mygreen}{m}}\right) + \frac{\textcolor{mygreen}{K}}{2} \left(J_{\textcolor{mygreen}{m}}-1\right)^2,\label{eq:bertoldiboyce}
	\end{equation}
	where $J_{\textcolor{mygreen}{m}} = \det\bt F_m$, $I_1 = \tr \bt C_m$, $\bt C_m$ is the microscopic right Cauchy-Green tensor $\bt C_m = \bt F_m^T \cdot \bt F_m$, and $\textcolor{mygreen}{c_1}$, $\textcolor{mygreen}{c_2}$ and $\textcolor{mygreen}{K}$ are \textcolor{mygreen}{bulk material parameters}, which take the values $\textcolor{mygreen}{c_1}=\SI{0.55}{MPa}$, $\textcolor{mygreen}{c_2}=\SI{0.3}{MPa}$, $\textcolor{mygreen}{K}=\SI{55}{MPa}$.\footnote{In 3D, the first two terms would have $I_1-3$ instead of $I_1-2$.}

	\label{par:governing_equation}
	The mechanical behavior is modeled using a standard Cauchy continuum.
	Assuming no body forces and neglecting inertia, this amounts to solving
	\begin{equation}
		\vec{\nabla}_{\textcolor{mygreen}{0},m} \cdot \bt P_m^{\textcolor{mygreen}{T}}(\bt F_m(\vec{u})) = \vec 0 \quad \forall \vec{x}^{\textcolor{mygreen}{\;\textrm{ref}}} \in \Omega \label{eq:PDE}
	\end{equation}
	for the microscopic displacement field $\vec{u}$, \textcolor{mygreen}{where $\vec{\nabla}_{0,m}$ is} the gradient operator with respect to the \emph{initial} microstructural position $\vec{x}^{\;\textrm{ref}}$ on the domain $\Omega$.
	This equation and its solutions respect the in- and equivariances briefly mentioned in the Introduction and further discussed in Section~\ref{sec:symmetries}.
	The new position $\vec{x}$ is decomposed as
	\begin{equation}
		\vec{x} = \vec{x}^{\;\textrm{ref}} + \vec{u}(\vec{x}^{\;\textrm{ref}}) = \bt F \cdot \vec{x}^{\;\textrm{ref}} + \vec{w}(\vec{x}^{\;\textrm{ref}}),\label{eq:xdecomp}
	\end{equation}
	where $\vec{u}(\vec{x})$ is the displacement field, $\bt F \cdot \vec{x}^{\;\textrm{ref}}$ is the affine part dictated by the prescribed macroscopic deformation gradient $\bt F$ and $\vec{w}$ denotes the fluctuation part of the solution.

	\label{par:homogenized_quantities}
	From this solution, the homogenized quantities can be obtained as follows:
	\begin{align}
		\mathfrak{W} &= \frac{1}{\Omega}\int_\Omega \mathfrak{W}_m \left(\bt F_m \left(\vec{u}\right)\right)\dd \vec{x}\label{eq:homoW},\\
		\bt P &= \pdv{\mathfrak{W}}{\bt F} = \frac{1}{\Omega}\int_\Omega \bt P_m \left(\bt F_m \left(\vec{u}\right)\right)\dd \vec{x}\label{eq:homoP},\\
		{}^4 \bt D &= \frac{\partial^2{\mathfrak{W}}}{\partial{\bt F}\partial{\bt F}} = \pdv{\bt P}{\bt F}. \label{eq:homoD}
	\end{align}
	We do not provide the full definition for ${}^4 \bt D$, but it can be found in \cite[p. 227]{Tadmor2012} and the method of computing it is adopted from \cite[\textsection{}2.4.4]{Kouznetsova2002}, see also \cite{Miehe2003}.

	\label{par:boundary_conditions}
	The periodic boundary conditions enforce the top and bottom boundaries of the RVE to deform in the same way, and likewise for the left and right boundaries. This implies that the microfluctuation field $\vec{w}$ (i.e., the displacement field $\vec u$ minus its affine part) is equal for corresponding points on these boundaries \cite{Miehe2002}, see Figure~\ref{fig:geometry}:
	\begin{align}
		\vec{w}\left(\Gamma_{\textrm{left}}\right) &= \vec{w}\left(\Gamma_{\textrm{right}}\right),\\
		\vec{w}\left(\Gamma_{\textrm{top}}\right) &= \vec{w}\left(\Gamma_{\textrm{bottom}}\right).
	\end{align}

	\label{par:control_points}
	The macroscale deformation gradient $\bt F$ is imposed on this RVE by applying it as an affine transformation through the three corner control nodes (the fourth one is dependent on the others due to periodicity)	in multiple adaptive load increments.

	\label{par:discretization}
	The RVE's geometry shown in Figure~\ref{fig:geometry} is discretized with quadratic triangular elements, using Gmsh (\url{https://gmsh.info/}) \cite{Geuzaine2009} to create the mesh shown in Figure~\ref{fig:mesh}. This mesh is fairly coarse, which means the results are not as accurate as desired for FE\textsuperscript{2}, but good enough to capture the overall behavior and to quickly generate a large data set to test the proposed method. The FE simulations were performed using an in-house Matlab code \cite{VanBree2020}.

	\label{par:buckling_perturbation}
	Because the neural network cannot handle the ambiguity in buckling (i.e., left vs right, clockwise vs counterclockwise), we ensure that the material always buckles in the same direction, by steering the bifurcation through a minor perturbation of the holes' initial geometry. To this end, the circles are flattened into ellipses by 1\% (see Figure~\ref{fig:mesh} for the resulting finite element mesh, where the foci of the four ellipses are depicted as black dots). Moreover, the ellipses are tilted such that both lateral and rotational bifurcations are triggered consistently.
	A stability analysis was performed to verify that this adequately controls the bifurcation ambiguity.
	\textcolor{mygreen}{Of course, this perturbation is only possible because we already know what modes may emerge. In a real-world application, where one would like to predict the behavior of unseen geometries, this would not be possible and other approaches must be considered.}

	\subsection{Sampling and Data Augmentation}\label{subsec:sampling}
	\label{par:sampling}
	For the purpose of training GNNs, a training dataset was generated by uniformly sampling the RVE's response to a prescribed macroscopic deformation gradient $\bt F$.
	Using the polar decomposition, $\bt F$ can be decomposed as  $\bt F = \bt R \cdot \bt U$ into the symmetric macroscopic right-stretch tensor $\bt U$ and a rotation tensor $\bt R$. Because $\bt R$ encodes only a rigid-body rotation, which does not meaningfully affect the results, we only sampled $\bt F = \bt U$. This simplification left us with only three components $U_{11}$, $U_{12}$ and $U_{22}$.

\begin{figure}
		\centering
        \includegraphics[width=0.85\textwidth]{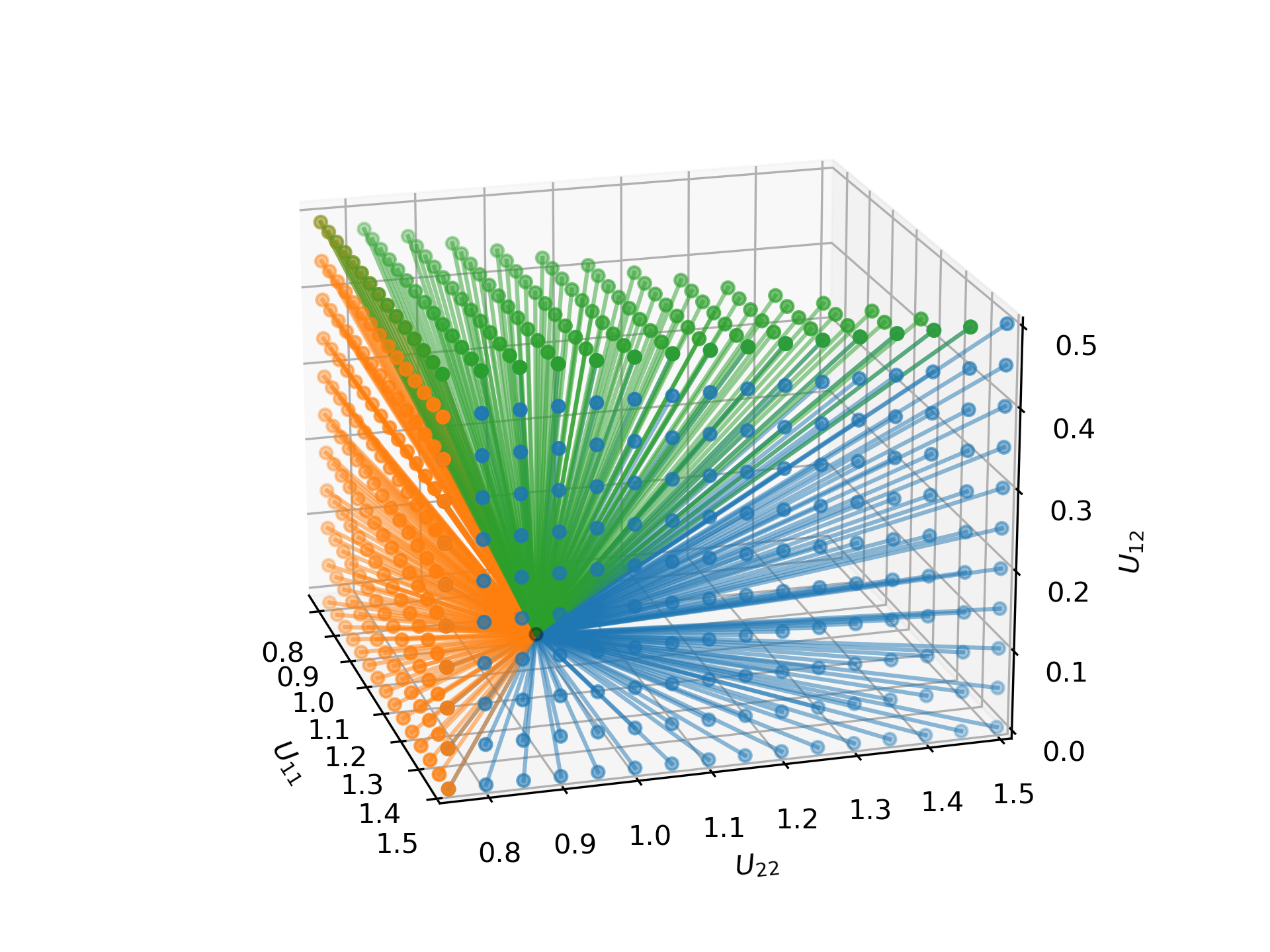}
        \caption{
			\textcolor{mygreen}{
			The sampled loading paths in $U_{11}$, $U_{22}$ and $U_{12}$ space. The lines indicate the loading paths, all starting from the black dot which indicates $\bt U =\bt I$. The colored dots represent the sampled values of final $\bt U^{\textrm{fin}}$, with the blue dots indicating cases with $U_{11}^{\textrm{fin}}$ constant, orange dots indicating $U_{22}^{\textrm{fin}}$ constant and green dots indicating $U_{12}^{\textrm{fin}}$ constant.}
		}
        \label{fig:samplingU}
    \end{figure}

	\textcolor{mygreen}{The sampling was carried out according to Figure~\ref{fig:samplingU}, where a visual representation of the sampling is presented.} Sampling was not uniform, because each simulation starts from $\bt U = \bt I$, and all intermediate configurations on the loading path to the prescribed final $\bt U$ were also saved and used for training, which is why \textcolor{mygreen}{when sampling the final $\bt U = \bt U^{\textrm{fin}}$, always at least one component was at a minimum or maximum value of the chosen range. For the intermediate values of $\bt U$, the loading was by default increased in 20 uniform steps from $\bt I$ to $\bt U^{\textrm{fin}}$, but because of the adaptive time stepping, this sometimes differed.}

	\textcolor{mygreen}{To sample different values of $\bt U^{\textrm{fin}}$, in turn one of the three components was kept at its extreme value while the other two were uniformly sampled from a specified range. These ranges were $[0.75, 1.5]$ in steps of 0.05 for $U_{11}^{\textrm{fin}}$ and $U_{22}^{\textrm{fin}}$, and [0, 0.5]  in steps of 0.05 for $U_{12}^{\textrm{fin}} = U_{21}^{\textrm{fin}}$.}
	This is a wide enough range to capture the different types of behavior. Because of the symmetry of the RVE (disregarding the small perturbation),
	we only consider $\bt U^{\textcolor{mygreen}{\textrm{fin}}}$ where $U_{11}^{\textcolor{mygreen}{\textrm{fin}}} > U_{22}^{\textcolor{mygreen}{\textrm{fin}}}$, and $U_{12}^{\textcolor{mygreen}{\textrm{fin}}} \geq 0$.

	Whenever a computed response featured overlapping elements, indicating the presence of contact (not considered in the numerical model), the corresponding $\bt U$ was removed from the dataset. This resulted in a data set of 446 trajectories, wich in total contain 8451 distinct loadcases. \textcolor{mygreen}{As indicated in the data statement at the end of the paper, all training data is included in a Zenodo repository accompanying this paper.}

	\label{par:data_augmentation}
	In the results Section~\ref{sec:results}, the newly developed machine learning model that respects all symmetries is compared to two models that have fewer symmetries incorporated.
	To train models with less symmetries to a higher degree of accuracy, a method called data augmentation can be used. Here, one trains the model on transformed versions of the data to promote learning of these symmetries (e.g., \textcolor{mygreen}{teaching the model} rotation by including rotated versions of each data point).
	Therefore, to make the comparison more fair, \textcolor{mygreen}{during training, we create} rotated, reflected and scaled versions of the data points. To this end, we transformed each \textcolor{mygreen}{training batch, using random uniform sampling} for the rotation and scaling, and random booleans for the $x$-reflection and $y$-reflection. 
	The bounds on rotation were set between $0$ and $2\pi$ radians and the scaling factor was chosen between 0.1 and 10 (sampling as $10^x$, with linear sampling of $x \in [-1, 1]$).

	\section{In- and Equivariance}
	\label{sec:symmetries}
	Inductive biases are a set of assumptions about the data, built into machine learning models, that do not need to be learned from data. Incorporating domain knowledge into a model \textcolor{mygreen}{can improve} the model's efficiency with respect to the amount of training data needed and \textcolor{mygreen}{can improve} its generalization capabilities. The symmetries (invariances and equivariances) mentioned in the introduction are examples of such inductive biases \cite{Bronstein2021}.

\begin{figure}
		\centering
		\begin{subfigure}[b]{0.49\textwidth}
			\centering
			\includegraphics[width=\textwidth]{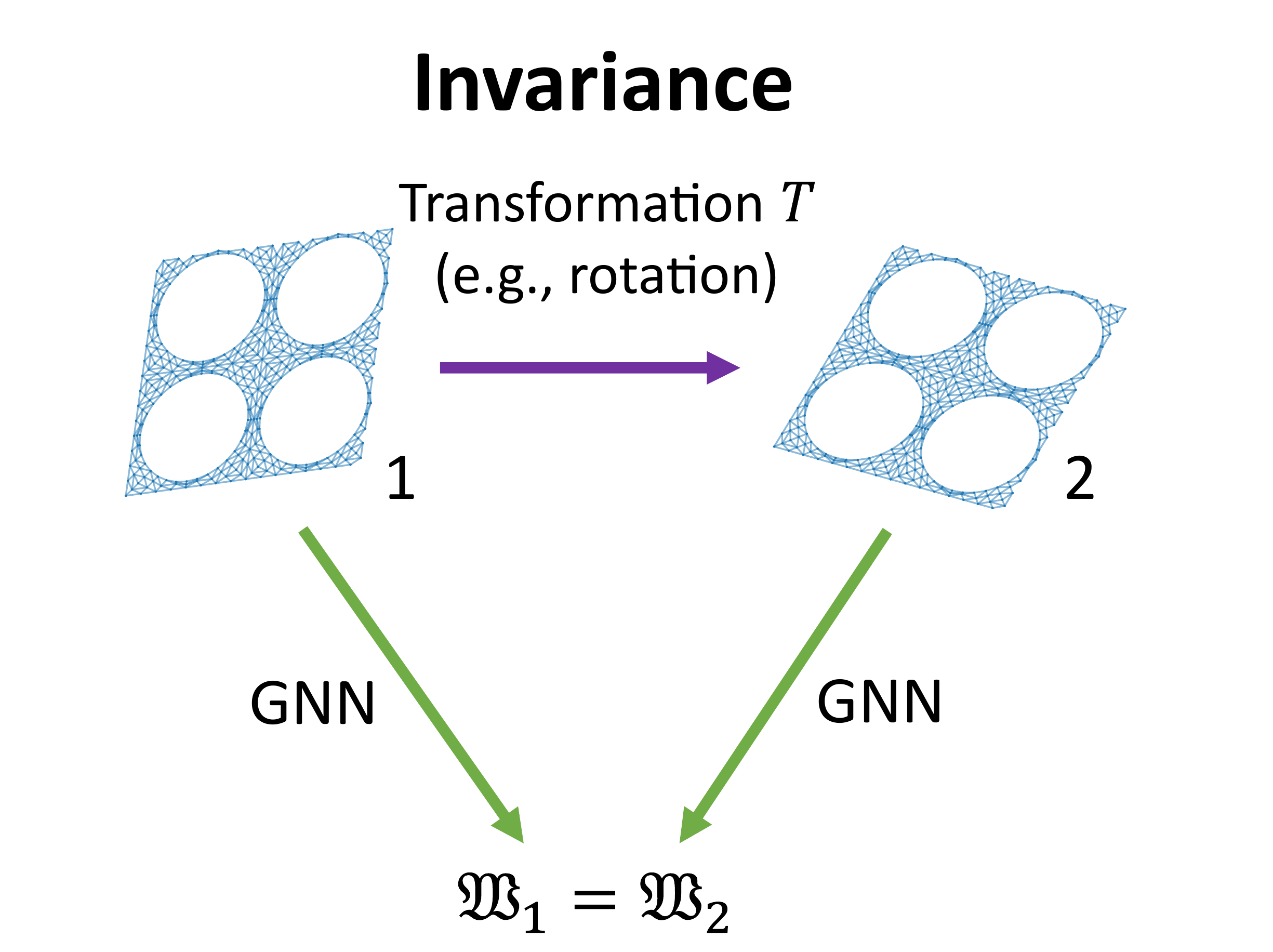}
			\caption{}
			\label{fig:inv}
		\end{subfigure}
		\hfill
		\begin{subfigure}[b]{0.49\textwidth}
			\centering
			\includegraphics[width=\textwidth]{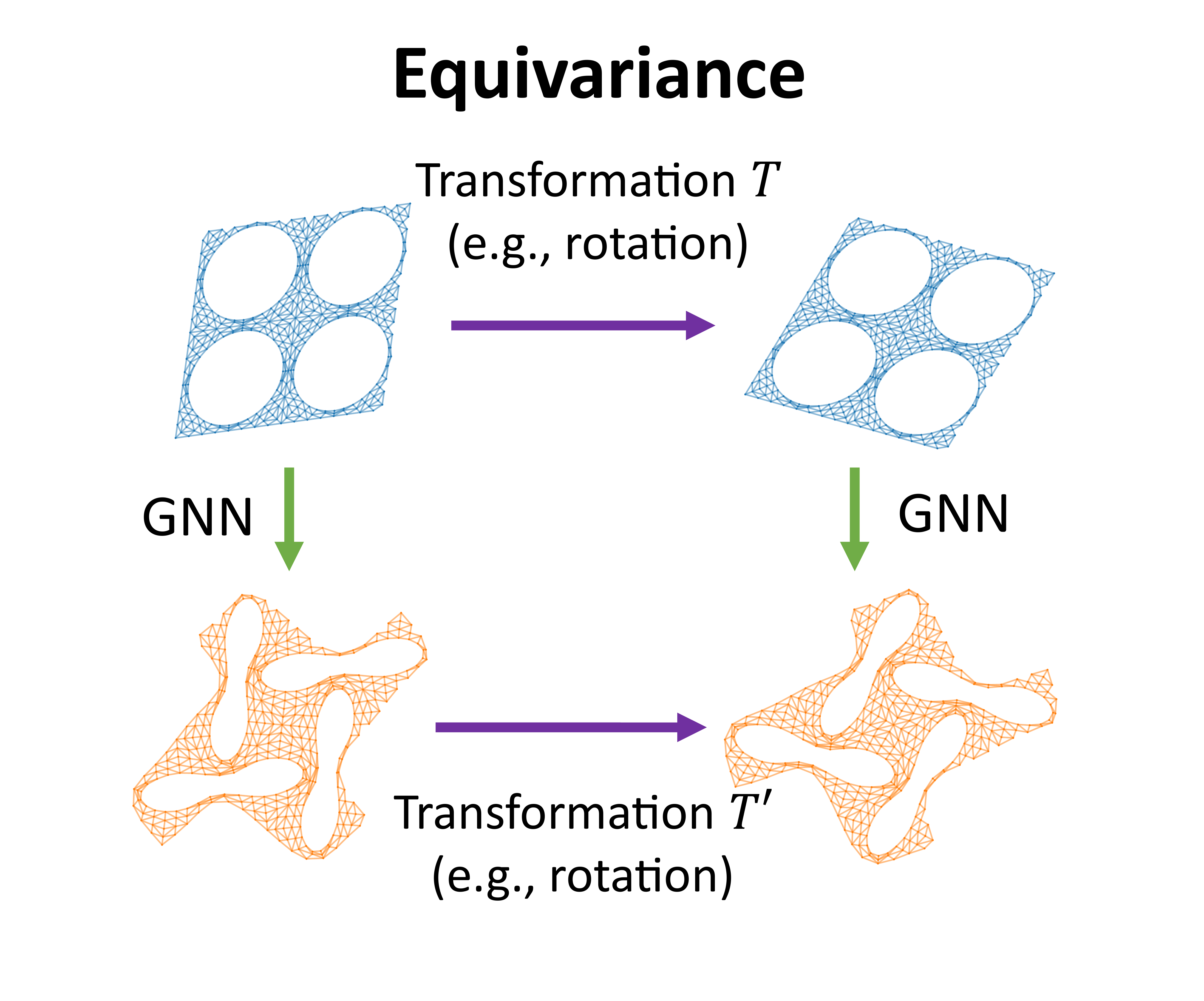}
			\caption{}
			\label{fig:equiv}
		\end{subfigure}
		\caption{The concepts of (a) invariance and (b) equivariance illustrated, using a GNN predicting the strain energy density and deformation as examples, respectively.}
		\label{fig:invequiv}
	\end{figure}

    Invariance with respect to a certain transformation $T$ means that when this transformation is applied to the input $\bt x$, the output $\bt y$ of the model $\phi$ stays the same
	\begin{equation}
		\phi(\bt x) = \phi(T \bt x).
	\end{equation}
	Equivariance with respect to a transformation means that there is a corresponding transformation $T'$ \textcolor{mygreen}{(often $T = T'$)} that can be performed on the model's output to get the same result
	\begin{equation}
		T'\phi(\bt x) = \phi(T \bt x).
	\end{equation}
	For example, if the input configuration is rotated (because the coordinate system is chosen differently, or because $\bt F$ contains a rotational part), then the energy density $\mathfrak{W}$ stays the same (invariance), while the output configuration $\vec{x}$ is also rotated (equivariance), and the homogenized stress and stiffness tensors are transformed as tensors under a rotation of the basis vectors (also equivariance). See Figure~\ref{fig:invequiv} for a visual explanation of these concepts. So, rather than giving many examples of such transformations in the training data and letting the model learn these invariances and equivariances, we aim to incorporate these symmetries directly into the model architecture and guarantee the model's performance under such transformations.

	As mentioned in the introduction, in addition to the node label permutation in-/equivariance that GNNs already respect, we consider similarity in-/equivariance (translation, rotation, reflection, scaling) and RVE in-/equivariance (periodicity). These in-/equivariances result directly from the governing balance equation of the system, recall Equation~\eqref{eq:PDE}.

\begin{figure}
		\centering
        \includegraphics[width=0.7\textwidth]{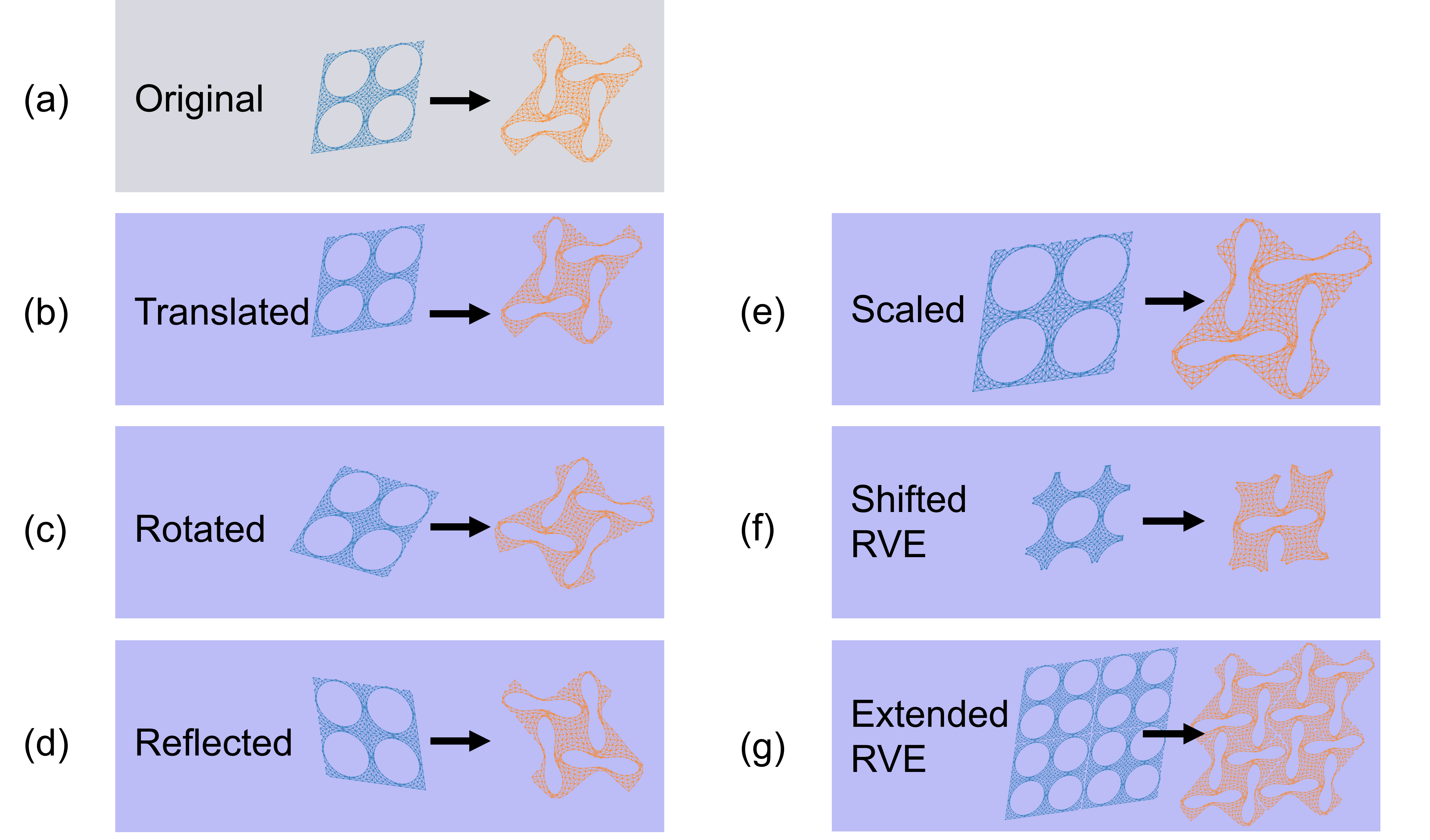}
        \caption{Different types of equivariance that should be respected by the model. Each case shows that the output configuration (orange) is transformed similarly to the input (blue), which is the meaning of equivariance.
        }
		\label{fig:equivariances}
    \end{figure}

	Figure~\ref{fig:equivariances} illustrates these equivariances for the deformation, and Table~\ref{tab:inequivariances} indicates which of the two (invariance or equivariance) applies for each type for each target variable. These are the node positions $\vec{x}$, the macroscopic strain energy density $\mathfrak{W}$, the first Piola-Kirchhoff stress tensor $\bt P$ and the stiffness ${}^4 \bt D$.

	\begin{table}
		\centering
		\caption{Overview of the invariance ({\color{Dandelion}I}) or equivariance ({\color{NavyBlue}E}) of target variables with respect to different types of transformations\label{tab:inequivariances}}
		\begin{tblr}{
		}
		Transformation & $\vec{x}$ & $\mathfrak{W}$ & $\bt P$ & ${}^4 \bt D$           \\
		\hline
		Translation  & {\color{NavyBlue}E} & {\color{Dandelion}I}     &  {\color{Dandelion}I}            & {\color{Dandelion}I}  \\
		Rotation     & {\color{NavyBlue}E}        & {\color{Dandelion}I}     &  {\color{NavyBlue}E}                     & {\color{NavyBlue}E}       \\
		Reflection   & {\color{NavyBlue}E}        & {\color{Dandelion}I}     &  {\color{NavyBlue}E}                     & {\color{NavyBlue}E}        \\
		Scaling      & {\color{NavyBlue}E}        & {\color{Dandelion}I}     &  {\color{Dandelion}I}                      & {\color{Dandelion}I}         \\
		Shifting RVE & {\color{NavyBlue}E}        & {\color{Dandelion}I}     &  {\color{Dandelion}I}                      & {\color{Dandelion}I}         \\
		Extending RVE   & {\color{NavyBlue}E} & {\color{Dandelion}I}     &  {\color{Dandelion}I}                      & {\color{Dandelion}I}
		\end{tblr}
	\end{table}

	Incorporating more in-/equivariances makes it easier for the model to construct and use features in a general way \cite{Bronstein2021}. Otherwise, it would have to learn them from the data, which can be encouraged by using data augmentation (see section \ref{subsec:sampling}).
	\textcolor{mygreen}{However, in general, obtaining the same performance from a model with fewer in-/equivariances will require more data, longer training and/or more parameters.}

	\section{GNN Approach}\label{sec:GNN}
	To achieve the goals set out in the introduction, we have developed a graph neural network (GNN)-based approach as detailed in this section.
	The final model -- the SimEGNN -- respects all the in-/equivariances discussed in Table~\ref{sec:symmetries}. Respecting these symmetries is important for data and parameter efficiency and generalizability \cite{Bronstein2021}.

	We first discuss the data structure encoding, i.e., graph encoding, and then present and compare three GNN architectures that were created specifically for metamaterial simulations. Each successive network has more symmetries built in, to quantify the added value of incorporating these symmetries.
	The first model, discussed in Subsection \ref{subsec:baseGNN}, is a base GNN model that can work with the data representation,
	and respects the periodicity (i.e., the `Shifted RVE' and `Extended RVE' in-/equivariance in Table~\ref{tab:inequivariances}), which is explained in Subsection \ref{subsec:periodicBC} as well as translation in-/equivariance.
	The second architecture, discussed in Subsection \ref{subsec:EGNN}, extends this base GNN to also incorporate $E(n)$-in-/equivariance.
	Finally, the third architecture, discussed in Subsection \ref{subsec:SimEGNN}, is our newly proposed SimEGNN, which also adds scale in-/equivariance. This means it respects all in-/equivariances listed in Table~\ref{tab:inequivariances}.

	Details on the implementation, including the training process, can be found in \ref{sec:experimental_setup}.

	\subsection{Graph Encoding}\label{subsec:graphenc}
	A graph consists of a set of nodes $\mathcal{V}$ and a set of edges $\mathcal{E}$ connecting pairs of the nodes.
	\textcolor{mygreen2}{When creating a graph that represents the microstructure, we desire the following properties of the graph representation:
	\begin{itemize}
	\item connected, since otherwise the message-passing steps cannot propagate information throughout the entire graph;
	\item small graph diameter (the maximum degree of separation of any two nodes), to reduce the number of message-passing steps needed;
	\item a sufficiently accurate representation of the physical structure of the material;
	\item easy to automate;
	\item not too large number of nodes and edges.
	\end{itemize}
	The element edges and nodes of the FE mesh (the one used for data generation, shown in Figure~\ref{fig:mesh}) are an obvious option for the input graph. This graph definitely offers an accurate representation of the physical structure of the material, however, it has a large number of nodes and edges and a large graph diameter.
	We therefore use it as our starting point}, but we only keep the nodes and edges at the boundary of the holes (i.e., we remove all the bulk nodes).
	Because the deformation of the boundary nodes of the mesh is enough to see how the material deforms, and the rest of the quantities of interest are global ($\mathfrak{W}$, $\bt P$, ${}^4 \bt D$), we do not require any local fields, which makes the bulk nodes unnecessary in the first place. Although it is possible to use the full FE mesh, our numerical experiments showed that it is a slower and less accurate option, presumably due to the higher number of message passing steps needed, leading to problems such as oversmoothing \cite{Chen2020a}.\footnote{When fully detailed local fields \emph{are} of interest, it may be possible to limit the number of necessary message passing steps by using graph pooling strategies.}

\begin{figure}
		\centering
        \includegraphics[width=0.8\textwidth]{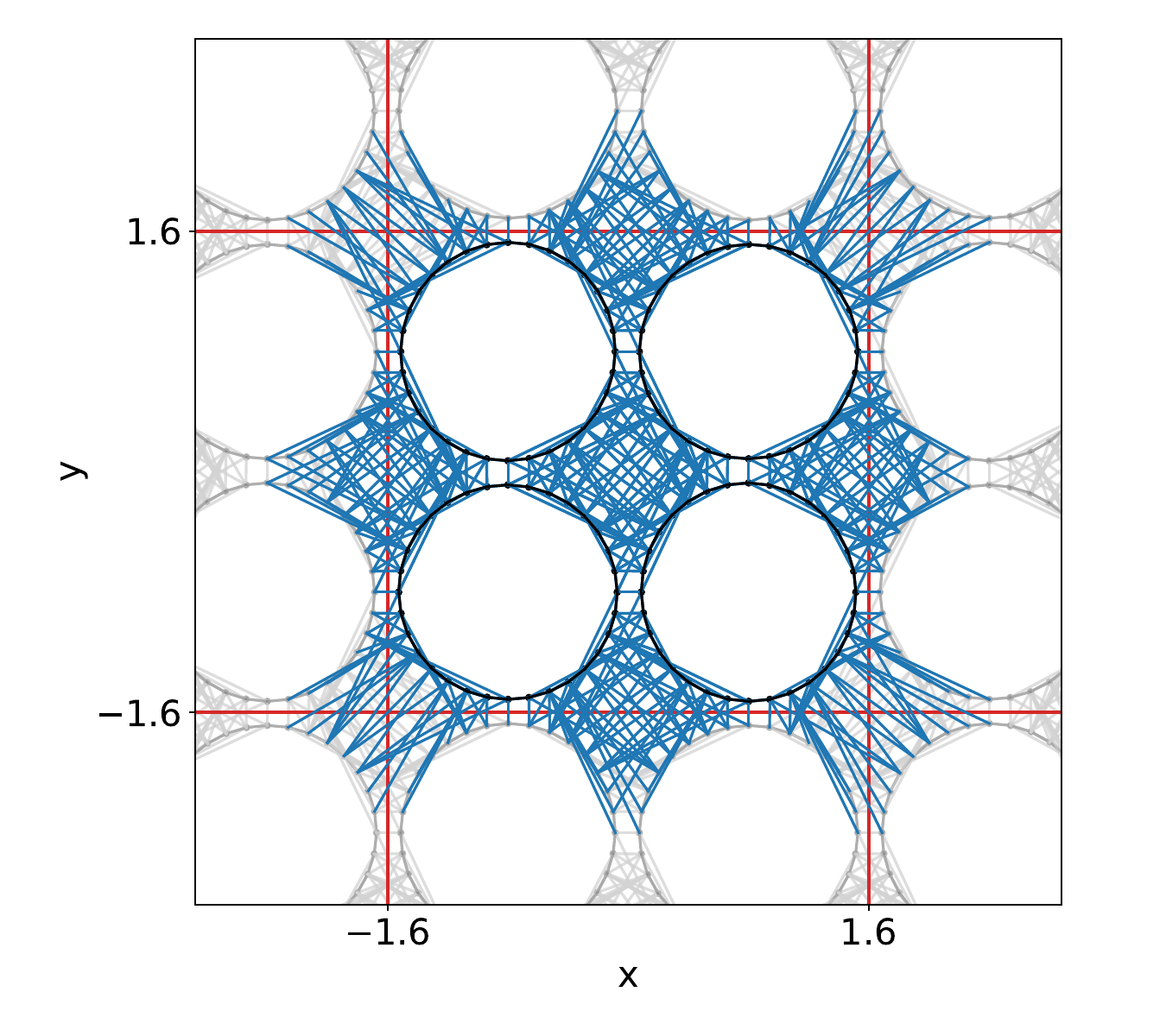}
        \caption{The input graph used for all GNNs considered in this paper, shown in blue and black. Periodic copies are shown in gray. \textcolor{mygreen}{The red lines indicate the RVE boundaries.} The graph includes only nodes at the boundaries of the holes and edges connecting the nodes to the nearest nodes at the boundaries of the other holes. \label{fig:removebulk}
		}
    \end{figure}

	When constructing the new graph, we make no distinction between nodes at the corner of the quadratic triangular elements and the mid-edge nodes. The elimination of the bulk nodes leaves us with four disconnected components, each corresponding to one of the four holes \textcolor{mygreen}{and consisting of the hole boundary nodes and edges}. \textcolor{mygreen2}{To ensure connectivity,} we join these components by connecting each node to the closest node on \textcolor{mygreen}{each of the other three hole boundaries} (i.e., adding three new edges for each node, and then deduplicating the set of edges). \textcolor{mygreen}{This results in at most $4N$ \textcolor{mygreen2}{undirected} edges, with $N$ the number of remaining hole boundary nodes, but in practice less due to the deduplication. In the final input graph, each undirected edge is represented by two directed edges.} Due to the RVE's periodicity, an edge may start on one side of the RVE and `wrap around' to the other side of the RVE (i.e., a Pac-Man World geometry). The resulting graph, derived from the FE mesh shown in Figure~\ref{fig:mesh}, is shown in Figure~\ref{fig:removebulk}. The original mesh contained 413 nodes and 1056 edges, while the generated input graph has 128 nodes and 480 edges.
	This fairly simple approach to constructing the graph representation \textcolor{mygreen2}{is easily automated and} already works adequately, even though it is possible to further optimize the graph representation, especially when the number of nodes and edges is very large, \textcolor{mygreen2}{or when the number of holes in the RVE becomes larger. In the latter case, it might be better not to connect each node to every other hole, but to only a subset of the holes.} Other options could be a small-world graph or a hierarchical graph for more efficient transfer of information throughout the graph, but we kept the question of optimizing the graph representation out of our scope for now.

\begin{figure}
		\centering
		\begin{subfigure}[b]{0.3\textwidth}
			\centering
			\includegraphics[width=\textwidth]{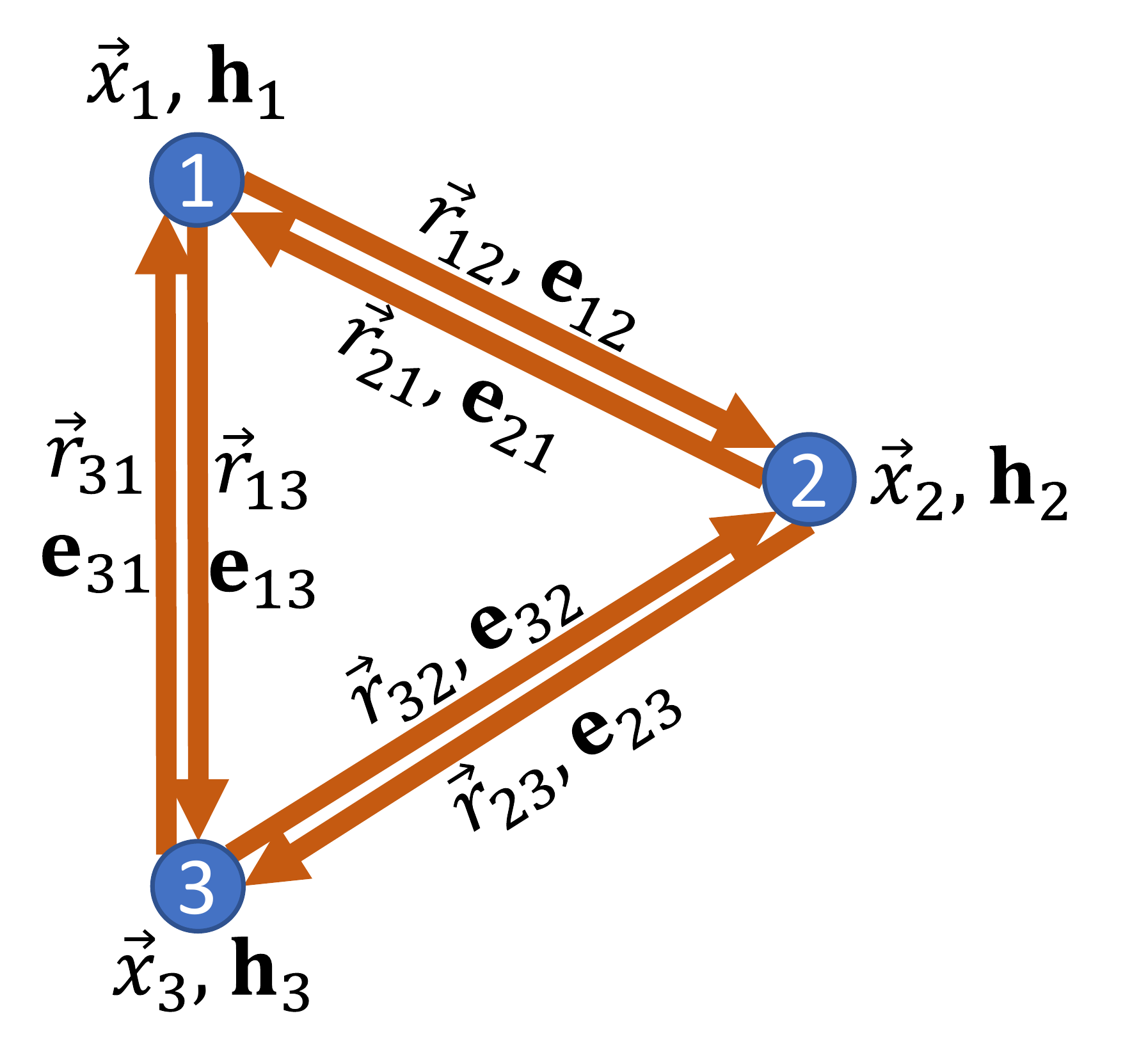}
			\caption{}
			\label{fig:graph_quantities}
		\end{subfigure}
		\begin{subfigure}[b]{0.3\textwidth}
			\centering
			\includegraphics[width=\textwidth]{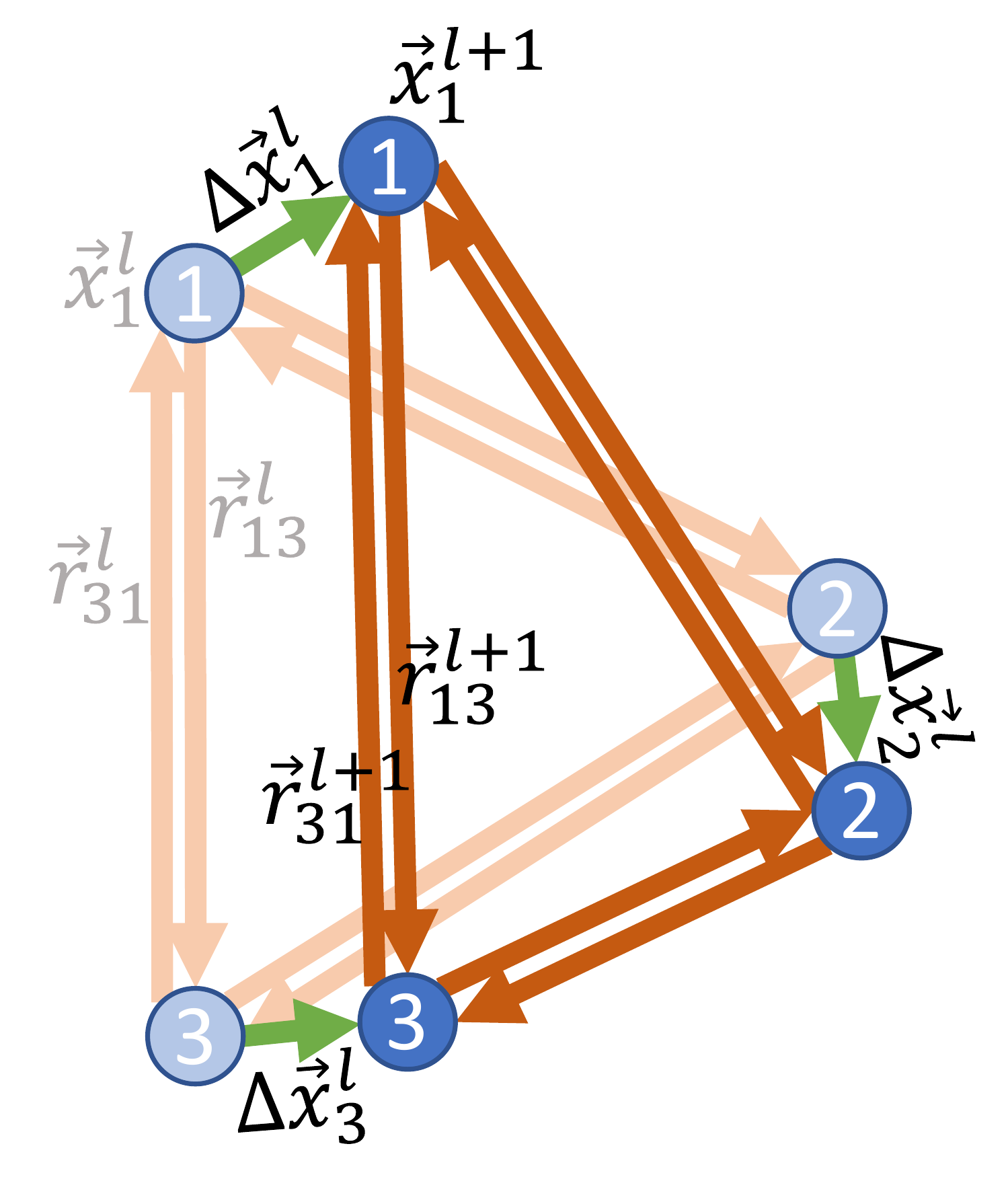}
			\caption{}
			\label{fig:graph_shift_quantities}
		\end{subfigure}
		\caption{
			Definition of the graph quantities: (a) each node $i\in\mathcal{V}$ has a position $\vec{x}_i$ (a 2D or 3D vector) and an abstract embedding $\bt h_i$ associated with it. Each edge $ij$ has a vector $\vec{r}_{ij}$ pointing along it and an abstract embedding $\bt e_{ij}$. (b) During a message passing step $l$, a shift $\Delta \vec{x}_i^{\;l}$ is applied to each node, which moves the node from its old position $\vec{x}_i^{\;l}$ to the new position $\vec{x}_i^{l+1} = \vec{x}_i^{\;l} \textcolor{mygreen}{+} \Delta \vec{x}_i^{\;l}$. The edge vectors are then updated correspondingly.
		}
		\label{fig:gq}
	\end{figure}

	In the constructed input graph, each node $i$ has a position $\vec{x}_i$ (a 2D or 3D vector) and an abstract embedding $\bt h_i$ associated with it. In addition, every edge $ij$ has a vector $\vec{r}_{ij}$ pointing along it and an abstract embedding $\bt e_{ij}$; see Figure~\ref{fig:gq} for a visual explanation of these graph quantities. \textcolor{mygreen2}{Initially, the edge embedding $\bt e_{ij}^{0}$ carries a Boolean, informing whether the edge corresponds to a hole boundary or not. In our setup, we encode this Boolean by assigning $\bt e_{ij}=-1$ if at a hole boundary and $+1$ otherwise. Swapping these values, or using other distinct values except zero is possible, since this information simply needs to indicate to the network that these edges differ. A transformation to other values would be straightforward for a neural network to learn. We chose -1 and +1 because by default, neural networks are initialized to handle values in this range the best.}

	The initial node embedding $\bt h_i^{0}$ is empty.

	\textcolor{mygreen}{
	Table~\ref{tab:datalayout} in \ref{sec:datalayout} shows what this graph representation actually looks like in the data.
	}

	\subsection{General approach}\label{subsec:generalapproach}
	GNNs incorporate permutation symmetry by updating the node and edge attributes by message passing, which is a way of propagating information throughout the graph \cite{Gilmer2017}.
	For all three models we compare, the message passing steps update the node positions $\vec{x}_i$ and generate abstract node and edge embeddings $\bt h_i$ and $\bt e_{ij}$. The node positions after the last message passing steps are the predicted positions.\footnote{Because any predicted displacement that contains a rigid body translation is still a valid solution, we compare the final GNN prediction for $\vec x$ to the FEM ground truth after setting the mean position such that it matches the FEM ground truth.}

	Message passing involves constructing a vector $\bt m_{ij}$ (called a `message') for each edge $ij \in \mathcal{E}$, and then aggregating at each node $i \in \mathcal{V}$ the `incoming' messages (the messages associated with edges that connect to the node) to update the node embedding $\bt h_i$. In its most general form, this looks like
	\textcolor{mygreen}{
	\begin{align}\label{eq:mpgnn}
		\bt m_{ij}^l &= \phi_m^{l} \left(\bt h_i^{l}, \bt h_j^{l}, \bt e_{ij} \right)
		 ,\\
		\bt h_i^{l+1} &= \phi_h^{l} \left( \bt h_i^{l}, \square_{j\in \mathcal{N}(i)}\bt m_{ij}^l \right),
	\end{align}
	}
	where $\square$ is the aggregation function (e.g. sum or average), $\bt m_{ij}^l$ is the message from node $j$ to $i$ in message passing step $l$,
	$\mathcal{N}(i)$ is the set of neighbors of node $i$, and $\phi_m^l$ and $\phi_h^l$ are learnable functions (in our case, one linear layer plus an activation function)\textcolor{mygreen}{, that are re-used for each node or edge, but which can differ per message passing step $l$}.
	We also update the edge embeddings from the messages:
	\begin{equation}
		\bt e_{ij}^{l+1} = \phi_e^l (\bt m_{ij}^l),
	\end{equation}
	where $\phi_e^l$ is another learnable function\textcolor{mygreen}{, which is re-used for each edge, but again can differ per message passing step $l$.}
	A graph neural network generally consists of multiple message passing steps.
	In our case, this message passing scheme needs to be modified to also take into account geometric quantities like the distances $r_{ij}$ between nodes.
	The exact form of the resulting message passing scheme differs for each of the three models we developed, but all have in common that they update the position $\vec{x}_i^{l}$ of the nodes in each message passing step $l$, by computing for each node a shift in position $\Delta \vec{x}_i^{\;l}$
	\begin{equation}
		\vec{x}_i^{\;l+1} = \vec{x}_i^{\;l} \textcolor{mygreen}{+} \Delta \vec{x}_i^{\;l}
	\end{equation}
	The way $\Delta \vec{x}_i^{\;l}$ is computed differs per model.

	At the end of the messages passing steps, still only local abstract quantities are available: node and edge embeddings, and the messages from the last message passing step. The way these quantities are converted into global predictions of the strain energy density $\mathfrak{W}$, stress $\bt P$ and stiffness ${}^4 \bt D$ also differs per model, but they all have in common that some local quantity (either an existing one or a newly created one) is being averaged. We use averaging, because it is permutation invariant with respect to the order of the nodes and edges, and is also the most common approach to homogenize local quantities to global ones in computational homogenization. \textcolor{mygreen}{It also possible to do weighted averaging, which essentially means adding an attention layer. This also applies to the message aggregation. However, we chose to keep our model architecture simple.}

	The mean square error in the prediction of each of the four target quantities ($\vec{w}$, $\mathfrak{W}$, $\bt P$, ${}^4 \bt D$) is used in the loss. Details about training and balancing these loss terms are given in \ref{sec:experimental_setup}.

	\subsubsection*{Periodic Boundary Conditions} \label{subsec:periodicBC}
	In order to properly implement periodic boundary conditions as described in Section~\ref{sec:dataset}, we need to construct and process the graph in such a way that the `Shifted RVE' and `Extended RVE' in-/equivariance shown in Figure~\ref{fig:equivariances} are respected. For this purpose, when constructing the input graph, we modify the computation of the edge vectors $\vec{r}_{ij}$ that are associated with the `wraparound' edges. Instead of pointing from $\vec{x}_i$ to $\vec{x}_j$, $\vec{r}_{ij}$ then points from $\vec{x}_i$ to the periodic image $\vec{x}_j$ on the correct side. Because we use only $\vec r_{ij}$ and the distance $r_{ij}$, which are independent of the actual location of the nodes, in updating the node embeddings, node positions and edge embeddings, the graph is updated independently of how the RVE is chosen.

	However, these wraparound edges entail a problem if not handled carefully; their edge vectors $\vec{r}_{ij}$ cannot be directly recalculated from the node positions $\vec x_i$. These positions $\vec x_i$ are updated multiple times during the message passing steps, and in the standard implementations of the MeshGraphNets and EGNNs, the updated $\vec{r}_{ij}$ are then recalculated from $\vec{x}_i$ as $\vec{r}_{ij} = \vec{x}_j - \vec{x}_i$, which will be inaccurate for the wraparound edges. In order to avoid using periodic images all the time, we only calculate $\vec{r}$ once in the beginning, after that, we keep  $\vec x_i$ and $\vec{r}_{ij}$ separate, and update $\vec{r}_{ij}$ directly using
	\begin{equation}
		\vec r^{l+1}_{ij} = \vec r^{l}_{ij} - \Delta\vec x^{\;l} + \Delta\vec x_{j},\label{eq:delta_r}
	\end{equation}
	where $\Delta\vec x_{i}$ is the shift in node position $\vec x_{i}^{l+1} = \vec x_{i}^{l} + \Delta\vec x_{i}$ in message passing step $l$. This ensures that the updates of $\vec{r}_{ij}$ are still independent of the absolute positions.

	To ensure that in-/equivariance with respect to the transformations `Extended RVE' and `Shifted RVE' in Figure~\ref{fig:equivariances} is still preserved after applying $\bt F$, we apply $\bt F$ as an affine transformation to all reference node positions $\vec{x}_i^{\;\textrm{ref}}$ and edge vectors $\vec{r}_{ij}^{\;\textrm{ref}}$
	at the same time. This provides the initial positions and edge vectors fed into the GNN:
	\begin{align}
		\vec{x}_i^{0} &= \bt F \cdot \vec{x}_i^{\;\textrm{ref}},\\
		\vec{r}_{ij}^{0} &= \bt F \cdot \vec{r}_{ij}^{\;\textrm{ref}}.
	\end{align}

	\begin{figure}
		\centering
        \includegraphics[scale=0.2]{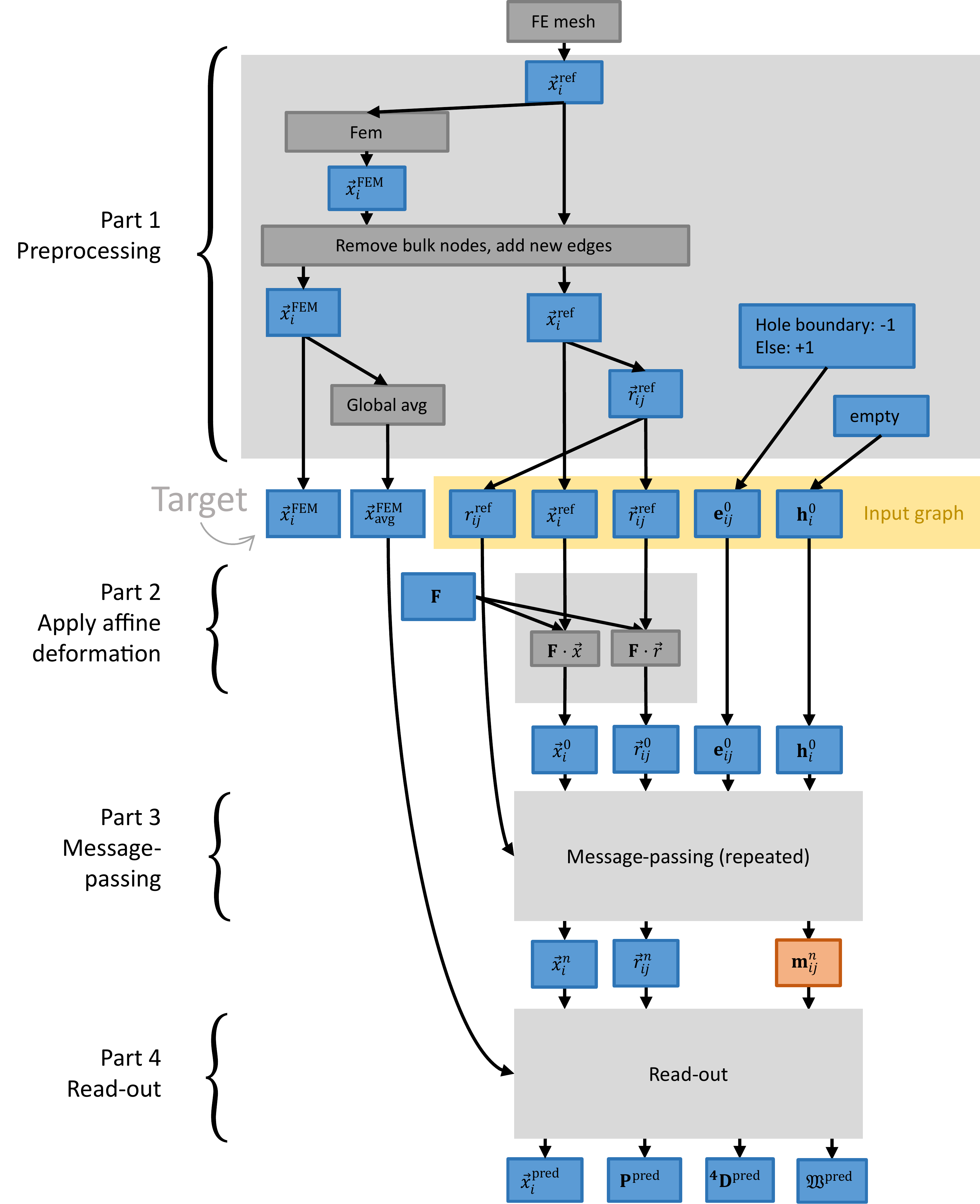}
        \caption{Diagram of the structure of all models, including the preprocessing. All blue quantities have a physical meaning, the orange quantity is abstract and generated by the neural network. The dark gray boxes indicate operations.
        \label{fig:workflow}}
    \end{figure}
	Figure~\ref{fig:workflow} shows the general structure of all three models, including the preprocessing. In the following paragraphs, the `message passing' and `read-out' blocks of this diagram will be described per model.

	\subsection{Base GNN}\label{subsec:baseGNN}
	For the base GNN used as a baseline in this paper, we use an approach similar to MeshGraphNets \cite{Pfaff2020}. For that, we use the following message computation:
	\begin{equation}
		\bt m_{ij}^l = \phi_m^l
		\left(
			\bt h_i^{l},
			\bt h_j^{l},
			\vec{r}_{ij}^{\;\textrm{ref}},
			r_{ij}^{\textrm{ref}},
			\vec{r}_{ij}^{\;l},
			r_{ij}^{l},
			\bt e_{ij}^{l}
		\right),
		\label{eq:GNNmessage}
	\end{equation}
	with $\phi_m^l$ a learnable function (in our case, one linear layer plus an activation function).
	This means the message computation for edge $ij$ uses both the reference edge length $r_{ij}^{\textrm{ref}}=\lVert\vec{r}_{ij}^{\;\textrm{ref}}\rVert_2$
	and all components of the edge vector $\vec{r}_{ij}^{\;\textrm{ref}}$ (before any deformation) and the current edge length $r_{ij}^l = \lVert\vec{r}_{ij}^{\;l}\rVert_2$ and edge vector $\vec{r}_{ij}^{\;l}$
	(after affine deformation and the message passing steps up to $l$), as well as the current node embeddings $\bt h_i^{l}, \bt h_j^{l}$ of both nodes, and the edge embedding $\bt e_{ij}^{l}$.

	The messages $\bt m_{ij}^l$ are then used to update the node and edge embeddings
	\begin{align}
		\bt h_{i}^{l+1} &= \phi_h^l
		\left(
			\bt h_i^{l},
			\frac{1}{|\mathcal{N}(i)|} \sum_{j\in\mathcal{N}(i)} \bt m_{ij}^l
		\right),\\
		\bt e_{ij}^{l+1} &= \phi_e^l (\bt m_{ij}^l),
	\end{align}
	where $|\mathcal{N}(i)|$ is the number of neighbors of node $i$ and $\phi_e^l$ and $\phi_h^l$ are, again, learnable functions. \textcolor{mygreen}{We chose not to include $\bt e_{ij}^l$ in the last equation, because it is already included in the message $\bt m_{ij}^l$.}

\begin{figure}
		\centering
		\begin{subfigure}[b]{0.43\textwidth}
			\centering
			\includegraphics[scale=0.2]{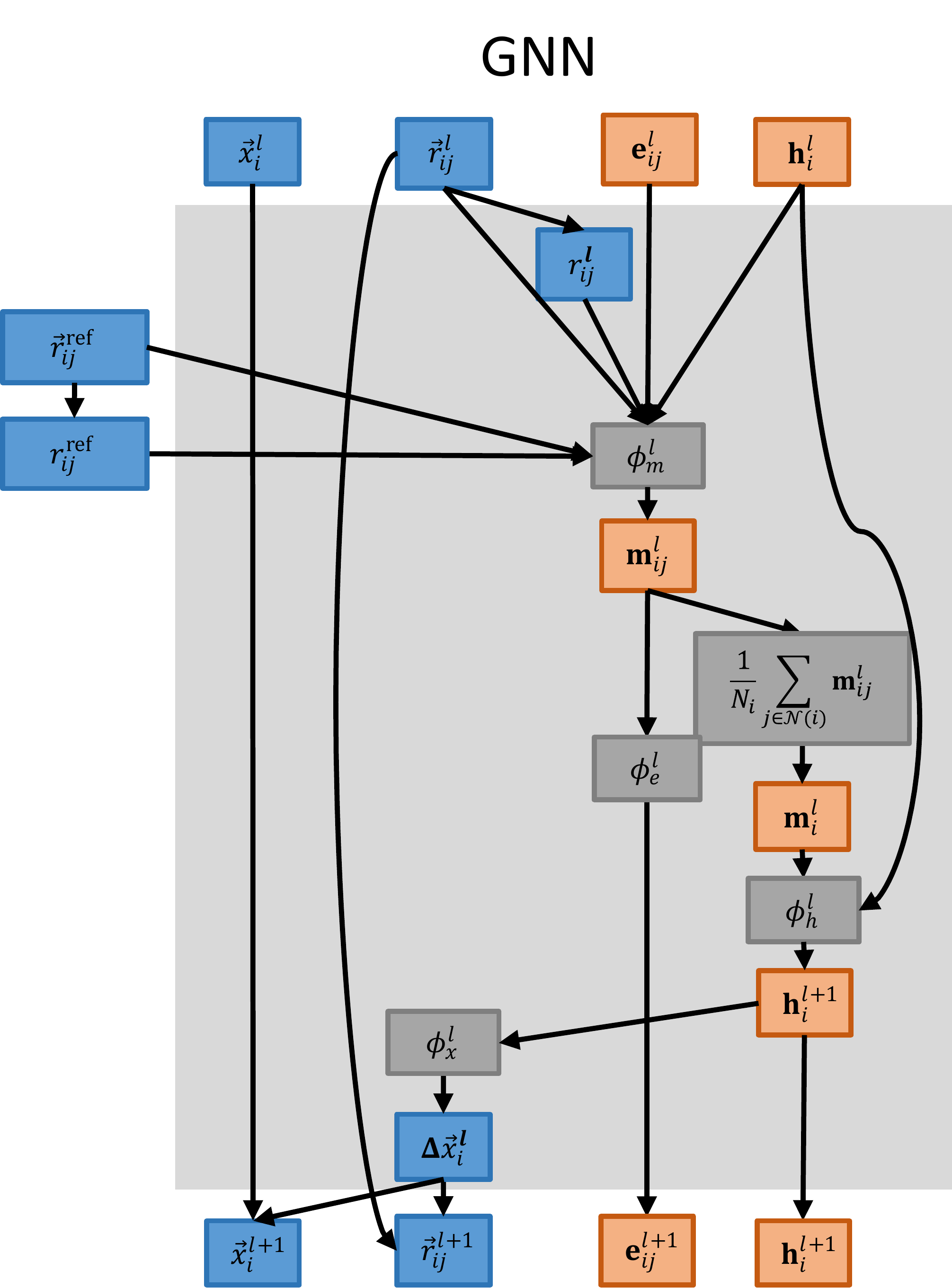}
			\caption{
				\label{fig:GNN_MP}
			}
		\end{subfigure}
		\hfill
		\begin{subfigure}[b]{0.55\textwidth}
			\centering
			\includegraphics[scale=0.2]{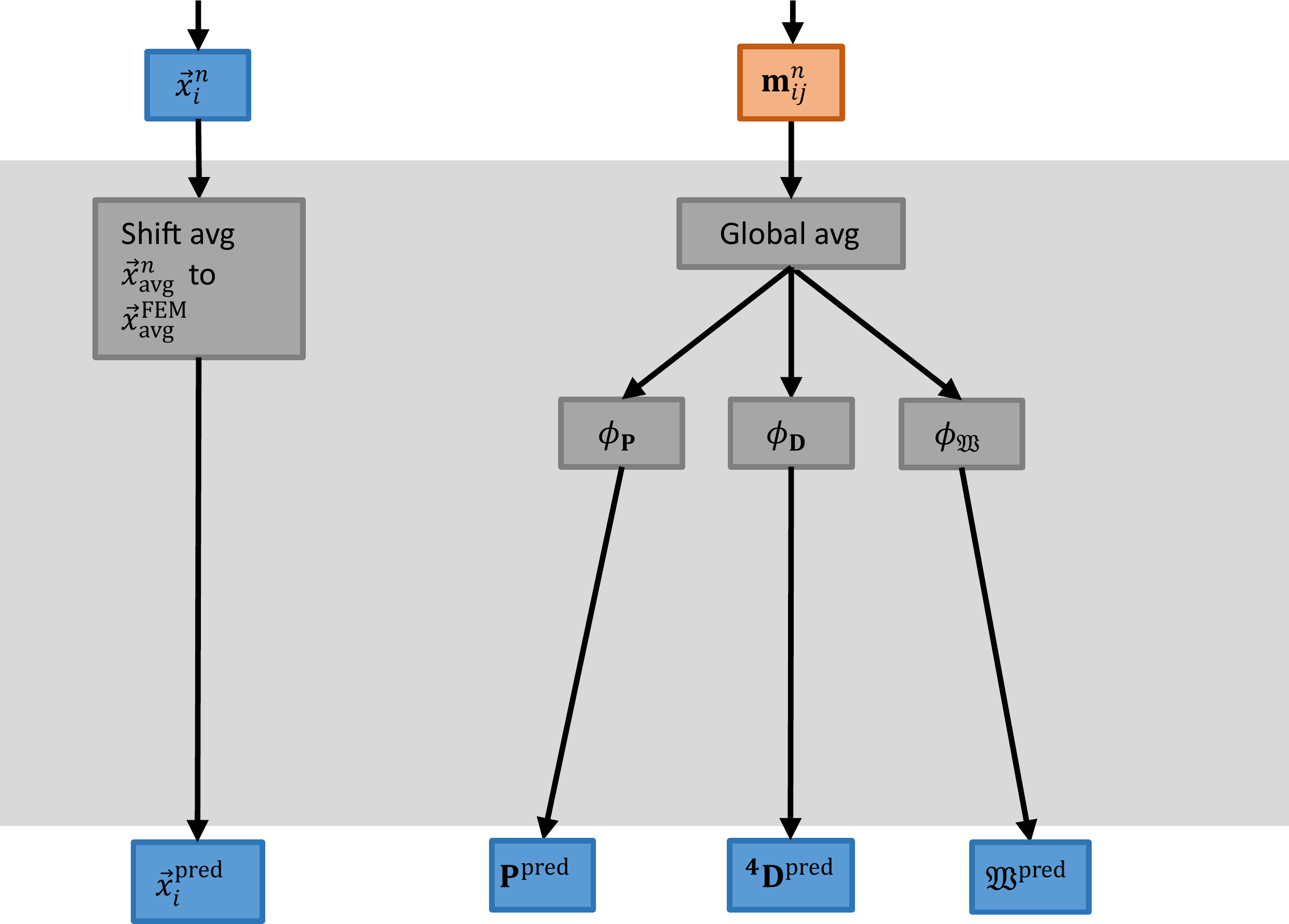}
			\caption{
				\label{fig:GNN_readout}
			}
		\end{subfigure}
		\caption{Diagrams of the base GNN as applied to metamaterial simulations. (a) One message passing step. (b) The read-out part, which computes final predictions for all quantities. The blue quantities have a physical meaning whereas the orange quantities are abstract and generated by the neural network (except for the initial edge and node embeddings $\bt e_{ij}^0$ and $\bt h_i^0$, which may also have a physical meaning since they are part of the input). The dark gray boxes indicate operations, either explicitly stated or through an underlying neural network.
		\label{fig:GNN}
		}
	\end{figure}

	From the new node embedding $\bt h_i^{l+1}$, during every message passing step the shift $\Delta\vec{x}_i$ in the position $\vec{x}_i^{\;l}$, as illustrated in Figure~\ref{fig:graph_shift_quantities}, is calculated as
	\begin{equation}
		\Delta\vec{x}_i^{\;l} = \phi_x^l\left(\bt h_i^{l+1}\right),\label{eq:GNNdx}
	\end{equation}
	with another learnable function $\phi_x^l$.
	The edge vectors are then updated correspondingly according to Equation~\eqref{eq:delta_r}.
	Figure~\ref{fig:GNN_MP} provides a flowchart of the message passing scheme of this base GNN.

	To predict the global quantities $\mathfrak{W}$, $\bt P$ and ${}^4 \bt D$, we aggregate the final messages $\bt m_{ij}^n$ (after all $n$ message passing steps) into a global abstract quantity, to which we then apply three final learnable functions $\phi_{\mathfrak{W}}$, $\phi_{P}$, $\phi_{D}$. Therefore, we predict $\mathfrak{W}$, $\bt P$ and ${}^4 \bt D$ with
	\begin{align}
		\mathfrak{W} = \phi_{\mathfrak{W}}\left(\frac{1}{\textcolor{mygreen}{N_e}} \sum_{i,j} \bt m_{ij}^n\right),\label{eq:GNNW}\\
		\bt P = \phi_P\left(\frac{1}{\textcolor{mygreen}{N_e}} \sum_{i,j} \bt m_{ij}^n\right),\label{eq:GNNP}\\
		{}^4 \bt D = \phi_D\left(\frac{1}{\textcolor{mygreen}{N_e}} \sum_{i,j} \bt m_{ij}^n\right),\label{eq:GNND}
	\end{align}
	where $\textcolor{mygreen}{N_e}$ is the total number of edges and $\phi_{\mathfrak{W}}$, $\phi_P$ and $\phi_D$ are learnable functions.
	See Figure~\ref{fig:GNN_readout} for a flowchart of this last part of the GNN.

	In Equation~\eqref{eq:GNNmessage}, the model treats different components of the edge vectors simply as different inputs. In Equations \eqref{eq:GNNdx}, \eqref{eq:GNNP} and \eqref{eq:GNND} different components of $\Delta \vec{x}$, $\bt P$ and ${}^4 \bt D$ are treated as different outputs. This implies that the network treats these components as unrelated, which means the model does not respect rotation, reflection or scaling in-/equivariance. It respects only translation in-/equivariance, because then all the inputs and outputs stay the same, as well as RVE in-/equivariance because of the periodic boundary conditions. Hence, this model will struggle with generalizing over rotations, reflections and scaling.

    \subsection{EGNN}\label{subsec:EGNN}
\begin{figure}
		\centering
		\begin{subfigure}[b]{0.45\textwidth}
			\centering
			\includegraphics[scale=0.2]{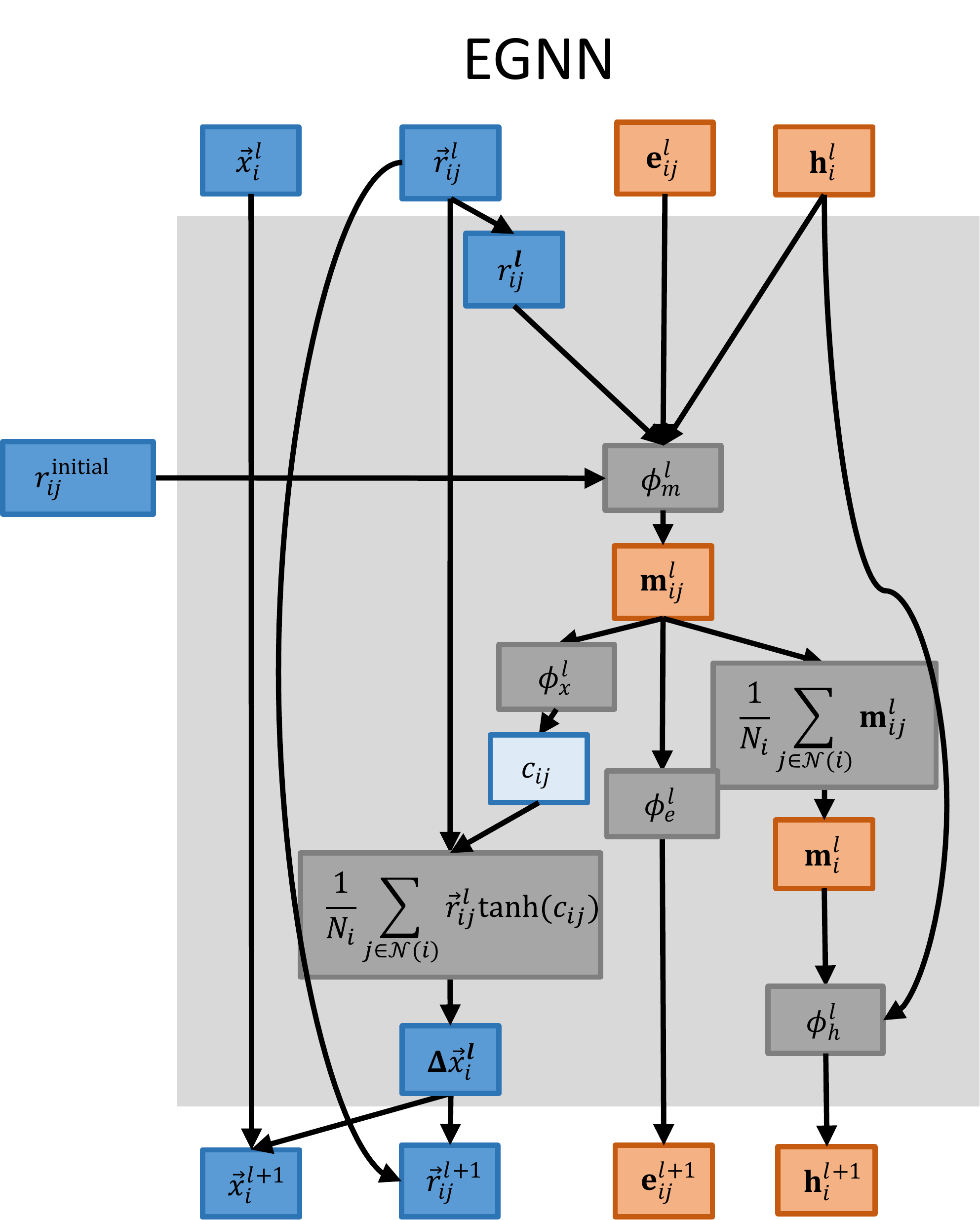}
			\caption{
			\label{fig:EGNN_MP}}
		\end{subfigure}
		\hfill
		\begin{subfigure}[b]{0.53\textwidth}
			\centering
			\includegraphics[scale=0.2]{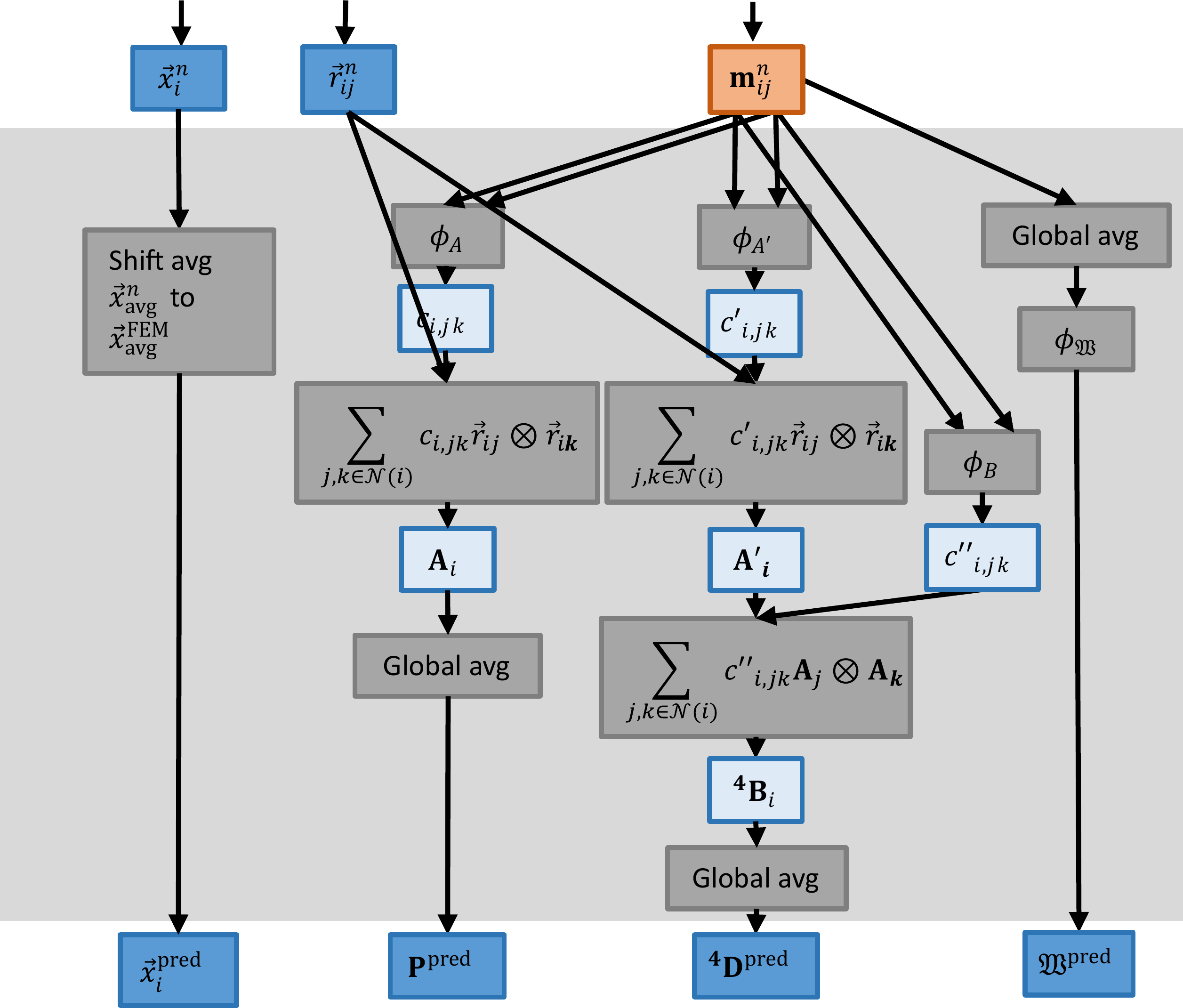}
			\caption{
			\label{fig:EGNN_readout}}
		\end{subfigure}
		\caption{
			Diagrams of the EGNN as applied to metamaterial simulations. (a) One message passing step, which iteratively updates the node positions and node and edge embeddings, (b) the read-out part, which computes final predictions for all quantities. Blue quantities have a physical meaning, and the orange quantities are abstract quantities generated by the neural network (except for the initial edge and node embeddings $\bt e_{ij}^0$ and $\bt h_i^0$, which may also have physical meaning as they are part of the input); finally, light blue quantities are not abstract but do not have a direct physical interpretation either. The dark gray boxes indicate operations, either explicitly stated or through an underlying neural network.
		}
		\label{fig:EGNN}
	\end{figure}

	Even though all positions must be expressed in a coordinate system, the way they evolve is independent of this coordinate system. Therefore, for our second GNN, we aim to update the positions in the chosen coordinate system, in a way that is independent of that coordinate system. To this end, we use $E(n)$-equivariant graph neural networks (EGNNs) by Satorras et al. \cite{Satorras2021a}.
	\footnote{The original paper \cite{Satorras2021a} also includes velocities $\vec v_i$. However, these are not relevant for our purposes, because we only consider a quasi-static system.} This is achieved by pulling or pushing the nodes along their edges, i.e., the shift $\Delta \vec{x}_i$ in the node position $\vec{x}_i$ is computed by using vectors $\vec{r}_{ij}$ as an overcomplete basis set.\footnote{Meaning, $\vec{x}_i$ is computed as a linear combination of $\vec{r}_{ij}$. This basis set is overcomplete because the number of edge vectors $\vec{r}_{ij}$ connected to node $i$ will always be larger than the number of dimensions.} The message passing scheme is described by the following relations:
	\begin{align}
		\bt m^l_{ij} &= \phi_m\left(\bt h_i^{l}, \bt h_j^{l}, r_{ij}^{\;l}, \bt e_{ij}\right) \label{eq:satorrasmessage}, \\
		\Delta \vec x_i &= \frac{1}{|\mathcal{N}(i)|}\sum_{j\in \mathcal{N}(i)} \vec r_{ij}^{l} \tanh{\left(\phi_x(\bt m^l_{ij})\right)}\label{eq:shiftx}, \\
		\bt m_i^l &= \frac{1}{|\mathcal{N}(i)|}\sum_{j\in \mathcal{N}(i)} \bt m^l_{ij}, \\
		\bt e_{ij}^{l+1} &= \phi_e\left(\bt h_i^{l}, \bt m_i^l\right), \\
		\bt h_i^{l+1} &= \phi_h\left(\bt h_i^{l}, \bt m_i^l\right),\label{eq:satorrash}
	\end{align}
	and summarized in the flowchart of Figure~\ref{fig:EGNN_MP}.
	In the expressions above, $\bt m_i^l$ denotes the aggregated message that node $i$ receives.
	\textcolor{mygreen}{
	The hyperbolic tangent in Equation~\eqref{eq:shiftx} stops the network from shifting the nodes over large distances. Without this operator, we observed that for some configurations the distances would grow exponentially with the message-passing steps. This happens because the node movements are based on a linear combination of the edge vectors, making the movements proportional to the distances. As a result, the distances can increase exponentially with the message-passing steps.
	}
	In this approach, when there is a change in the coordinate system, the edge vectors $\vec{r}^{\;l}_{ij}$ are changed correspondingly. Consequently, since the predicted shifts in nodal positions are linear combinations of these edge vectors and the scalars $\tanh{\left(\phi^l_x(\bt m^l_{ij})\right)}$ are independent of the coordinate system, the shifts $\Delta\vec{x}^{\;l}_i$ also change with the coordinate system, resulting in the desired $E(n)$-equivariance. For a proof of this, see \cite{Satorras2021a}.

	To predict second- and fourth-order tensors (stress and stiffness, respectively) in an $E(n)$-equivariant way, we take inspiration from the calculation of $\Delta \vec{x}^{\;l}_i$ and use the edge vectors $\vec{r}_{ij}$ as an overcomplete basis set again, although the final ones $\vec{r}^{\;n}_{ij}$ this time. A second-order tensor for each node $i$ can be constructed using dyadic products of $\vec{r}^{\;n}_{ij}$, i.e.,
	\begin{align}
	\bt A_i &= \frac{1}{|\mathcal{N}(i)|^2}\sum_{j\in \mathcal{N}(i)}\sum_{k\in \mathcal{N}(i)} c_{i,jk} \vec{r}_{ij} \otimes \vec{r}_{ik},\label{eq:tensor2}\\
	c_{i,jk} &= \phi_A \left(\bt m^n_{ij}, \bt m^n_{ik}\right)\label{eq:tensor2co},
	\end{align}
	where $c_{i,jk}$ is a scalar, computed with a learnable layer $\phi_A$, for each triplet $i,j,k$ where node $j$ and $k$ are from the neighborhood $\mathcal{N}(i)$ of node $i$. For the maximum number of neighbors of the graph in Figure~\ref{fig:removebulk} $|\mathcal{N}(i)| = 8$, Equation~\eqref{eq:tensor2} has $|\mathcal{N}(i)|^2 = 64$ terms, which is reasonable.

	Theoretically, we can extend this approach to predict a fourth-order tensor as follows
	\begin{align}
		{}^4\bt{B}_i &= \frac{1}{|\mathcal{N}(i)|^4}\sum_{j\in \mathcal{N}(i)}\sum_{k\in \mathcal{N}(i)}\sum_{l\in \mathcal{N}(i)}\sum_{m\in \mathcal{N}(i)} c_{i,jklm} \vec{r}_{ij} \otimes \vec{r}_{ik} \otimes \vec{r}_{il} \otimes \vec{r}_{im},\label{eq:tensor4}\\
		c_{i,jklm} &= \phi_B \left(\bt m^n_{ij}, \bt m^n_{ik}, \bt m^n_{il}, \bt m^n_{im}\right)
	\end{align}
	where $c_{i,jklm}$ is a scalar, computed with a learnable layer $\phi_B$, for each combination of edge vectors. However, for the maximum number of neighbors $|\mathcal{N}(i)| = 8$ this results in $|\mathcal{N}(i)|^4 = 4096$ terms, which becomes computationally too expensive.
	For that reason, we first predict a second-order tensor $\bt A'_i$ for each node $i$ using the approach in Equations \eqref{eq:tensor2} and \eqref{eq:tensor2co}, and then use these tensors of neighboring nodes as a basis to predict a fourth-order tensor
	\begin{align}
		{}^4\bt{B}_i &= \frac{1}{|\mathcal{N}(i)|^2}\sum_{j\in \mathcal{N}(i)}\sum_{k\in \mathcal{N}(i)} c'_{i,jk} \bt{A}'_{j} \otimes \bt{A}'_{k},\label{eq:tensor4_v2}\\
		c'_{i,jk} &= \phi_B \left(\bt m^n_{ij}, \bt m^n_{ik}\right).
	\end{align}

	In this approach, because the predicted second and fourth-order tensors are linear combinations of dyadic/tetradic products of the edge vectors and the scalars $c_{i,jk}$ are independent of the coordinate system, the $\bt A_i$ and ${}^4\bt{B}_i$ are transformed as tensors under a change in the coordinate system, resulting in the desired $E(n)$-equivariance for these tensors as well.

	For the final prediction of the global $\bt P$ and ${}^4 \bt D$, we take a global average of the per-node second- and fourth-order tensors, such that
	\begin{align}
		\bt P_{\textrm{pred}} &= \frac{1}{N} \sum_i^N \bt A_i \\
		{}^4\bt D_{\textrm{pred}} &= \frac{1}{N} \sum_i^N \bt {}^4\bt{B}_i.
	\end{align}
	The whole read-out part of the EGNN is illustrated with the flowchart in Figure~\ref{fig:EGNN_readout}.

	All the predictions of this model respect $E(n)$-in-/equivariance, as well as RVE in-/equivariance because of the periodic boundary conditions. Only scale in-/equivariance is missing. Hence, this model will struggle with generalizing over scaling.

	Since we are modeling a conservative system, an obvious alternative approach to predict tensors would be using Sobolev training \cite{Czarnecki} (i.e., training not just for a certain output, but also for one or more of its derivatives). This means the network would predict only the predicted energy density $\mathfrak{W}$. The macroscopic first Piola-Kirchhoff stress tensor and the stiffness tensor would then be calculated from $\mathfrak{W}$ according to Equations \eqref{eq:homoP} and \eqref{eq:homoD}. This means calculating $\bt P = \pdv{\mathfrak{W}}{\bt F}$ and ${}^4 \bt D = \pdv[2]{\mathfrak{W}}{\bt F}$ using the autodifferentiation that any package for neural network training such as PyTorch is capable of, and then training for $\mathfrak{W}$ as well as $\bt P$ and ${}^4 \bt D$.
	Many molecular GNNs similarly train for the energy as well as the forces \cite{Schutt2018,Gasteiger}.
	This approach would also guarantee that the differentiability relations in Equations \ref{eq:homoP} and \ref{eq:homoD} are exactly satisfied, which is not the case with the current approach.
	However, we found that for the second derivative ${}^4 \bt D$ this is prohibitively slow and unstable.
	Presumably \textcolor{mygreen}{the instability} is due to the repeated differentiation, which means a small change in $\mathfrak{W}$ can lead to a large change in ${}^4 \bt D$. 
	For the first derivative $\bt P$ this approach is nevertheless feasible, but still not as fast or accurate as predicting $\bt P$ directly.

	\subsection{Similarity-Equivariant GNN (SimEGNN)}\label{subsec:SimEGNN}
\begin{figure}
		\centering
		\begin{subfigure}[b]{0.45\textwidth}
			\centering
			\includegraphics[scale=0.2]{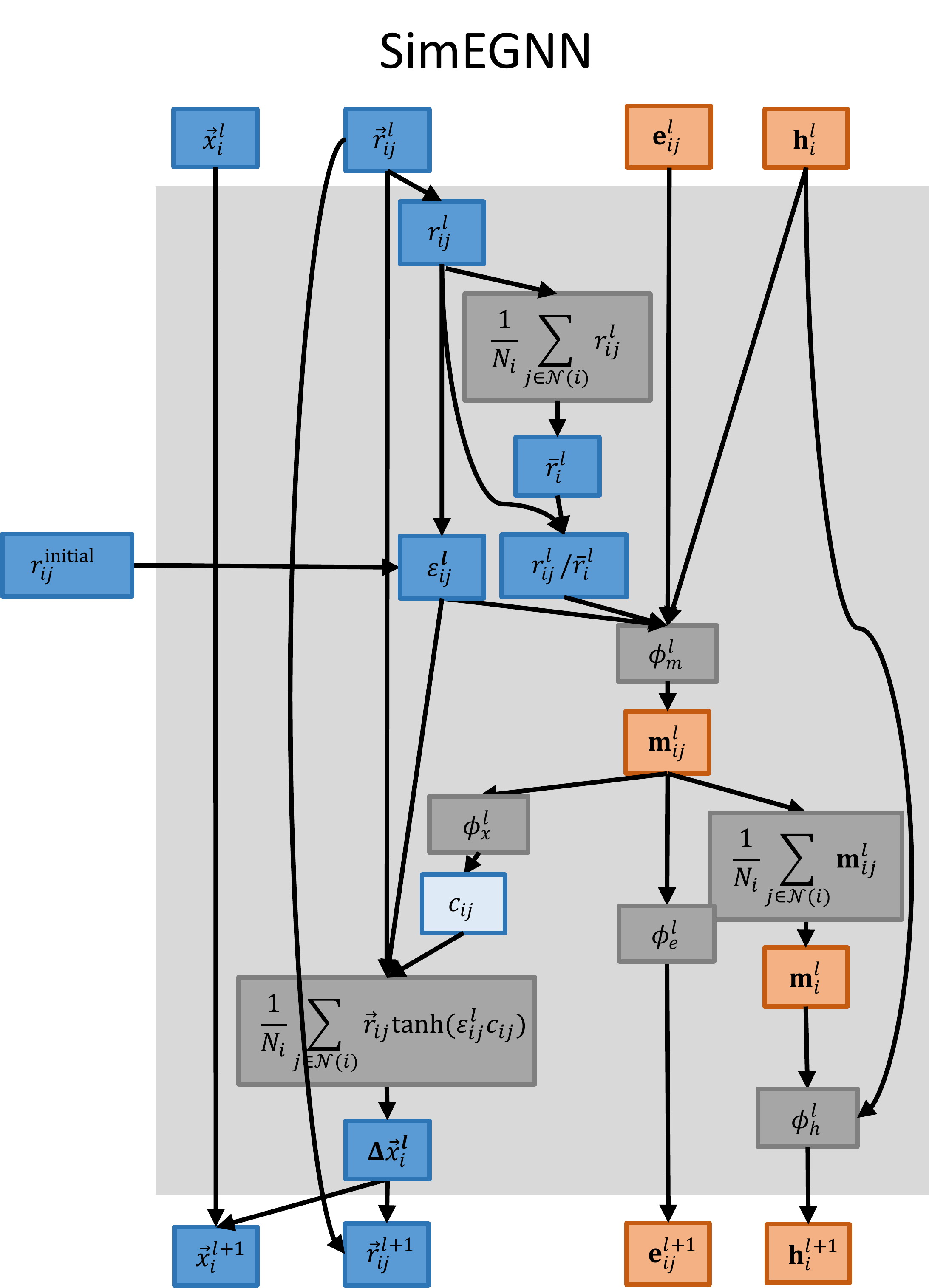}
			\caption{
			\label{fig:SimEGNN_MP}}
		\end{subfigure}
		\hfill
		\begin{subfigure}[b]{0.53\textwidth}
			\centering
			\includegraphics[scale=0.2]{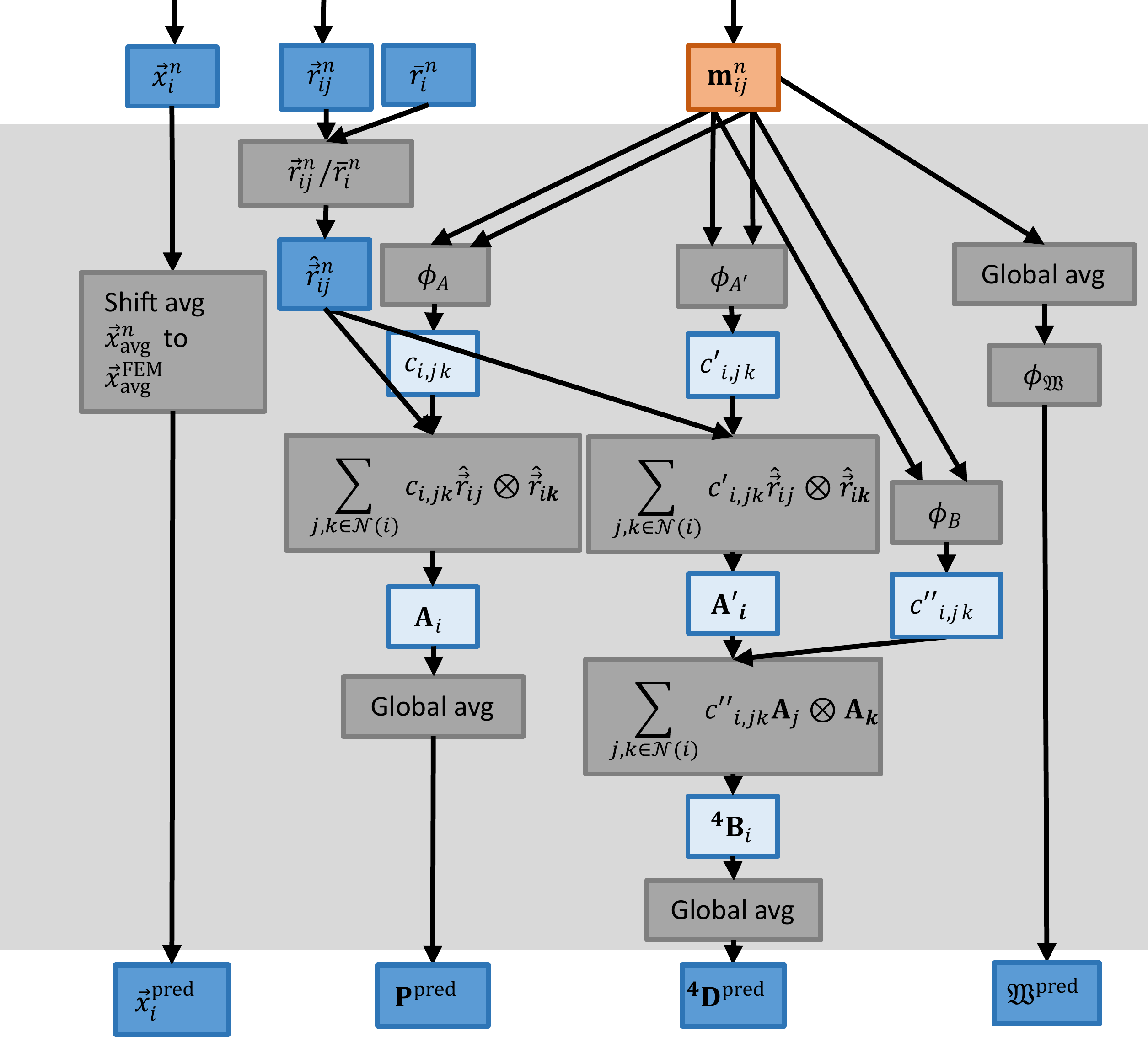}
			\caption{
			\label{fig:SimEGNN_readout}}
		\end{subfigure}
		\caption{
			Diagrams of the SimEGNN as applied to metamaterial simulations. (a) One message passing step, which iteratively updates the node positions and node and edge embeddings, (b) the read-out part, which computes final predictions for all quantities. Blue quantities have a physical meaning, and the orange quantities are abstract quantities generated by the neural network (except for the initial edge and node embeddings $\bt e_{ij}^0$ and $\bt h_i^0$, which may also have a physical meaning since they are part of the input); finally, light blue quantities are not abstract but do not have a direct physical interpretation either. The dark gray boxes indicate operations, either explicitly stated or through an underlying neural network.
		}
		\label{fig:SimEGNN}
	\end{figure}

	Finally, the last architecture developed and tested in this work is \textcolor{mygreen}{the Similarity-Equivariant GNN (SimEGNN)}, possessing all required in-/equivariances of Table~\ref{tab:inequivariances}.
	The Euclidean group $E(n)$ appearing in the previous architecture includes transformations that preserve Euclidean distances (i.e., rotations, reflections, translations). However, in order to apply it to first-order homogenization problems, we also require scale in-/equivariance. The group equivariance we need is thus with respect to the similarities group $S$, which also includes isotropic scaling and preserves distance ratios. To achieve this, instead of the distance between nodes $r_{ij}^l$ in the message computation of Equation~\eqref{eq:satorrasmessage}, we use the strain measure defined as $\varepsilon_{ij}^{l} = (r_{ij}^{l} - r_{ij}^{\textrm{ref}})/r_{ij}^{\text{ref}}$, \textcolor{mygreen}{as well as a relative distance $r_{ij}^l/\bar{r}_i^{\;l}$, with $\bar{r}_i^{\;l}$ defined as
	\begin{align}
		\bar{r}_i^{\;l} &= \frac{1}{|\mathcal{N}(i)|}\sum_{j\in \mathcal{N}(i)} r_{ij}^{\;l}.\\
	\end{align}
	This feature means that although each node does not know the absolute lengths of its incoming edges, it does know which incoming edges are longer or shorter than the others.	The message computation then looks as follows:}
	\begin{equation}\label{eq:SimEGNN_m}
		\bt m^l_{ij} = \phi_m^l\left(\bt h_i^{l}, \bt h_j^{l}, \varepsilon_{ij}^{l}, r_{ij}^l/\bar{r}_i^{\;l}, \bt e_{ij}^l\right).
	\end{equation}
	Strains are also used in the computation of the shift in Equation~\eqref{eq:shiftx}, i.e.,
	\begin{equation}\label{eq:SimEGNN_dx}
		\Delta \vec x_i^{\;l} = \frac{1}{|\mathcal{N}(i)|}\sum_{j\in \mathcal{N}(i)} \vec r_{ij}^{\;l} \tanh{\left(\varepsilon_{ij}^l \phi_x^l(\bt m^l_{ij})\right)}, \\
	\end{equation}
	which guarantees that there is no position shift for $\bt F = \bt I$, because then all strains $\varepsilon^l_{ij}$ will be zero, which will make all $\Delta \vec{x}^{\;l}_i$ equal to zero. The resulting message passing scheme is shown in the flowchart of Figure~\ref{fig:SimEGNN_MP}.
	Because $\mathfrak{W}$, $\bt P$ and ${}^4 \bt D$ are scale \emph{invariant}, we normalize the final edge vectors $\vec{r}^{\;n}_{ij}$ used as an overcomplete basis set to predict node-wise tensors $\bt A_i$ (recall Equation~\eqref{eq:tensor2}):
	\begin{equation}
		\bt A_i = \frac{1}{|\mathcal{N}(i)|^2}\sum_{j\in \mathcal{N}(i)}\sum_{k\in \mathcal{N}(i)} c_{i,jk} \hat{\vec{r}}^{\;n}_{ij} \otimes \hat{\vec{r}}^{\;n}_{ik},
	\end{equation}
	where \textcolor{mygreen}{$\hat{\vec{r}}^{\;n}_{ij} = \vec{r}^{\;n}_{ij}/\bar{r}^{\;n}_{i}$} are the normalized \textcolor{mygreen}{final} edge vectors.
	This read-out part of the SimEGNN is illustrated with the flowchart in Figure~\ref{fig:SimEGNN_readout}.

	\section{Results}\label{sec:results}
	The model evaluation consists of \textcolor{mygreen}{2 parts: The first part in Section~\ref{subsec:effect_symmetries} focuses on the effect of embedding the symmetries on test cases that reflect those symmetries. The second part in Section~\ref{subsec:generalization} focuses on the generalization of the models to different geometries, by testing the models on microstructures with different hole sizes. Finally a discussion on efficiency and scaling is included in Section~\ref{subsec:scaling}.}

	\textcolor{mygreen}{\subsection{Effect of Embedding Symmetries}\label{subsec:effect_symmetries}
	We test} 7 different versions of the 3 graph neural networks described in Section~\ref{sec:GNN}, from the smallest to the largest number of symmetries, on 6 different test cases, for 4 target quantities. We test them on the homogenization problem introduced in Section~\ref{sec:dataset}.
	\textcolor{mygreen}{
		We split that data set by trajectory, such that the training and validation data are from different trajectories, with 80\% of the trajectories in the training data and 20\% in the validation data. Because we do 5-fold cross-validation, each time we choose a different 20\%. The data set is described in more detail in \ref{sec:datalayout}.
	}
	In addition to the three types of GNN architectures, for the GNN and EGNN, we tested augmenting the training data as described in Section~\ref{par:data_augmentation}, as a different way to enforce the symmetries. The models trained on augmented data are marked `DA' for `Data Augmentation'.
	This resulted in \textcolor{mygreen}{5 models in total: 2 GNNs, 2 EGNNs and 1 SimEGNN. As mentioned, we perform 5-fold cross-validation for each model, resulting in 5 sets of results per model type.}
	The results cover 6 different test cases:
	\begin{enumerate}
		\item the validation data (labeled `untransformed', where none of the configurations have been rotated, reflected, translated or scaled),
		\item the validation data reflected in the $y$-axis (`reflected'),
		\item rotated by $\pi/4$ radians (`rotated'),
		\item with a shifted RVE, which means the material is shifted up and left by one quarter the size of the RVE, such that the RVE is centered on one hole (labeled `shifted RVE'),
		\item with a larger RVE that consists of four original RVEs merged together (`extended RVE'),
		\item and an RVE which has been scaled by a factor 1.5 (`scaled').
	\end{enumerate}
	These test cases represent the various in-/equivariances of Figure~\ref{fig:equivariances} and Table~\ref{tab:inequivariances} that we built into the model, and were chosen to evaluate the effect of embedding these in-/equivariances.
	The 4 different target quantities are the microfluctuation field $\vec{w} = \vec{x} - \bt F \cdot \vec{x}^{\;\textrm{ref}}$ (see Equation~\ref{eq:xdecomp}), homogenized strain energy density $\mathfrak{W}$, homogenized stress $\bt P$ and tangent stiffness tensor ${}^4 \bt D$.

\begin{figure}
        \includegraphics[width=\textwidth]{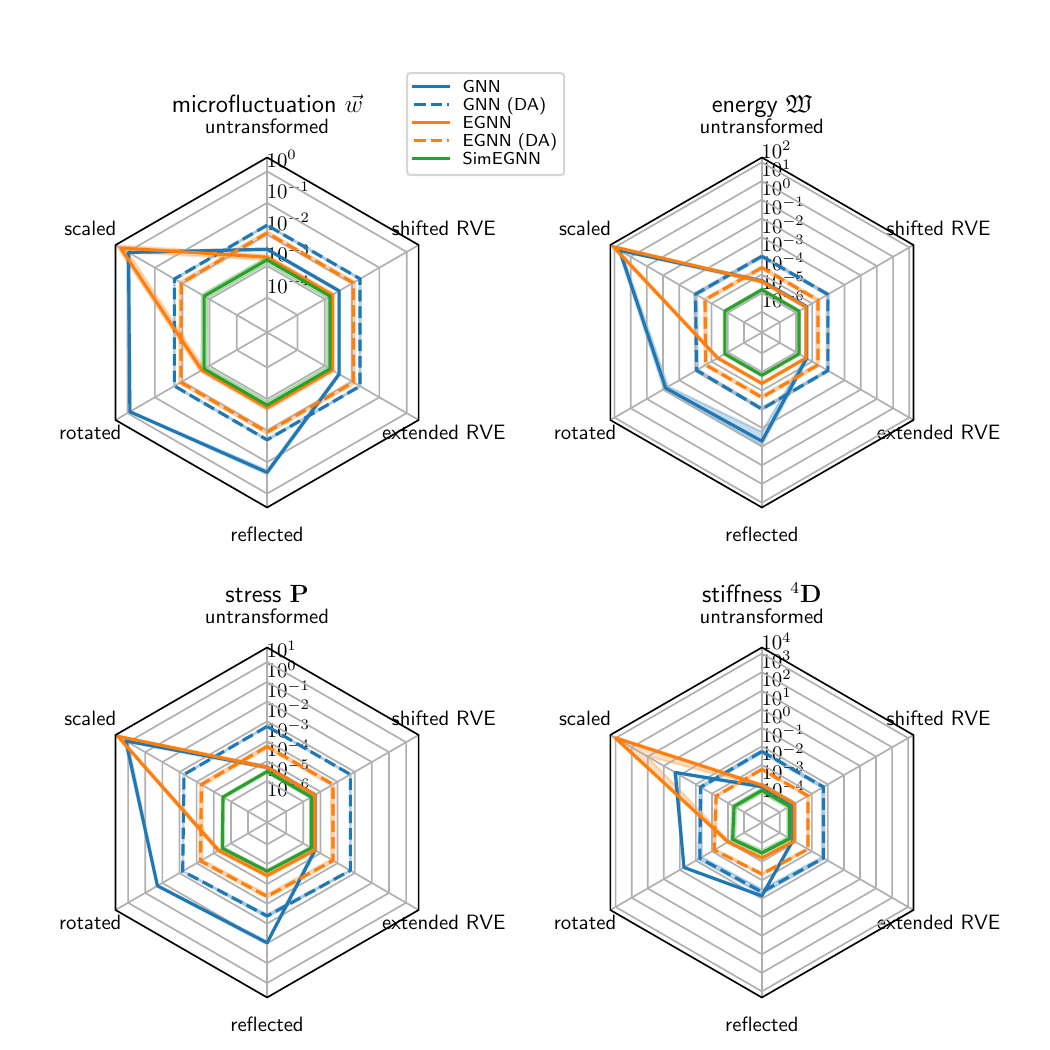}
        \caption{\textcolor{mygreen}{Mean} fraction of variance unexplained of the four target quantities ($\vec{w}$, $\mathfrak{W}$, $\bt{P}$, $\bt{D}$) achieved by the investigated models (shown in different line colors and styles) for the six test cases (untransformed, reflected, rotated, shifted RVE, larger RVE, scaled). \textcolor{mygreen}{The mean is calculated over the 5 folds from the cross-validation. The standard error is indicated by a shaded area. However, it is visible mainly for $\mathfrak{W}$ in the reflected case by the GNN and for ${}^4 \bt D$ in the scaled case by the EGNN.} The SimEGNN is the only model that respects all symmetries, and it outperforms the other models.
		\label{fig:spiderplots}}
    \end{figure}

	The spider web charts in Figure~\ref{fig:spiderplots} show the \textcolor{mygreen}{mean} fraction of variance unexplained (FVU) obtained by each neural network architecture, \textcolor{mygreen}{averaged over the 5 cross-validation models. Shaded areas indicate the standard error (although it is difficult to see because it is very narrow).}
	\textcolor{mygreen}{
		The FVU of the prediction of a quantity $y$ is defined as
		\begin{equation}
			\text{FVU} = \frac{\sum_i^N (y_i - \hat{y}_i)^2}{\sum_i^N (y_i - \bar{y})^2} = \frac{\text{MSE}(y, \hat{y})}{\text{Var}(y)},
		\end{equation}
		where $\hat{y}$ is the prediction, $\bar{y}$ is the average value of the target, and Var$(y)$ is the variance of the target and MSE the mean squared error between the real and predicted values.
		This means the FVU is the variance in the errors of quantities predicted by the models divided by the variance in the data of the target quantity\footnote{i.e., $\text{FVU} = 1 - R^2$, where $R^2$ is the coefficient of determination}.
	}
	Each chart corresponds to one target quantity: $\vec{w}$, $\mathfrak{W}$, $\bt{P}$ or $\bt{D}$. The investigated models are shown in different line colors and styles, where the color indicates the type of neural network (GNN, EGNN or SimEGNN) and the line style (solid or dashed) indicates the \textcolor{mygreen}{whether data augmentation was used during training}. The different spokes of the spider plots correspond to the six test cases (reference, reflected, rotated, shifted RVE, larger RVE, scaled).

\begin{figure}
		\centering
        \includegraphics[width=0.8\textwidth]{"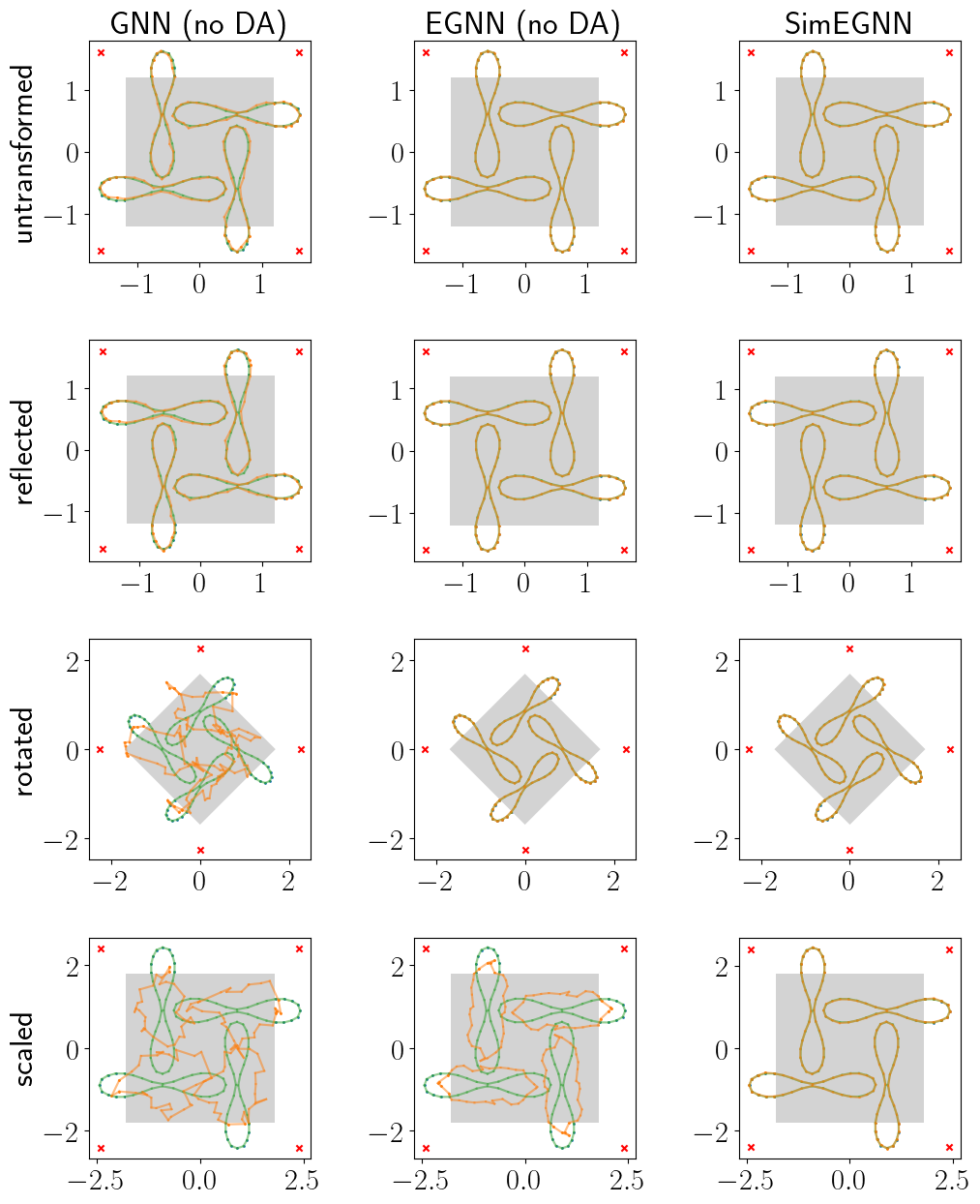"}
        \caption{Predicted deformation, using the three models without data augmentation, of the hole boundaries (orange) compared to the FEM ground truth (green), for $\bt F=\begin{bmatrix}0.75 & 0\\ 0 & 0.75\end{bmatrix}$ (biaxial compression, resulting in a rotational pattern).\label{fig:deformations1}
		}
    \end{figure}

	Additionally, for four of these six test cases, we show a comparison of the predicted hole deformations from three of the networks against FEM outputs -- considered the ground truth in this work -- for biaxial compression in Figure~\ref{fig:deformations1}.  We exclude the `Shifted RVE' and `Extended RVE' cases, because all models respect periodicity. The Figure only shows the networks trained without data augmentation. \textcolor{mygreen}{For the sake of completeness, we show figures that include all models in \ref{sec:fullresultsapp}, including the ones trained with data augmentation. These figures also show the results for different $\bt F$, such that all bifurcation patterns are included.}
	Tables with the data behind Figure~\ref{fig:spiderplots} are \textcolor{mygreen}{also} provided in \ref{sec:fullresultsapp}, as well as tables with the relative error.

	The results of Figure~\ref{fig:spiderplots} show that the periodicity is correctly implemented in all models, because all of them can predict the `shifted RVE' and `larger RVE' cases with an FVU equal to the untransformed case. By construction, the EGNN indeed additionally respects $E(n)$-in-/equivariance (`reflected' and `rotated', translated not shown in Figure~\ref{fig:spiderplots}), and the SimEGNN additionally respects scale in-/equivariance.\footnote{The EGNN and SimEGNN still show a small deviation in the FVU for the rotated and reflected cases for the stress $\bt P$ and stiffness ${}^4 \bt D$; this is because those transformations slightly change the variance in the ground truth data. The mean square error does remain the same, however.}
	The figure also shows that the microfluctuation $\vec{w}$ and stiffness ${}^4 \bt D$ are generally more difficult to predict than the strain energy density $\mathfrak{W}$ and the stress $\bt P$; in the plots for $\vec{w}$ and ${}^4 \bt D$, the FVU stays above $10^{-4}$, whereas in the plots for $\mathfrak{W}$ and $\bt P$ the FVU gets close to $10^{-6}$.

	For GNNs (the blue lines in Figure~\ref{fig:spiderplots}), augmenting the data with reflected, rotated and scaled cases improves the predictions for those cases, i.e., it makes the result more similar to the `untransformed' result. However, it worsens the `untransformed' result at the same time. This indicates that the model is trying to learn all the equivalent cases separately, which effectively reduces the capacity of the network. A similar effect can be seen for the EGNN (orange lines in Figure~\ref{fig:spiderplots}) and the `scaled' case.

	Figure~\ref{fig:spiderplots} shows that with all symmetries incorporated directly into the network's architecture, the FVU improves by about an order of magnitude or more, compared to the GNN with 2× data augmentation, on all target quantities. This holds for all test cases, even for the untransformed case, where none of the configurations have been rotated, reflected, translated or scaled.
	Incorporating more symmetries thus makes it easier for the model to construct and use features in a general way. For example, it can more easily transfer learned features from one edge to predict the behavior of another longer edge with scale in-/equivariance. Embedding the symmetries thus allows the network to focus on constructing actually meaningful features, because it does not first have to filter out the aspects that are irrelevant (scale, rotation, etc.).
	Consequently, the newly proposed SimEGNN outperforms the other models in all aspects.

	\textcolor{mygreen}{
	\subsection{Generalization to Different Geometries}}
	\label{subsec:generalization}

	\textcolor{mygreen}{
	As a first test of the generalization capabilities of the above-introduced models, we consider microstructures with hole size diameters of $0.4\ell$, $0.425\ell$, $0.45\ell$ and $0.475\ell$ (corresponding to minimum ligament thicknesses of $0.1\ell$, $0.075\ell$, $0.05\ell$ and $0.025\ell$, respectively). We test the three architectures, and additionally train with data augmentation for the GNN and the EGNN.}

\begin{figure}
		\centering
        \includegraphics[width=\textwidth]{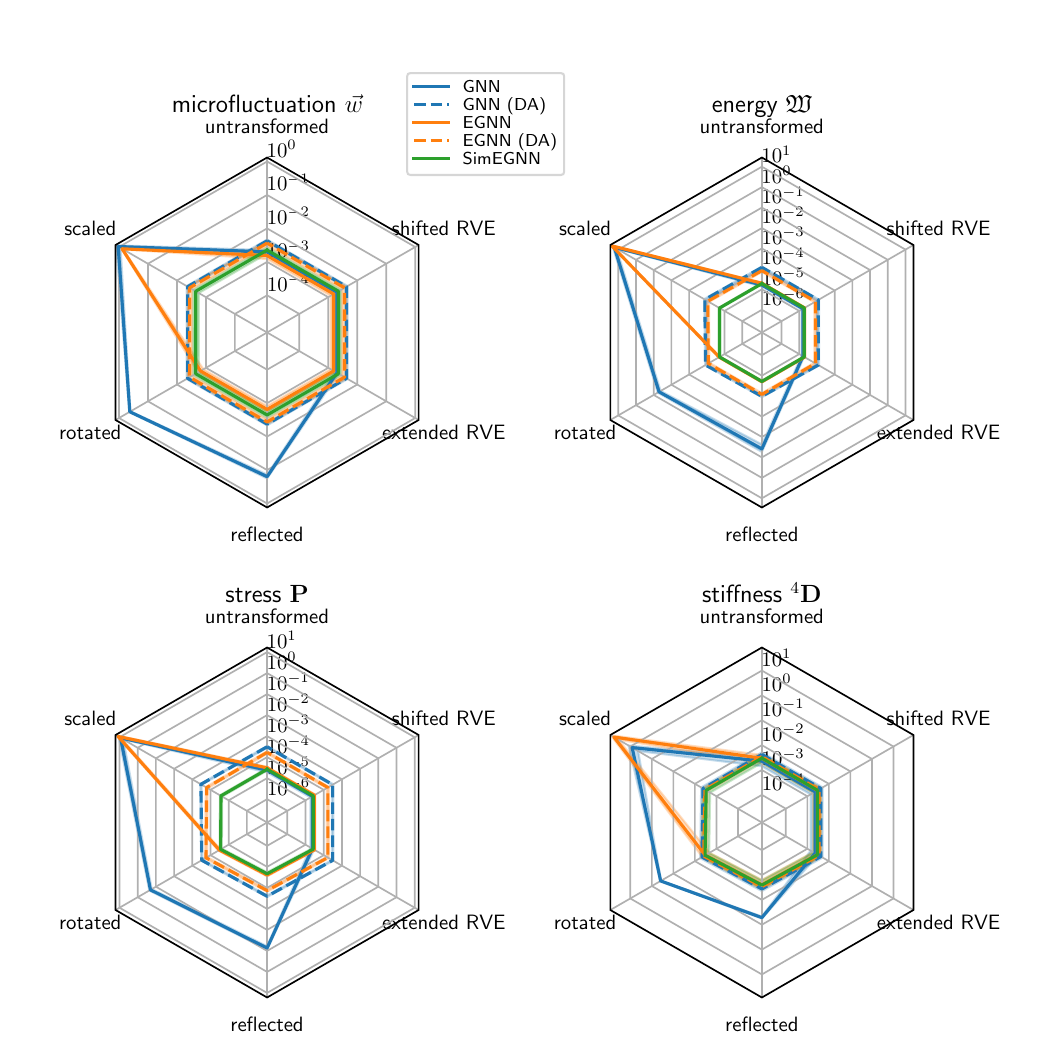}
        \caption{\textcolor{mygreen}{Mean fraction of variance unexplained of the four target quantities ($\vec{w}$, $\mathfrak{W}$, $\bt{P}$, $\bt{D}$) achieved by the investigated models (shown in different line colors and styles) for the six test cases (untransformed, reflected, rotated, shifted RVE, larger RVE, scaled), on the dataset with 4 different diameters. The mean is calculated over the 5 folds from the cross-validation. The standard error is indicated by a shaded area.}
		\label{fig:spiderplots4diams}}
    \end{figure}

	\textcolor{mygreen}{
	The results displayed in Figure~\ref{fig:spiderplots4diams} show that on microstructures with 4 different diameters, the models can still get performance close to the performance on the data set with just one microstructure. It also shows that the SimEGNN is still the only model that performs well on all test cases. The GNN performs really poorly on the scaled, rotated and reflected cases, and the EGNN performs badly on the scaled case. With data augmentation, the performance on these cases can be improved. However, especially for the energy $\mathfrak{W}$ and the stress $\bt P$, we see the same trade-off for data augmentation as in the previous section: without the data augmentation, the performance is very bad on cases that reflect the symmetries that are not included in the model, but with it, the performance is worse overall. For the microfluctuation $\vec{w}$ the data augmentation also makes it worse overall, but the effect is less dramatic. For the stiffness ${}^4 \bt D$, the data augmentation on average makes it worse, but in the case of the EGNN on the reference case, the standard errors overlap, which means it is unclear if this difference is significant.}

	\bigskip

	\subsection{Efficiency and Scaling} \label{subsec:scaling}
	\textcolor{mygreen}{
	Generating one prediction with the finite element method takes on average 0.93 seconds on a single-threaded CPU (an Intel(R) Core(TM) i7-9750H), and 0.49 seconds on a 12-threaded CPU. Preprocessing the results for GNNs takes 9.0 milliseconds per graph, of which 5.2 milliseconds are for removing cases with contact, which would be necessary as post-processing for FEM anyway. In inference, the SimEGNN takes 0.13 seconds to generate one prediction on a single-threaded CPU, and 0.06 seconds on a 12-threaded CPU, which gives a speed-up of 7.10 and 7.56 times, respectively. If the SimEGNN is evaluated on a GPU (NVIDIA Quadro P2000 GPU), it takes \SI{3.6}{milliseconds} to generate one prediction, which is a speed-up of 135 relative to the 12-threaded FEM.}
	\textcolor{mygreen}{However, because the training time is quite long at 11.4 hours, the initial investment into the SimGNN is quite high, and will currently only pay off if many predictions are needed. Fortunately, we expect that this trade-off can still improve in several ways in the future.
	For example,} it is likely that our SimEGNN approach can also be made faster in the most obvious way: by lowering our standards. The current model is remarkably accurate. If we require a maximum relative error of about 5\% in the homogenized target quantities (instead of the current $0.31\% - 0.90\%$), a much smaller model will probably also suffice.
	We also expect our method to scale much better when applied to finer meshes and more complicated or 3D geometries, because we only use the boundary nodes. This means the number of used nodes scales linearly with the mesh size, while the number of nodes in the finite element mesh scales quadratically (in 2D). In 3D the number of boundary nodes scales quadratically and the number of bulk nodes scales to the power 3, hence we expect the SimEGNN to scale much better than the finite element method.
	The main limitation of the presented approach is the need for at least as many message passing steps as the graph diameter, although this can possibly be addressed by, e.g., using pooling strategies and adding long-distance edges to the graph.

	\section{Conclusion}\label{sec:conclusion}
	For the development of new hyperelastic, flexible, porous, 2D mechanical metamaterials with complex microstructures, fast simulations of the homogenized response are needed. Conventional finite element simulations are often too slow for this purpose. In this paper, we present a graph neural network architecture called a Similarity-Equivariant Graph Neural Network (SimEGNN) to speed up simulations of these materials. The model is trained on finite element simulations and incorporates all relevant symmetries.
	The network predicts the final node positions $\vec{x}$ by repositioning the nodes during message passing. From the final graph, the homogenized strain energy density $\mathfrak{W}$, homogenized stress $\bt P$ and tangent stiffness tensor ${}^4 \bt D$ are predicted, which serve as fast surrogates in computational homogenization schemes.

	All necessary in-/equivariances pertinent to computational homogenization are incorporated \textcolor{mygreen}{in the proposed architecture}. To achieve this, we use $E(n)$-equivariant graph neural networks \cite{Satorras2021a} as the initial stepping stone and extend them to also
	(i) respect periodicity (i.e., RVE in-/equivariance), which we achieve by using a periodic graph (i.e., a Pac-Man world geometry) and coordinate-independent updates,
	(ii) respect scale in-/equivariance, for which we use strains \textcolor{mygreen}{and relative distances} instead of distances and normalize the edge vectors $\vec{r}_{ij}$ where necessary,
	and (iii) equivariantly output higher-order tensors, by using normalized edge vectors $\hat{\vec{r}}_{ij}$ as an overcomplete basis for predicting $\bt P$ and ${}^4 \bt D$.

	We demonstrated that the SimEGNN approximates the results of material simulations (both micro- and macroscopic response) in first-order homogenization with very high accuracy (i.e. FVU $<10^{-3}$, or $R^2>0.999$ in all cases), even for cases with large deformations.
	Our numerical experiments also confirm that SimEGNN outperforms the architectures with fewer built-in in-/equivariances (base GNN and EGNN). Even though some symmetries can also be trained for in the GNN and EGNN   explicitly incorporating them (using data augmentation), training for them is achieved at the expense of a decreased performance, \textcolor{mygreen}{which suggests the only reason the GNN and EGNN manage to get a satisfactory performance at all on the reference case -- i.e., relative error $<10\%$ for all target quantities --  is because this data set can be `standardized' in the sense one can always chose a unit cell with the same orientation and scale. Such standardization is not possible for more complicated data sets.}

	\textcolor{mygreen}{
	This is also the case for varying geometries, as shown by the results on the dataset with 4 different diameters. For a data set with even more geometries, the SimEGNN should be able to generalize with respect to geometry, i.e., predict the response of unseen geometries.
	} More detailed investigations \textcolor{mygreen}{regarding the generalization capabilities} are needed, but are left for future study. The SimEGNN's high accuracy and potential to generalize with respect to geometry make this model highly promising for the future development of new soft porous mechanical metamaterials.

	We will also address the main limitation of our current approach, which is the need for at least as many message passing steps as the graph diameter.  
	The current model can only use one type of vector as input (the edge vectors). To extend the model such that it can handle more vectors and/or tensors as input, we investigate ways to increase model expressiveness, or even higher-order body terms (e.g. \cite{Batatia2022a}),
	by allowing the features to be similarity-equivariant instead of similarity-invariant (similar to \cite{Jing2021, Gasteiger2021}). See \cite{Joshi2023} for a more in-depth look at graph neural network expressiveness.

	We also investigate approaches \textcolor{mygreen}{that can model the bifurcations without needing a perturbation that depends on the buckling mode. The symmetry-breaking involved in the pattern transformation will then cause there to be multiple correct possible outputs, and one cannot simply train on a training dataset of input/output pairs. Therefore, we are investigating approaches} based on generative models to predict all possible bifurcations.

\section{Acknowledgements}
This project has received funding from the Eindhoven Artificial Intelligence Institute (EAISI).
MD acknowledges the support of the Czech Science Foundation through Project No. 19-26143X in 2023 and co-funding by the European Union under the ROBOPROX project (reg. no. CZ.02.01.01/00/22 008/0004590) in 2024.
The authors furthermore acknowledge the High Performance Computing Lab of Eindhoven University of Technology for their support related to numerical experiments, and the NWO grant (number EINF-5845) for compute resources on the Dutch National Supercomputer Snellius.

\textcolor{mygreen}{
\section{Data Statement}
The datasets used in this study are available at \url{https://zenodo.org/records/14229619}. This repository contains the raw FEM outputs as well as the prepocessed data in the form of graphs, which can be used as input for the GNNs. We share this data under an CC BY license, such that it can be used for further research.}

\newpage
\appendix

	\section{Data Set}\label{sec:datalayout}
\begin{table}[h]
\centering
\caption{\textcolor{mygreen}{Number of trajectories and loadcases in the datasets. The dataset containing only diameter 0.45 is the data set we do most of the experiments with.}}
\label{tab:datasetdescription}
\textcolor{mygreen}{
\begin{tabular}{llr}
Total          & Trajectories & 1784   \\
               & Load cases   & 33708  \\
Diameter 0.4   & Trajectories & 446    \\
               & Load cases   & 8364   \\
Diameter 0.425 & Trajectories & 446    \\
               & Load cases   & 8404   \\
Diameter 0.45  & Trajectories & 446    \\
               & Load cases   & 8451   \\
Diameter 0.475 & Trajectories & 446    \\
               & Load cases   & 8489
\end{tabular}
}
\end{table}

	\begin{table}[h]
	\caption{\label{tab:datalayout} \textcolor{mygreen}{Data layout of an input graph for the GNNs. $N$ is the number of nodes, $N_e$ the number of edges.}}
\textcolor{mygreen}{
	\begin{tabular}{lp{8cm}ll}
	Quantity    & Description                                                                                                                    & Shape       & Data type    \\
	\hline
	\emph{Inputs} & & &\\
	\hline
	x           & Node attributes, initially empty                                                                                               & $ (N, 0) $    & float     \\
	edge\_index & Indices of nodes that are connected by an edge                                                                                 & $ (2, N_e) $     & integer    \\
	edge\_attr  & Edge attributes, indicates if edge is a hole boundary edge (-1), or an extra added edge that connects to a different hole (+1) & $ (N_e, 1) $  & float       \\
	pos         & Node positions $\vec{x}_i$ in the reference configuration                                                                      & $ (N, 2) $    & float     \\
	r           & Edge vectors $\vec{r}_{ij}$ in the reference configuration                                                                     & $ (N_e, 2) $    & float     \\
	d           & Edge lengths $r_{ij}$ in the reference configuration                                                                           & $ (N_e, 1) $    & float     \\
	mean\_pos   & Average position $\vec{x}_{avg}$ of the nodes in the reference configuration                                                   &  $(1, 2) $  & float       \\
	F           & Applied macroscale deformation gradient tensor $\bt F$                                                                     &  $(1, 2, 2) $  & float    \\
	traj        & Index indicating to which trajectory this load case belongs                                                                 &  $(1,) $ & integer \\
	\hline
	\emph{Target outputs} & & &\\
	\hline
	y           & Final node positions (the target positions)                                                                                              & $ (N, 2) $   & float      \\
	W           & Strain energy density $\mathfrak{W}$                                                                                       &  $(1,)$     & float       \\
	P           & First Piola-Kirchhoff stress tensor $\bt P$                                                                                & $ (1, 2, 2) $  & float    \\
	D           & Effective stiffness tensor ${}^4 \bt D$                                                                                    & $ (1, 2, 2, 2, 2) $ & float
	\end{tabular}
}
	\end{table}
	\newpage

	\section{Implementation}
	\label{sec:experimental_setup}
	All GNN architectures involve message passing steps; recall Figures \ref{fig:GNN_MP}, \ref{fig:EGNN_MP} and \ref{fig:SimEGNN_MP}. One can reuse the parameters (i.e., the ones parametrizing $\phi_m$, $\phi_e$, $\phi_h$ and $\phi_x$ in aforementioned Figures) of one message passing step for another step. In other words, one message passing `layer' can be used for multiple message passing steps. In this work, each network has five distinct message passing layers, out of which the middle three layers are repeated three times each. This provides an adequate balance between the number of parameters and the expressiveness of the network. This results in eleven message passing steps in total. This is close to the graph diameter, which is nine for our input graph\textcolor{mygreen}{, which means that in nine message passing steps, all nodes can communicate with all other nodes in the graph.}
	The message size is set to 64 in all networks and the embedding sizes (edge and node embedding) at 32. $\phi_m$, $\phi_e$, $\phi_h$ and $\phi_x$ each only have one layer and use the softplus activation function. This setup resulted in the following number of learnable parameters: 76351 for base GNN, 91465 parameters for EGNN, and 91145 for SimEGNN.

	The networks are implemented in PyTorch (\url{https://pytorch.org/}) and PyTorch Geometric (\url{https://pytorch-geometric.readthedocs.io/en/latest/}).
	For the EGNN and SimEGNN, each message passing step, described by Equations \eqref{eq:satorrasmessage} to \eqref{eq:satorrash} and Equations \eqref{eq:SimEGNN_m} and \eqref{eq:SimEGNN_dx}, can be implemented as two sequential message passing steps. In the first message passing step, node embeddings $\bt h_i$ are updated using computed messages $\bt m_{ij}$, which are aggregated into $\bt m_i$. In the second message passing step, $\vec x_i^{\;l}$ is updated using `messages' $\vec r_{ij}^{\;l} \tanh\left( \phi_x^l \left(\bt m^l_{ij}\right)\right)$ or $\vec r_{ij}^{\;l} \tanh\left( \phi_x^l \left(\varepsilon_{ij} \bt m^l_{ij}\right)\right)$ that are aggregated into $\Delta \vec x_i^{\;l}$. Accordingly, the networks are implemented such that each message passing layer contains two different PyTorch Geometric message passing layers that are alternated. (For the GNN, one type of message-passing layer is sufficient.)

\begin{figure}
        \includegraphics[width=\textwidth]{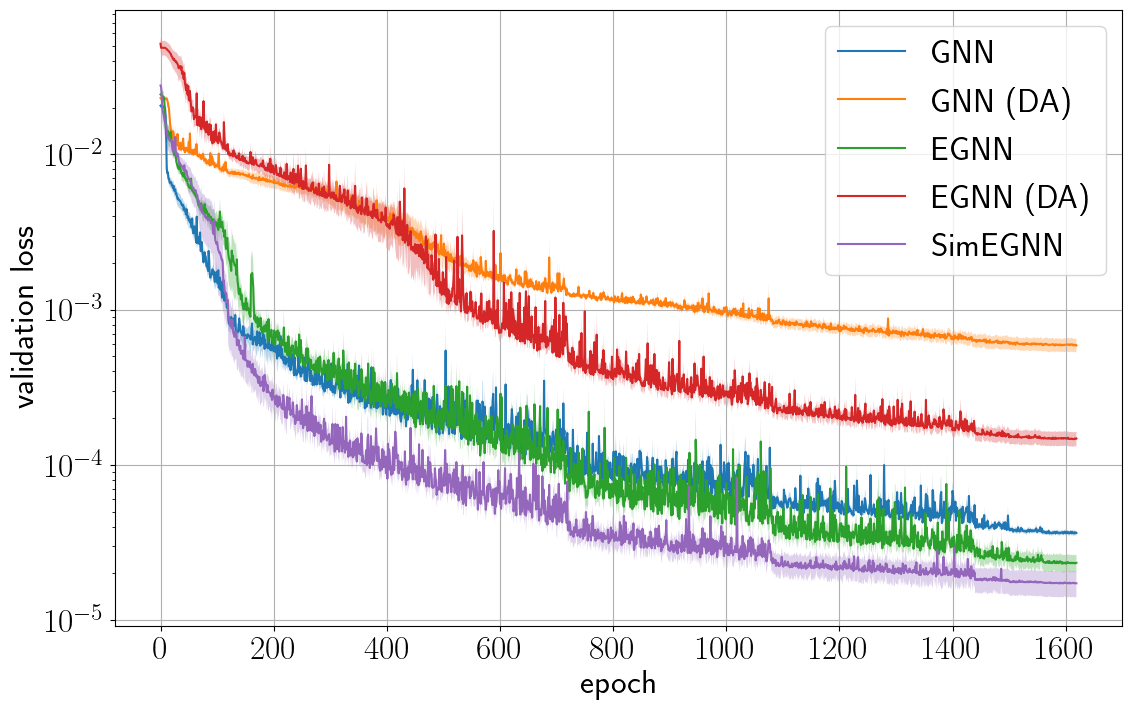}
        \caption{The \textcolor{mygreen}{mean} validation loss during training for each architecture. \textcolor{mygreen}{The mean is calculated over the 5 models resulting from the cross-validation. The standard error is indicated with a shaded area.}\label{fig:lossplot}}
    \end{figure}

	For training all learnable parameters, we use the Adam optimizer, with a schedule that reduces the learning rate in steps from \SI{2.5e-4} to \SI{2.5e-6}{}\textcolor{mygreen}{, see Table~\ref{tab:learningschedule}}. In total, we train for 1620 epochs, with a batch size of 12, which takes 6 to 10 hours per model on an NVIDIA A100-SXM4-40GB GPU. All outputs except the position of the nodes are scaled such that their target values have an average squared value of 1.
	The mean square error of each of the four outputs (position $\vec{w}$,  $\mathfrak{W}$, $\bt P$ and ${}^4 \bt D$) is used in the loss. The loss terms are weighted proportionally to the inverse of the square value of the target values (which for all loss terms except $\vec{w}$ was 1, because of the aforementioned scaling). \textcolor{mygreen}{We apply gradient clipping at 0.5 to prevent exploding gradients.}
	See Figure~\ref{fig:lossplot} for a plot of the validation loss during training for each model.

\begin{table}[h]
	\centering
	\caption{Learning rate schedule for training the models.}
	\label{tab:learningschedule}
	\textcolor{mygreen}{
	\begin{tabular}{ll}
		\toprule
		Epochs & Learning rate \\
		\midrule
		0-119 &     \SI{2.5e-4}{} \\
		120-719 &   \SI{1.0e-4}{} \\
		720-1079 &  \SI{5.0e-5}{} \\
		1080-1439 & \SI{2.5e-5}{} \\
		1440-1499 & \SI{1.0e-5}{} \\
		1500-1559 & \SI{5.0e-6}{} \\
		1560-1619 & \SI{2.5e-6}{} \\
		\bottomrule
	\end{tabular}
	}
\end{table}

\section{Full results}
\label{sec:fullresultsapp}
Tables \ref{tab:resultsy}, \ref{tab:resultsW}, \ref{tab:resultsP}, \ref{tab:resultsD} show, for each target quantity respectively, the fraction of variance unexplained (FVU) obtained by each neural network architecture for each test case.
Tables \ref{tab:resultsfroby}, \ref{tab:resultsfrobW}, \ref{tab:resultsfrobP}, \ref{tab:resultsfrobD} show the relative error, which we define as the mean Frobenius norm of the error divided by the mean Frobenius norm of the target
\begin{equation}
	\frac{
		\frac{1}{N} \sum_i^N
		\left\|\begin{matrix} \bt A_{\textrm{pred},i}- \bt A_{\textrm{target},i} \end{matrix}\right\|_F
	}{
		\frac{1}{N} \sum_i^N
		\left\|\begin{matrix} \bt A_{\textrm{target},i} \end{matrix}\right\|_F
	} \times 100\%,
\end{equation}
where $\bt A_{\textrm{pred},i}$ is some predicted tensor, $\bt A_{\textrm{target},i}$ is the ground truth value of this tensor, $\left\|\begin{matrix} ... \end{matrix}\right\|_F$ is the Frobenius norm and $N$ is the number of targets in the validation data, which for the global quantities $\mathfrak{W}$, $\bt P$, ${}^4 \bt D$ is equal to the number of graphs, but for the local quantity $\vec{w}$ is equal to the total number of nodes (number of graphs $\times$ number of nodes per graph). For a scalar quantity such as $\mathfrak{W}$, this reduces to the root mean squared error divided by the root mean square target value.

\begin{figure}
        \includegraphics[width=\textwidth]{"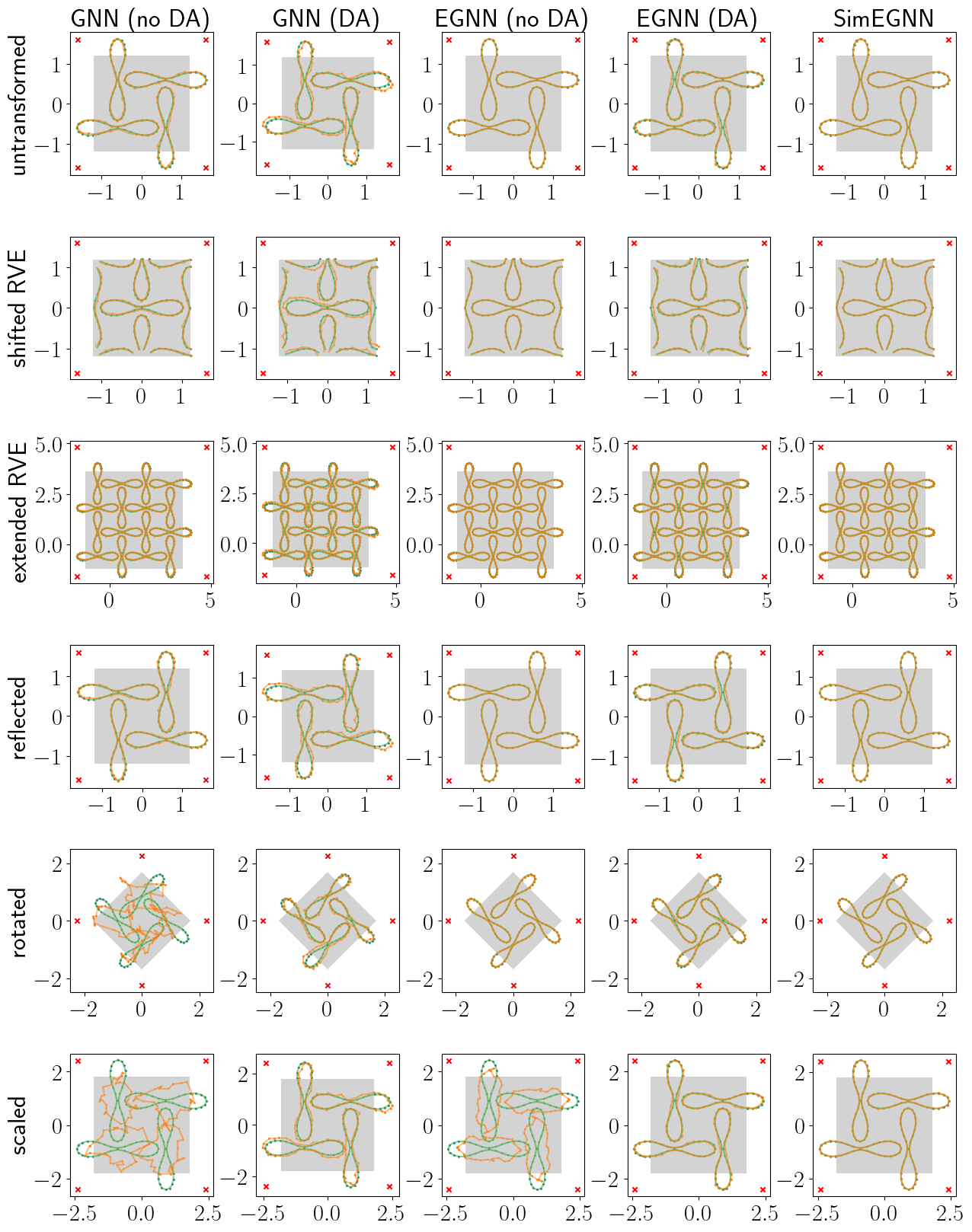"}
        \caption{Predicted deformation of the hole boundaries (orange) compared to the FEM ground truth (green), for $\bt F=\begin{bmatrix}0.75 & 0\\ 0 & 0.75\end{bmatrix}$ (biaxial compression, resulting in a rotational pattern). The grey square indicates how $\bt F$ deforms the square RVE through an affine transformation. The red crosses indicate the original positions of the corners.\label{fig:deformations0}}
    \end{figure}

\begin{figure}
        \includegraphics[width=\textwidth]{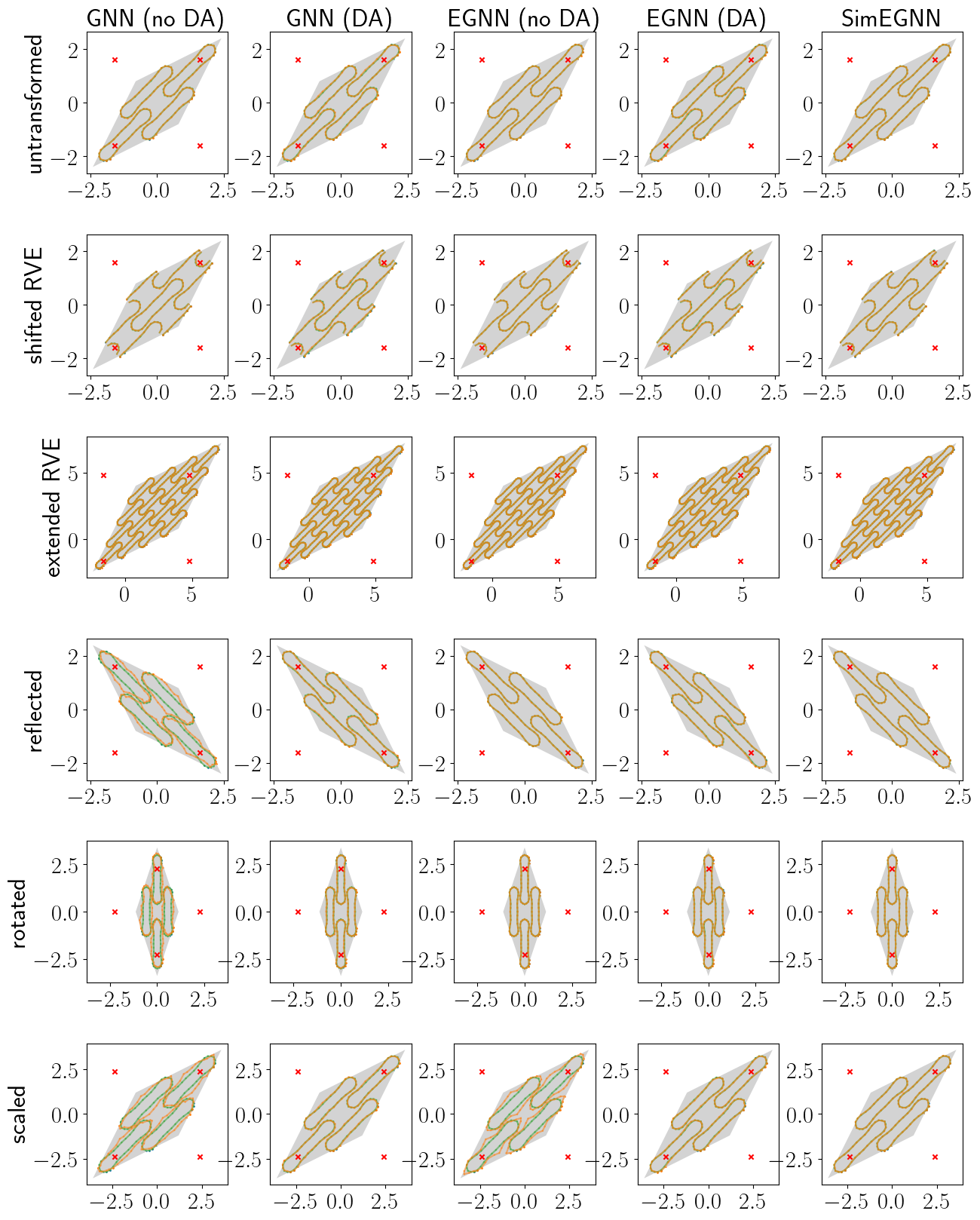}
        \caption{Model-predicted deformation (orange) compared to the FEM ground truth (green), for $\bt F=\begin{bmatrix}1 & 0.5\\ 0.5 & 1\end{bmatrix}$ (pure shear). The grey square indicates how $\bt F$ deforms the square RVE through an affine transformation. The red crosses indicate the original positions of the corners.
		\label{fig:deformations2}}
    \end{figure}

\begin{figure}
        \includegraphics[width=\textwidth]{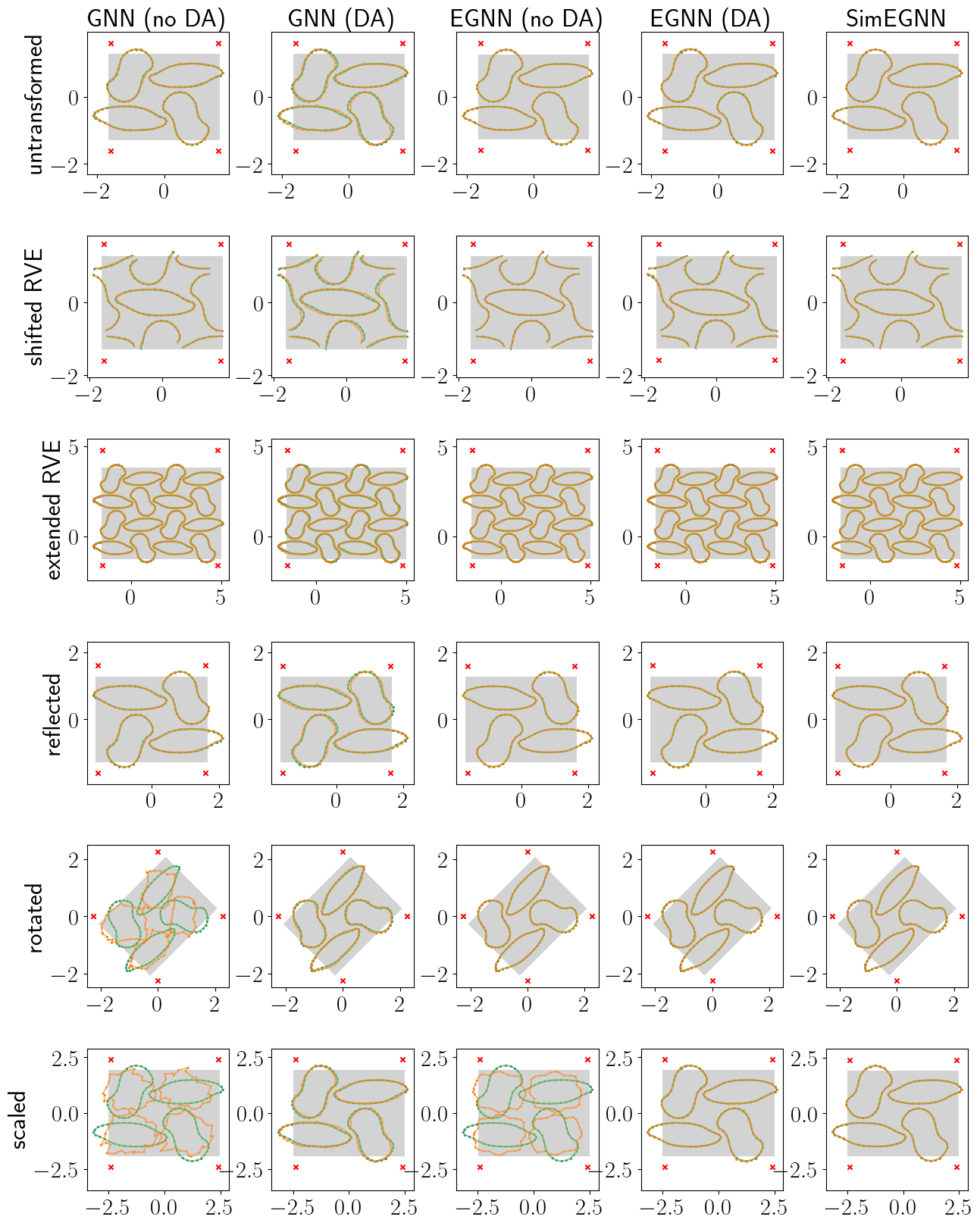}
        \caption{Model-predicted deformation (orange) compared to the FEM ground truth (green), for $\bt F=\begin{bmatrix}1.04 & 0\\ 0 & 0.8\end{bmatrix}$ (combination of tension and compression, resulting in left/right bifurcation as well as rotational bifurcation). The grey square indicates how $\bt F$ deforms the square RVE through an affine transformation. The red crosses indicate the original positions of the corners.\label{fig:deformations3}}
    \end{figure}

\begin{figure}
        \includegraphics[width=\textwidth]{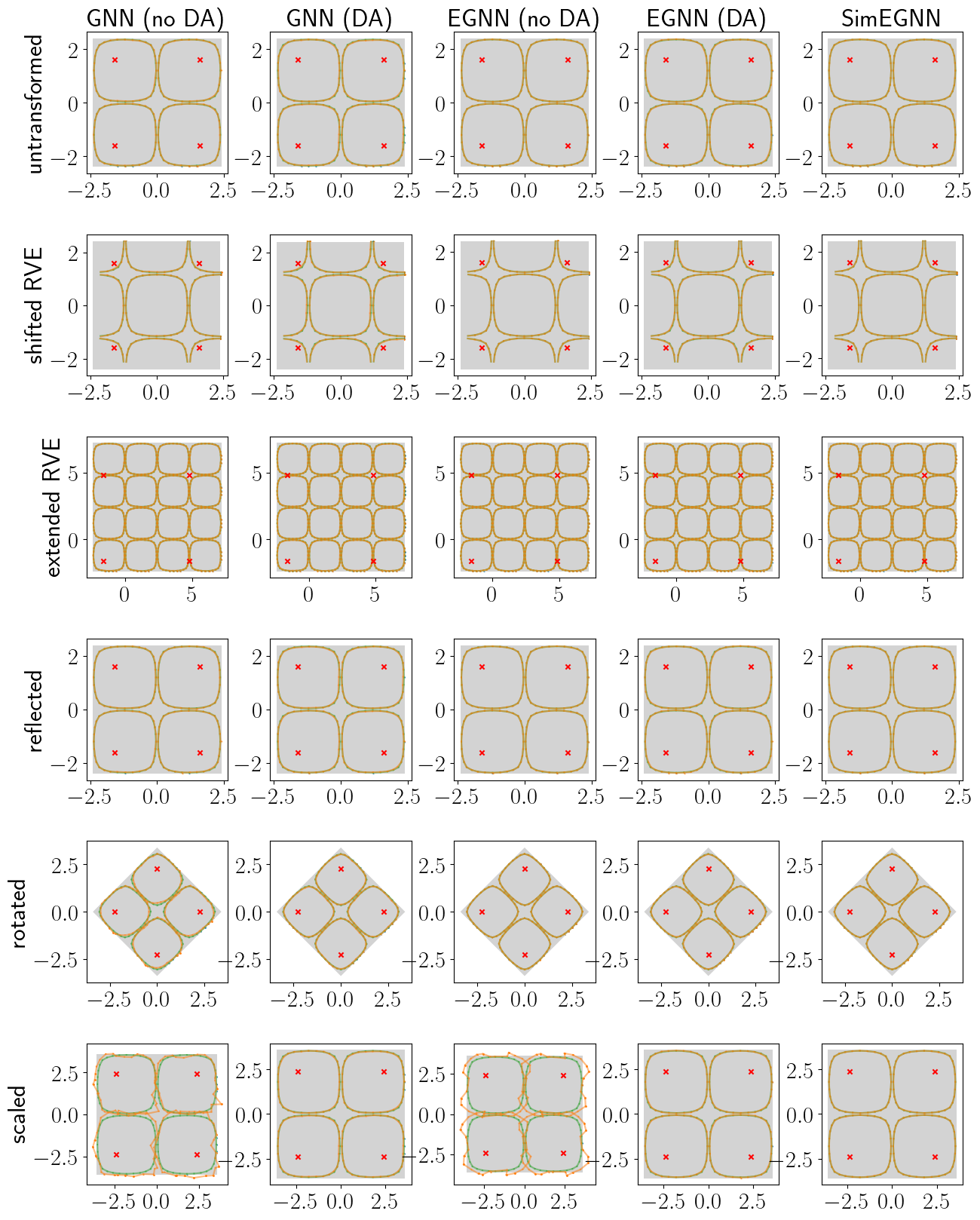}
        \caption{Model-predicted deformation (orange) compared to the FEM ground truth (green), for $\bt F=\begin{bmatrix}1.5 & 0\\ 0 & 1.5\end{bmatrix}$ (biaxial tension). The grey square indicates how $\bt F$ deforms the square RVE through an affine transformation. The red crosses indicate the original positions of the corners.\label{fig:deformations4}}
    \end{figure}

\begin{figure}
        \includegraphics[width=\textwidth]{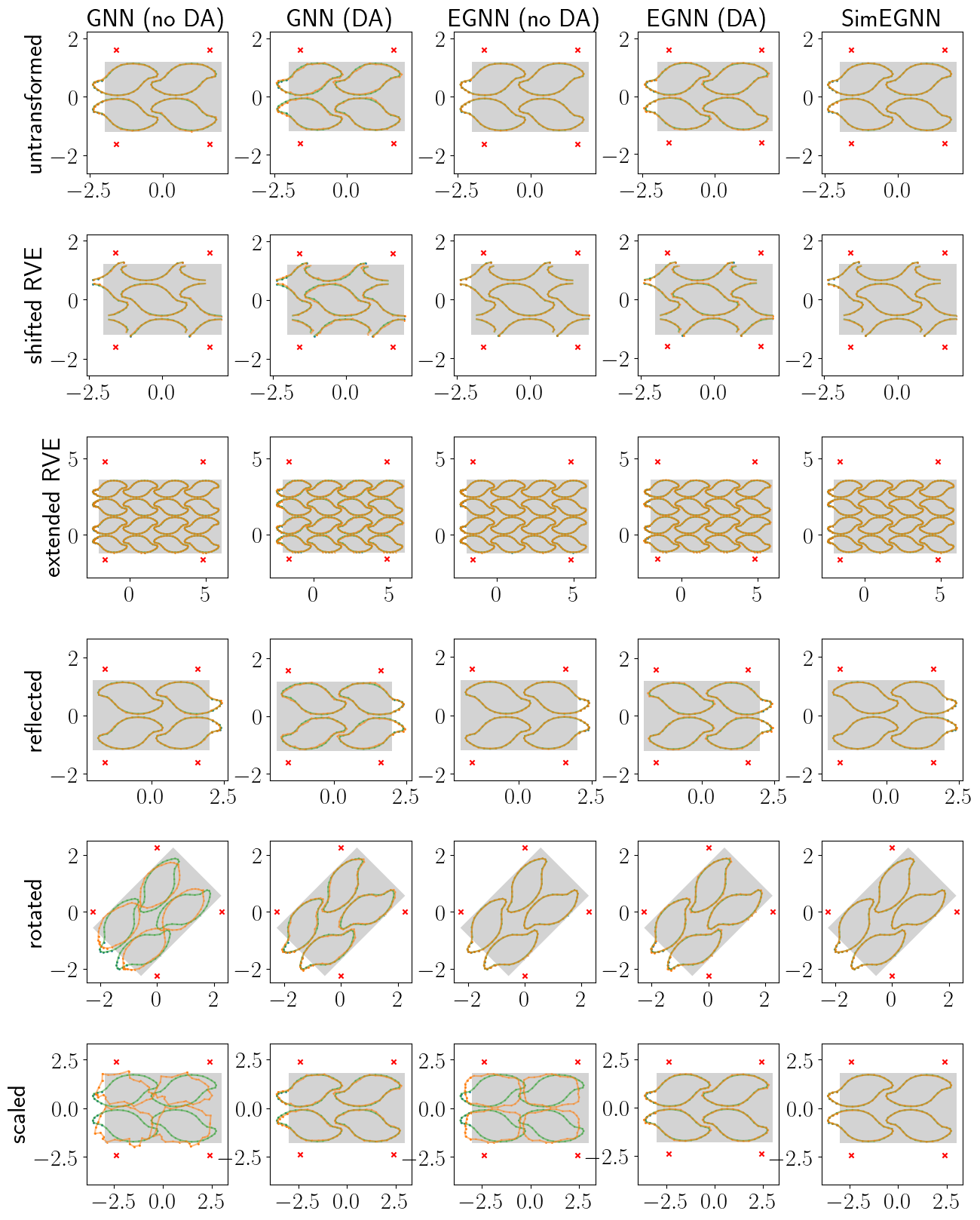}
        \caption{Model-predicted deformation (orange) compared to the FEM ground truth (green), for $\bt F=\begin{bmatrix}1.25 & 0\\ 0 & 0.75\end{bmatrix}$ (combination of tension and compression, resulting in left/right bifurcation). The grey square indicates how $\bt F$ deforms the square RVE through an affine transformation. The red crosses indicate the original positions of the corners.\label{fig:deformations5}}
    \end{figure}

In addition to Figure~\ref{fig:deformations1} in Section~\ref{sec:results}, Figures \ref{fig:deformations0}, \ref{fig:deformations2}, \ref{fig:deformations3}, \ref{fig:deformations4} and \ref{fig:deformations5} show
for all test cases a comparison of predicted deformations from all of the networks against FEM outputs. Each figure shows a different loading $\bt F$, such that all bifurcation patterns are included.

\begin{table}
\centering
\tiny
\caption{Mean FVU of the microfluctuation $\vec{w}$\label{tab:resultsy}}
\begin{tabular}{llllll}
\toprule
{} &                  GNN &           GNN (DA) &                 EGNN &          EGNN (DA) &                       SimEGNN \\
\midrule
untransformed &  $0.0034 \pm 0.0003$ &  $0.020 \pm 0.001$ &  $0.0019 \pm 0.0004$ &  $0.011 \pm 0.002$ &  $\mathbf{0.0016 \pm 0.0004}$ \\
shifted RVE   &  $0.0034 \pm 0.0003$ &  $0.020 \pm 0.001$ &  $0.0019 \pm 0.0004$ &  $0.011 \pm 0.002$ &  $\mathbf{0.0016 \pm 0.0004}$ \\
extended RVE  &  $0.0034 \pm 0.0003$ &  $0.020 \pm 0.001$ &  $0.0019 \pm 0.0004$ &  $0.011 \pm 0.002$ &  $\mathbf{0.0016 \pm 0.0004}$ \\
reflected     &      $0.21 \pm 0.03$ &  $0.020 \pm 0.001$ &  $0.0019 \pm 0.0004$ &  $0.011 \pm 0.002$ &  $\mathbf{0.0016 \pm 0.0004}$ \\
rotated       &      $0.82 \pm 0.04$ &  $0.019 \pm 0.001$ &  $0.0019 \pm 0.0004$ &  $0.011 \pm 0.002$ &  $\mathbf{0.0016 \pm 0.0004}$ \\
scaled        &      $0.93 \pm 0.11$ &  $0.019 \pm 0.001$ &        $1.8 \pm 0.8$ &  $0.011 \pm 0.002$ &  $\mathbf{0.0016 \pm 0.0004}$ \\
\bottomrule
\end{tabular}

\bigskip

\centering
\tiny
\caption{Mean FVU of the strain energy density $\mathfrak{W}$\label{tab:resultsW}}
\begin{tabular}{llllll}
\toprule
{} &                                          GNN &                                     GNN (DA) &                                         EGNN &                                    EGNN (DA) &                                               SimEGNN \\
\midrule
untransformed &  $\left(4.50 \pm 0.87\right) \times 10^{-5}$ &  $\left(9.34 \pm 1.44\right) \times 10^{-4}$ &  $\left(4.16 \pm 0.57\right) \times 10^{-5}$ &  $\left(2.30 \pm 0.46\right) \times 10^{-4}$ &  $\mathbf{\left(1.51 \pm 0.25\right) \times 10^{-5}}$ \\
shifted RVE   &  $\left(4.50 \pm 0.87\right) \times 10^{-5}$ &  $\left(9.34 \pm 1.44\right) \times 10^{-4}$ &  $\left(4.16 \pm 0.57\right) \times 10^{-5}$ &  $\left(2.30 \pm 0.46\right) \times 10^{-4}$ &  $\mathbf{\left(1.51 \pm 0.25\right) \times 10^{-5}}$ \\
extended RVE  &  $\left(4.50 \pm 0.87\right) \times 10^{-5}$ &  $\left(9.34 \pm 1.44\right) \times 10^{-4}$ &  $\left(4.16 \pm 0.57\right) \times 10^{-5}$ &  $\left(2.30 \pm 0.46\right) \times 10^{-4}$ &  $\mathbf{\left(1.51 \pm 0.25\right) \times 10^{-5}}$ \\
reflected     &                            $0.051 \pm 0.035$ &  $\left(9.70 \pm 1.48\right) \times 10^{-4}$ &  $\left(4.16 \pm 0.57\right) \times 10^{-5}$ &  $\left(2.30 \pm 0.46\right) \times 10^{-4}$ &  $\mathbf{\left(1.51 \pm 0.25\right) \times 10^{-5}}$ \\
rotated       &                            $0.069 \pm 0.021$ &  $\left(8.45 \pm 1.40\right) \times 10^{-4}$ &  $\left(4.16 \pm 0.57\right) \times 10^{-5}$ &  $\left(2.30 \pm 0.46\right) \times 10^{-4}$ &  $\mathbf{\left(1.51 \pm 0.25\right) \times 10^{-5}}$ \\
scaled        &                                  $48 \pm 11$ &  $\left(9.87 \pm 2.12\right) \times 10^{-4}$ &     $\left(1.2 \pm 0.5\right) \times 10^{2}$ &  $\left(2.55 \pm 0.48\right) \times 10^{-4}$ &  $\mathbf{\left(1.51 \pm 0.25\right) \times 10^{-5}}$ \\
\bottomrule
\end{tabular}

\bigskip

\centering
\tiny
\caption{Mean FVU of the first Piola-Kirchhoff stress tensor $\textbf{\textrm{P}}$\label{tab:resultsP}}
\begin{tabular}{llllll}
\toprule
{} &                                          GNN &             GNN (DA) &                                         EGNN &                                    EGNN (DA) &                                               SimEGNN \\
\midrule
untransformed &  $\left(4.85 \pm 0.79\right) \times 10^{-5}$ &  $0.0055 \pm 0.0003$ &  $\left(4.79 \pm 0.87\right) \times 10^{-5}$ &  $\left(5.39 \pm 1.17\right) \times 10^{-4}$ &  $\mathbf{\left(2.90 \pm 0.48\right) \times 10^{-5}}$ \\
shifted RVE   &  $\left(4.85 \pm 0.79\right) \times 10^{-5}$ &  $0.0055 \pm 0.0003$ &  $\left(4.79 \pm 0.87\right) \times 10^{-5}$ &  $\left(5.39 \pm 1.17\right) \times 10^{-4}$ &  $\mathbf{\left(2.90 \pm 0.48\right) \times 10^{-5}}$ \\
extended RVE  &  $\left(4.85 \pm 0.79\right) \times 10^{-5}$ &  $0.0055 \pm 0.0003$ &  $\left(4.79 \pm 0.87\right) \times 10^{-5}$ &  $\left(5.39 \pm 1.17\right) \times 10^{-4}$ &  $\mathbf{\left(2.90 \pm 0.48\right) \times 10^{-5}}$ \\
reflected     &                            $0.093 \pm 0.016$ &  $0.0041 \pm 0.0007$ &  $\left(3.74 \pm 0.74\right) \times 10^{-5}$ &  $\left(4.22 \pm 0.99\right) \times 10^{-4}$ &  $\mathbf{\left(2.24 \pm 0.37\right) \times 10^{-5}}$ \\
rotated       &                              $0.19 \pm 0.01$ &  $0.0065 \pm 0.0002$ &  $\left(5.24 \pm 0.99\right) \times 10^{-5}$ &  $\left(5.90 \pm 1.31\right) \times 10^{-4}$ &  $\mathbf{\left(3.16 \pm 0.52\right) \times 10^{-5}}$ \\
scaled        &                                   $14 \pm 3$ &  $0.0055 \pm 0.0003$ &                                  $40 \pm 11$ &  $\left(5.33 \pm 1.16\right) \times 10^{-4}$ &  $\mathbf{\left(2.90 \pm 0.48\right) \times 10^{-5}}$ \\
\bottomrule
\end{tabular}

\bigskip

\centering
\tiny
\caption{Mean FVU of the stiffness tensor $\textbf{\textrm{D}}$\label{tab:resultsD}}
\begin{tabular}{llllll}
\toprule
{} &                                          GNN &           GNN (DA) &                                         EGNN &            EGNN (DA) &                                               SimEGNN \\
\midrule
untransformed &  $\left(6.60 \pm 0.89\right) \times 10^{-4}$ &  $0.053 \pm 0.010$ &  $\left(7.91 \pm 1.15\right) \times 10^{-4}$ &  $0.0057 \pm 0.0008$ &  $\mathbf{\left(4.19 \pm 1.27\right) \times 10^{-4}}$ \\
shifted RVE   &  $\left(6.60 \pm 0.89\right) \times 10^{-4}$ &  $0.053 \pm 0.010$ &  $\left(7.91 \pm 1.15\right) \times 10^{-4}$ &  $0.0057 \pm 0.0008$ &  $\mathbf{\left(4.19 \pm 1.27\right) \times 10^{-4}}$ \\
extended RVE  &  $\left(6.60 \pm 0.89\right) \times 10^{-4}$ &  $0.053 \pm 0.010$ &  $\left(7.91 \pm 1.15\right) \times 10^{-4}$ &  $0.0057 \pm 0.0008$ &  $\mathbf{\left(4.19 \pm 1.27\right) \times 10^{-4}}$ \\
reflected     &                            $0.071 \pm 0.006$ &  $0.044 \pm 0.013$ &  $\left(6.45 \pm 1.00\right) \times 10^{-4}$ &  $0.0047 \pm 0.0007$ &  $\mathbf{\left(3.38 \pm 1.00\right) \times 10^{-4}}$ \\
rotated       &                              $0.55 \pm 0.03$ &  $0.057 \pm 0.014$ &  $\left(9.80 \pm 1.53\right) \times 10^{-4}$ &  $0.0071 \pm 0.0011$ &  $\mathbf{\left(5.15 \pm 1.53\right) \times 10^{-4}}$ \\
scaled        &                                $1.9 \pm 0.4$ &  $0.051 \pm 0.009$ &   $\left(1.02 \pm 1.01\right) \times 10^{4}$ &  $0.0055 \pm 0.0008$ &  $\mathbf{\left(4.19 \pm 1.27\right) \times 10^{-4}}$ \\
\bottomrule
\end{tabular}

\bigskip

\end{table}

\begin{table}
\centering
\tiny
\caption{Mean relative error in percentage of the microfluctuation $\vec{w}$\label{tab:resultsfroby}}
\begin{tabular}{llllll}
\toprule
{} &                                       GNN &    GNN (DA) &                                      EGNN &      EGNN (DA) &                 SimEGNN \\
\midrule
untransformed &                             $5.4 \pm 0.2$ &  $14 \pm 0$ &                             $3.7 \pm 0.4$ &     $10 \pm 1$ &  $\mathbf{3.3 \pm 0.4}$ \\
shifted RVE   &                             $5.4 \pm 0.2$ &  $14 \pm 0$ &                             $3.7 \pm 0.4$ &     $10 \pm 1$ &  $\mathbf{3.3 \pm 0.4}$ \\
extended RVE  &                             $5.4 \pm 0.2$ &  $14 \pm 0$ &                             $3.7 \pm 0.4$ &     $10 \pm 1$ &  $\mathbf{3.3 \pm 0.4}$ \\
reflected     &                                $37 \pm 2$ &  $14 \pm 0$ &                             $3.7 \pm 0.4$ &     $10 \pm 1$ &  $\mathbf{3.3 \pm 0.4}$ \\
rotated       &                                $90 \pm 3$ &  $14 \pm 0$ &                             $3.9 \pm 0.4$ &     $11 \pm 1$ &  $\mathbf{3.4 \pm 0.4}$ \\
scaled        &  $\left(1.0 \pm 0.1\right) \times 10^{2}$ &  $14 \pm 0$ &  $\left(1.4 \pm 0.3\right) \times 10^{2}$ &  $9.9 \pm 0.6$ &  $\mathbf{3.3 \pm 0.4}$ \\
\bottomrule
\end{tabular}

\bigskip

\centering
\tiny
\caption{Mean relative error in percentage of the strain energy density $\mathfrak{W}$\label{tab:resultsfrobW}}
\begin{tabular}{llllll}
\toprule
{} &                                       GNN &       GNN (DA) &                                        EGNN &      EGNN (DA) &                   SimEGNN \\
\midrule
untransformed &                           $0.58 \pm 0.04$ &  $3.0 \pm 0.3$ &                             $0.60 \pm 0.05$ &  $1.5 \pm 0.2$ &  $\mathbf{0.36 \pm 0.04}$ \\
shifted RVE   &                           $0.58 \pm 0.04$ &  $3.0 \pm 0.3$ &                             $0.60 \pm 0.05$ &  $1.5 \pm 0.2$ &  $\mathbf{0.36 \pm 0.04}$ \\
extended RVE  &                           $0.58 \pm 0.04$ &  $3.0 \pm 0.3$ &                             $0.60 \pm 0.05$ &  $1.5 \pm 0.2$ &  $\mathbf{0.36 \pm 0.04}$ \\
reflected     &                                $14 \pm 5$ &  $3.2 \pm 0.4$ &                             $0.60 \pm 0.05$ &  $1.5 \pm 0.2$ &  $\mathbf{0.36 \pm 0.04}$ \\
rotated       &                                $24 \pm 3$ &  $3.0 \pm 0.4$ &                             $0.60 \pm 0.05$ &  $1.5 \pm 0.2$ &  $\mathbf{0.36 \pm 0.04}$ \\
scaled        &  $\left(7.4 \pm 0.9\right) \times 10^{2}$ &  $3.0 \pm 0.4$ &  $\left(1.10 \pm 0.32\right) \times 10^{3}$ &  $1.6 \pm 0.2$ &  $\mathbf{0.36 \pm 0.04}$ \\
\bottomrule
\end{tabular}

\bigskip

\centering
\tiny
\caption{Mean relative error in percentage of the first Piola-Kirchhoff stress tensor $\textbf{\textrm{P}}$\label{tab:resultsfrobP}}
\begin{tabular}{llllll}
\toprule
{} &                                       GNN &       GNN (DA) &                                      EGNN &      EGNN (DA) &                   SimEGNN \\
\midrule
untransformed &                           $0.64 \pm 0.04$ &  $7.3 \pm 0.3$ &                           $0.65 \pm 0.07$ &  $2.3 \pm 0.3$ &  $\mathbf{0.48 \pm 0.04}$ \\
shifted RVE   &                           $0.64 \pm 0.04$ &  $7.3 \pm 0.3$ &                           $0.65 \pm 0.07$ &  $2.3 \pm 0.3$ &  $\mathbf{0.48 \pm 0.04}$ \\
extended RVE  &                           $0.64 \pm 0.04$ &  $7.3 \pm 0.3$ &                           $0.65 \pm 0.07$ &  $2.3 \pm 0.3$ &  $\mathbf{0.48 \pm 0.04}$ \\
reflected     &                                $23 \pm 2$ &  $7.0 \pm 0.8$ &                           $0.65 \pm 0.07$ &  $2.3 \pm 0.3$ &  $\mathbf{0.48 \pm 0.04}$ \\
rotated       &                                $43 \pm 2$ &  $7.5 \pm 0.3$ &                           $0.65 \pm 0.07$ &  $2.3 \pm 0.3$ &  $\mathbf{0.48 \pm 0.04}$ \\
scaled        &  $\left(3.8 \pm 0.4\right) \times 10^{2}$ &  $7.2 \pm 0.3$ &  $\left(5.6 \pm 0.7\right) \times 10^{2}$ &  $2.2 \pm 0.3$ &  $\mathbf{0.48 \pm 0.04}$ \\
\bottomrule
\end{tabular}

\bigskip

\centering
\tiny
\caption{Mean relative error in percentage of the stiffness tensor $\textbf{\textrm{D}}$\label{tab:resultsfrobD}}
\begin{tabular}{llllll}
\toprule
{} &                                       GNN &    GNN (DA) &                                        EGNN &      EGNN (DA) &                   SimEGNN \\
\midrule
untransformed &                             $1.3 \pm 0.1$ &  $19 \pm 2$ &                               $1.5 \pm 0.1$ &  $5.0 \pm 0.4$ &  $\mathbf{0.97 \pm 0.10}$ \\
shifted RVE   &                             $1.3 \pm 0.1$ &  $19 \pm 2$ &                               $1.5 \pm 0.1$ &  $5.0 \pm 0.4$ &  $\mathbf{0.97 \pm 0.10}$ \\
extended RVE  &                             $1.3 \pm 0.1$ &  $19 \pm 2$ &                               $1.5 \pm 0.1$ &  $5.0 \pm 0.4$ &  $\mathbf{0.97 \pm 0.10}$ \\
reflected     &                                $17 \pm 1$ &  $18 \pm 3$ &                               $1.5 \pm 0.1$ &  $5.0 \pm 0.4$ &  $\mathbf{0.97 \pm 0.10}$ \\
rotated       &                                $58 \pm 2$ &  $17 \pm 2$ &                               $1.5 \pm 0.1$ &  $5.0 \pm 0.4$ &  $\mathbf{0.97 \pm 0.10}$ \\
scaled        &  $\left(1.2 \pm 0.1\right) \times 10^{2}$ &  $18 \pm 2$ &  $\left(1.36 \pm 1.08\right) \times 10^{3}$ &  $4.9 \pm 0.4$ &  $\mathbf{0.97 \pm 0.10}$ \\
\bottomrule
\end{tabular}

\bigskip

\end{table}

\newpage

\clearpage

	\bibliographystyle{elsarticle-num}
	\bibliography{library}{}






\end{document}